\documentclass[12pt, a4paper]{article}
\usepackage[margin=2cm]{geometry}

    \usepackage{graphicx}%
    \usepackage{multirow}%
    \usepackage{amsmath,amssymb,amsfonts}%
    \usepackage{amsthm}%
    \usepackage{mathrsfs}%
    \usepackage{xcolor}%
    \usepackage{textcomp}%
    \usepackage{booktabs}%
    \usepackage{algorithm}%
    \usepackage{algorithmicx}%
    \usepackage{algpseudocode}%
    \usepackage{listings}%
    \usepackage[numbers]{natbib}

    \usepackage{paper-style}
    \newenvironment{keyword}{\noindent\textbf{Keywords: }}{\par}
    
    \usepackage{hyperref}           
    \usepackage{authblk} 

    \usepackage{local-colors}
    \usepackage{stem-operators}

    \graphicspath{%
        {img/}
    }

\newglossaryentry{i}{
    type=hidden,
    name={\ensuremath{i}}
}
\newglossaryentry{j}{
    type=hidden,
    name={\ensuremath{j}}
}
\newglossaryentry{k}{
    type=hidden,
    name={\ensuremath{k}}
}
\newglossaryentry{p}{
    type=hidden,
    name={\ensuremath{p}}
}
\newglossaryentry{q}{
    type=hidden,
    name={\ensuremath{q}}
}

\newglossaryentry{radial}{
    type=hidden,
    name={\ensuremath{r}},
    description={Radial coordinate of the cylindric coordinate system}
}
\newglossaryentry{azimuthal}{
    type=hidden,
    name={\ensuremath{\theta}},
    description={Circunferential coordinate of the cylindric coordinate system}
}
\newglossaryentry{z}{
    type=hidden,
    name={\ensuremath{z}},
    description={Axial coordinate of the cylindric coordinate system}
}

\newglossaryentry{xAxis}{
    type=hidden,
    name={\ensuremath{x}},
    description={%
        First coordinate of the Cartesian system (normally, $x_1$)
    }
}
\newglossaryentry{yAxis}{
    type=hidden,
    name={\ensuremath{y}},
    description={%
        Second coordinate of the Cartesian system (normally, $x_2$)
    }
}
\newglossaryentry{zAxis}{
    type=hidden,
    name={\ensuremath{z}},
    description={%
        Third coordinate of the Cartesian system (normally, $x_3$)
    }
}
\newglossaryentry{xAxisUnitVector}{
    type=hidden,
    name={\ensuremath{\vec{i}}},
    description={%
    }
}
\newglossaryentry{yAxisUnitVector}{
    type=hidden,
    name={\ensuremath{\vec{j}}},
    description={%
    }
}
\newglossaryentry{zAxisUnitVector}{
    type=hidden,
    name={\ensuremath{\vec{k}}},
    description={%
    }
}

\newglossaryentry{genericUnitVector}{
    type=hidden,
    name={\ensuremath{\vec{e}}}
}

\newglossaryentry{average}{
    type=subscript,
    name={\ensuremath{avg}},
    description={Indicates a time and surface averaged variable}
}
\newglossaryentry{maximum}{
    type=subscript,
    name={\ensuremath{max}},
    description={Indicates the maximum value of a sample}
}
\newglossaryentry{minimum}{
    type=subscript,
    name={\ensuremath{min}},
    description={Indicates the minimum value of a sample}
}
\newglossaryentry{percentil95}{
    type=subscript,
    name={\ensuremath{95}},
    description={Indicates the 95th percentile of a distribution}
}
\newglossaryentry{percentil99}{
    type=subscript,
    name={\ensuremath{99}},
    description={Indicates the 99th percentile of a distribution}
}
\newglossaryentry{solid}{
    type=superscript,
    name={\ensuremath{\mathrm{s}}},
    description={Used for solid-related variables}
}
\newglossaryentry{fluid}{
    type=superscript,
    name={\ensuremath{\mathrm{f}}},
    description={Used for fluid-related variables}
}
\newglossaryentry{fs}{
    type=superscript,
    name={\ensuremath{\mathrm{fs}}},
    description={Used for variables related to the fluid-solid interface}
}
\newglossaryentry{zeroTime}{
    type=subscript,
    name={\ensuremath{\mathrm{0}}},
    description={Indicates variables at the initial instant $t = 0$}
}
\newglossaryentry{timeIndex}{
    type=superscript,
    name={\ensuremath{\mathrm{n}}},
    description={Time index}
}
\newglossaryentry{fsiIteration}{
    type=subscript,
    name={\ensuremath{k}},
    description={%
        Index for a coupling iteration of the outer loop of the IQN-ILS
        algorithm
    }
}
\newglossaryentry{iteration}{
    type=hidden,
    name={\ensuremath{m}},
    description={%
        Index for iterations when solving a linear system  $\matrix{A}
        \cvec{x} = \cvec{b}$ with an iterative method
    }
}
\newglossaryentry{normal}{
    type=subscript,
    name={\ensuremath{n}},
    description={%
        Indicate the normal component of a vector to a surface, i.e.
        $\vec{u}_{n} = \vec{n}(\vec{n} \dprod \vec{u})$
    }
}
\newglossaryentry{tangential}{
    type=subscript,
    name={\ensuremath{t}},
    description={%
        Indicate the tangential component of a vector to a surface, i.e.
        $\vec{u}_{t} = (\tensor{I} - \vec{n}\vec{n}) \dprod \vec{u}$
    }
}
\newglossaryentry{boundary}{
    type=subscript,
    name={\ensuremath{b}},
    description={%
        Indicate a property on a boundary of a discretized domain in the
        context of the Finite Volume Method.
    }
}

\newglossaryentry{time}{
    type=latin,
    name={\ensuremath{t}},
    description={Time},
    unit={\si{\second}}
}
\newglossaryentry{positionVector}{
    type=latin,
    name={\ensuremath{\vec{r}}},
    description={Generic position vector},
    unit={\si{\meter}}
}
\newglossaryentry{temperature}{
    type=latin,
    name={\ensuremath{T}},
    description={Temperature},
    unit={\si{\kelvin}}
}
\newglossaryentry{deltat}{
    type=hidden,
    name={\ensuremath{\Delta \gls{time}}},
    description={Time-step},
    unit={\si{\second}}
}
\newglossaryentry{mass}{
    type=latin,
    name={\ensuremath{m}},
    description={Mass},
    unit={\si{\kilo\gram}}
}
\newglossaryentry{length}{
    type=latin,
    name={\ensuremath{L}},
    description={Length},
    unit={\si{\meter}}
}
\newglossaryentry{curveLength}{
    type=latin,
    name={\ensuremath{\vec{s}}},
    description={Vector length along a curve in space},
    unit={\si{\meter}}
}
\newglossaryentry{curve}{
    type=latin,
    name={\ensuremath{C}},
    description={%
        Representation of a curve in space
    },
    unit={-}
}
\newglossaryentry{perimeter}{
    type=latin,
    name={\ensuremath{P}},
    description={Perimeter of a contour},
    unit={\si{\meter}}
}
\newglossaryentry{surfaceArea}{
    type=latin,
    name={\ensuremath{A}},
    description={Area of a surface or region},
    unit={\si{\square\meter}}
}
\newglossaryentry{volume}{
    type=latin,
    name={\ensuremath{V}},
    description={Volume of a region in spatial coordinates},
    unit={\si{\cubic\meter}}
}
\newglossaryentry{surface}{
    type=latin,
    name={\ensuremath{S}},
    description={%
        Surface of volume \gls{volume}, $\partial V$, in spatial coordinates
    },
    unit={\si{\square\meter}}
}
\newglossaryentry{density}{
    type=greek,
    name={\ensuremath{\rho}},
    description={Density},
    unit={\si{\kilo\gram\per\cubic\meter}}
}
\newglossaryentry{force}{
    type=latin,
    name={\ensuremath{\vec{f}}},
    description={Generic force vector},
    unit={\si{\newton}}
}
\newglossaryentry{linearMomentum}{
    type=latin,
    name={\ensuremath{\vec{p}}},
    description={Linear momentum vector},
    unit={\si{\kilo\gram\meter\per\second}}
}
\newglossaryentry{torque}{
    type=latin,
    name={\ensuremath{\vec{T}}},
    description={Torque},
    unit={\si{\newton\meter}}
}
\newglossaryentry{angularMomentum}{
    type=latin,
    name={\ensuremath{\vec{H}}},
    description={Angular momentum vector},
    unit={\si{\kilo\gram\square\meter\per\second}}
}
\newglossaryentry{forceComp}{
    type=hidden,
    name={\ensuremath{f}},
    description={Component of the force vector},
    unit={\si{\newton}}
}
\newglossaryentry{traction}{
    type=latin,
    name={\ensuremath{\vec{t}}},
    description={Generic traction vector},
    unit={\si{\newton\per\square\meter}}
}
\newglossaryentry{tractionComp}{
    type=hidden,
    name={\ensuremath{t}},
    description={Generic traction vector},
    unit={\si{\newton\per\square\meter}}
}
\newglossaryentry{CauchyStress}{
    type=greek,
    name={\ensuremath{\tensor{\sigma}}},
    description={Cauchy stress tensor},
    unit={\si{\pascal}}
}
\newglossaryentry{CauchyStressComp}{
    type=hidden,
    name={\ensuremath{\sigma}},
    description={Cauchy stress tensor component},
    unit={\si{\pascal}}
}
\newglossaryentry{bodyForcesPerVolume}{
    type=latin,
    name={\ensuremath{\vec{f}_b}},
    description={Body forces per unit volume},
    unit={\si{\newton\per\cubic\meter}}
}
\newglossaryentry{bodyForcesPerMass}{
    type=latin,
    name={\ensuremath{\vec{b}}},
    description={Body forces per unit mass},
    unit={\si{\newton\per\kilo\gram}}
}
\newglossaryentry{bodyForcesPerMassComp}{
    type=hidden,
    name={\ensuremath{{b}}},
    description={Body forces per unit mass component},
    unit={\si{\newton\per\kilo\gram}}
}
\newglossaryentry{gravity}{
    type=latin,
    name={\ensuremath{\vec{g}}},
    description={Gravitational acceleration vector},
    unit={\si{\meter\per\square\second}}
}
\newglossaryentry{gravityComp}{
    type=hidden,
    name={\ensuremath{g}},
    description={Gravitational acceleration vector},
    unit={\si{\meter\per\square\second}}
}

\newglossaryentry{particle}{
    type=greek,
    name={\ensuremath{\xi}},
    description={Representation of a material particle},
    unit={\si{-}}
}
\newglossaryentry{EulerCoord}{
    type=latin,
    name={\ensuremath{\vec{x}}},
    description={Eulerian or spatial coordinates system (Cartesian coordinates)},
    unit={\si{\meter}}
}
\newglossaryentry{meshCoord}{
    type=greek,
    name={\ensuremath{\gvec{\chi}}},
    description={%
        Referential coordinates system for the arbitrary Lagrangian-Eulerian
        description
    },
    unit={\si{\meter}}
}
\newglossaryentry{LagrangeCoord}{
    type=greek,
    name={\ensuremath{\gvec{\xi}}},
    description={Lagrangian or material coordinates system},
    unit={\si{\meter}}
}
\newglossaryentry{meshCoordComp}{
    type=hidden,
    name={\ensuremath{\chi}},
    description={Referential, ALE or mesh coordinates system components},
    unit={\si{\meter}}
}
\newglossaryentry{EulerCoordComp}{
    type=hidden,
    name={\ensuremath{x}},
    description={Eulerian or spatial coordinates components},
    unit={\si{\meter}}
}
\newglossaryentry{LagrangeCoordComp}{
    type=hidden,
    name={\ensuremath{\xi}},
    description={Lagrangian or material coordinates components},
    unit={\si{\meter}}
}

\newglossaryentry{domain}{
    type=greek,
    name={\ensuremath{\Omega}},
    description={Spatial solution domain},
    unit={\si{\cubic\meter}}
}
\newglossaryentry{domainBoundary}{
    type=greek,
    name={\ensuremath{\Lambda}},
    description={Boundary of the domain \gls{domain}},
    unit={\si{\square\meter}}
}
\newglossaryentry{contour}{
    type=greek,
    name={\ensuremath{\Gamma}},
    description={%
        Generic three-dimensional contour; boundary of an open surface
        \gls{surface}, $\partial \gls{surface}$, in spatial coordinates
    },
    unit={\si{\meter}}
}
\newglossaryentry{nablaSpatial}{
    type=other,
    name={\ensuremath{\nabla}},
    description={Nabla operator in Eulerian coordinates},
    unit={\si{\per\meter}}
}
\newglossaryentry{nablaMaterial}{
    type=other,
    name={\ensuremath{\nabla_0}},
    description={Nabla operator in the initial material coordinates},
    unit={\si{\per\meter}}
}
\newglossaryentry{materialDomain}{
    type=greek,
    name={\ensuremath{\Omega_{\xi}}},
    description={Solution domain as seen by the material coordinates},
    unit={\si{\cubic\meter}}
}
\newglossaryentry{meshDomain}{
    type=greek,
    name={\ensuremath{\Omega_{\chi}}},
    description={Solution domain as seen by the reference coordinates},
    unit={\si{\cubic\meter}}
}
\newglossaryentry{materialVolume}{
    type=latin,
    name={\ensuremath{V_{\xi}}},
    description={Volume of an amount of particles in material coordinates},
    unit={\si{\cubic\meter}}
}
\newglossaryentry{meshVolume}{
    type=latin,
    name={\ensuremath{V_{\chi}}},
    description={Control volume in ALE coordinates},
    unit={\si{\cubic\meter}}
}
\newglossaryentry{sNormalVector}{
    type=latin,
    name={\ensuremath{\versor{n}}},
    description={%
        Unit normal vector to the surface \gls{surface} pointing outwards
    },
    unit={\si{-}}
}
\newglossaryentry{sNormalVectorComp}{
    type=hidden,
    name={\ensuremath{n}},
    unit={\si{-}}
}

\newglossaryentry{funcMaterial}{
    type=latin,
    name={\ensuremath{f_{\xi}}},
    description={Function $f$ representing any scalar variable of the material},
    unit={\si{-}}
}
\newglossaryentry{funcSpatial}{
    type=latin,
    name={\ensuremath{f}},
    description={%
        Spatial function $f$ representing any scalar variable of the material
    },
    unit={\si{-}}
}
\newglossaryentry{funcMesh}{
    type=latin,
    name={\ensuremath{f_{\chi}}},
    description={%
        ALE description of a function $f$ representing any scalar variable of
        the material
    },
    unit={\si{-}}
}
\newglossaryentry{materialSpatialMap}{
    type=greek,
    name={\ensuremath{\gvec{\Phi}}},
    description={Mapping of material coordinates onto spatial coordinates},
    unit={\si{-}}
}
\newglossaryentry{meshSpatialMap}{
    type=greek,
    name={\ensuremath{\gvec{\Psi}}},
    description={Mapping of mesh coordinates onto spatial coordinates},
    unit={\si{-}}
}
\newglossaryentry{materialMeshMap}{
    type=greek,
    name={\ensuremath{\gvec{\Xi}}},
    description={Mapping of material coordinates onto mesh coordinates},
    unit={\si{-}}
}
\newglossaryentry{identityTensor}{
    type=latin,
    name={\ensuremath{\tensor{I}}},
    description={Second order identity tensor},
    unit={\si{-}}
}
\newglossaryentry{KroneckerDeltaComp}{
    type=hidden,
    name={\ensuremath{\delta}},
    description={Kronecker's delta symbol},
    unit={\si{-}}
}
\newglossaryentry{kroneckerDelta}{
    type=hidden,
    name={\ensuremath{\gls{KroneckerDeltaComp}_{ij}}},
    description={Kronecker's delta},
    unit={\si{-}}
}
\newglossaryentry{cartesianVersor}{
    type=hidden,
    name={\ensuremath{\versor{e}}},
    description={%
        Generic cartesian versor
    }
}
\newglossaryentry{permutationTensor}{
    type=hidden,
    name={\ensuremath{\epsilon_{ijk}}},
    description={Permutation or Levi-Civita third-order tensor},
    unit={\si{-}}
}
\newglossaryentry{permutationTensorComp}{
    type=hidden,
    name={\ensuremath{\epsilon}},
    description={Permutation's tensor symbol},
    unit={\si{-}}
}
\newglossaryentry{fourthIdentityTensor}{
    type=latin,
    name={\ensuremath{\fourthTensor{I}}},
    description={Fourth-order identity tensor},
    unit={\si{-}}
}

\newglossaryentry{Jacobian}{
    type=latin,
    name={\ensuremath{J}},
    description={Jacobian (determinant of the deformation gradient)},
    unit={\si{-}}
}
\newglossaryentry{materialJacobianMatrix}{
    type=latin,
    name={\ensuremath{\mathbf{J}_{\gls{materialSpatialMap}}}},
    description={Jacobian matrix of the material-spatial mapping},
    unit={\si{-}}
}
\newglossaryentry{meshJacobianMatrix}{
    type=latin,
    name={\ensuremath{\mathbf{J}_{\gls{meshSpatialMap}}}},
    description={Jacobian matrix of the mesh-spatial mapping},
    unit={\si{-}}
}
\newglossaryentry{materialJacobian}{
    type=latin,
    name={\ensuremath{\mathrm{J}_{\gls{materialSpatialMap}}}},
    description={Jacobian of the material-spatial mapping},
    unit={\si{-}}
}
\newglossaryentry{meshJacobian}{
    type=latin,
    name={\ensuremath{\mathrm{J}_{\gls{meshSpatialMap}}}},
    description={Jacobian of the mesh-spatial mapping},
    unit={\si{-}}
}

\newglossaryentry{velocity}{
    type=latin,
    name={\ensuremath{\vec{v}}},
    description={Material velocity},
    unit={\si{\meter\per\second}}
}
\newglossaryentry{acceleration}{
    type=latin,
    name={\ensuremath{\vec{a}}},
    description={Material acceleration},
    unit={\si{\meter\per\square\second}}
}
\newglossaryentry{angularVelocity}{
    type=latin,
    name={\ensuremath{\gvec{\omega}}},
    description={Angular velocity},
    unit={\si{\radian\per\second}}
}
\newglossaryentry{velocityComp}{
    type=hidden,
    name={\ensuremath{v}},
    description={Material velocity},
    unit={\si{\meter\per\second}}
}
\newglossaryentry{angularVelocityComp}{
    type=hidden,
    name={\ensuremath{\omega}},
    unit={\si{\radian\per\second}}
}
\newglossaryentry{accelerationComp}{
    type=hidden,
    name={\ensuremath{a}},
    description={Material acceleration},
    unit={\si{\meter\per\square\second}}
}
\newglossaryentry{xVelocity}{
    type=latin,
    name={\ensuremath{v_x}},
    description={%
        Velocity component on the x direction of the spatial coordinates
    },
    unit={\si{\meter\per\second}}
}
\newglossaryentry{yVelocity}{
    type=latin,
    name={\ensuremath{v_y}},
    description={%
        Velocity component on the y direction of the spatial coordinates
    },
    unit={\si{\meter\per\second}}
}
\newglossaryentry{zVelocity}{
    type=latin,
    name={\ensuremath{v_z}},
    description={%
        Velocity component on the z direction of the spatial coordinates
    },
    unit={\si{\meter\per\second}}
}
\newglossaryentry{meshVelocity}{
    type=greek,
    name={\ensuremath{\gvec{\omega}}},
    description={Mesh velocity},
    unit={\si{\meter\per\second}}
}
\newglossaryentry{xMeshVelocity}{
    type=hidden,
    name={\ensuremath{\omega_x}},
    description={%
        Mesh velocity component on the x direction of the spatial coordinates
    },
    unit={\si{\meter\per\second}}
}
\newglossaryentry{yMeshVelocity}{
    type=hidden,
    name={\ensuremath{\omega_y}},
    description={%
        Mesh velocity component on the y direction of the spatial coordinates
    },
    unit={\si{\meter\per\second}}
}
\newglossaryentry{zMeshVelocity}{
    type=hidden,
    name={\ensuremath{\omega_z}},
    description={%
        Mesh velocity component on the z direction of the spatial coordinates
    },
    unit={\si{\meter\per\second}}
}
\newglossaryentry{meshVelocityNode}{
    type=greek,
    name={\ensuremath{\gvec{\omega}_{node}}},
    description={%
        Mesh velocity at the geometric nodes of the Finite Volume mesh
    },
    unit={\si{\meter\per\second}}
}
\newglossaryentry{materialMeshVelocity}{
    type=latin,
    name={\ensuremath{\hat{\vec{v}}}},
    description={%
        Material velocity measured from the ALE coordinates point of view
    },
    unit={\si{\meter\per\second}}
}
\newglossaryentry{convectiveVelocity}{
    type=latin,
    name={\ensuremath{\vec{c}}},
    description={Convective velocity},
    unit={\si{\meter\per\second}}
}

\newglossaryentry{numberDoF}{
    type=latin,
    name={\ensuremath{V}},
    description={Degrees of freedom of a heterogeneous thermodynamical system},
    unit={-}
}
\newglossaryentry{inlet}{
    type=subscript,
    name={\ensuremath{e}},
    description={Indicates inlet variables}
}
\newglossaryentry{outlet}{
    type=subscript,
    name={\ensuremath{s}},
    description={Indicates outlet variables}
}
\newglossaryentry{reversible}{
    type=superscript,
    name={\ensuremath{\mathrm{rev}}},
    description={Indicates a reversible process}
}
\newglossaryentry{irreversible}{
    type=superscript,
    name={\ensuremath{\mathrm{irrev}}},
    description={Indicates an irreversible process}
}
\newglossaryentry{saturated}{
    type=subscript,
    name={\ensuremath{\mathit{SAT}}},
    description={Indicates a saturated state}
}
\newglossaryentry{satLiquid}{
    type=subscript,
    name={\ensuremath{l}},
    description={Indicates the liquid-satured state}
}
\newglossaryentry{satVapor}{
    type=subscript,
    name={\ensuremath{v}},
    description={Indicates the vapor-satured state}
}
\newglossaryentry{satSolid}{
    type=subscript,
    name={\ensuremath{s}},
    description={Indicates the solid-satured state}
}
\newglossaryentry{compressedLiquid}{
    type=subscript,
    name={\ensuremath{\mathit{LC}}},
    description={Indicates compressed liquid states}
}
\newglossaryentry{critical}{
    type=subscript,
    name={\ensuremath{C}},
    description={Indicates the critical state}
}
\newglossaryentry{triple}{
    type=subscript,
    name={\ensuremath{triplo}},
    description={Indicates the triple point}
}
\newglossaryentry{reduced}{
    type=subscript,
    name={\ensuremath{r}},
    description={Indicates reduced thermodinamic properties}
}
\newglossaryentry{substance}{
    type=subscript,
    name={\ensuremath{i}},
    description={Indicates a substance}
}
\newglossaryentry{closedSystem}{
    type=subscript,
    name={\ensuremath{\textsc{s}}},
    description={Indicates a closed system quantity}
}
\newglossaryentry{controlVolume}{
    type=subscript,
    name={\ensuremath{\textsc{vc}}},
    description={Indicates a control-volume quantity}
}
\newglossaryentry{environmentState}{
    type=subscript,
    name={\ensuremath{0}},
    description={%
        Indicates a the state of the environment or surrounding of a
        system
    }
}
\newglossaryentry{hotReservoir}{
    type=subscript,
    name={\ensuremath{H}},
    description={Indicates the properties of a \enquote{hot} reservoir}
}
\newglossaryentry{coldReservoir}{
    type=subscript,
    name={\ensuremath{L}},
    description={Indicates the properties of a \enquote{cold} reservoir}
}
\newglossaryentry{cyclicMachine}{
    type=subscript,
    name={\ensuremath{M}},
    description={Indicates a cyclic thermal machine}
}
\newglossaryentry{refrigerator}{
    type=subscript,
    name={\ensuremath{R}},
    description={Indicates a refrigerator}
}
\newglossaryentry{heatPump}{
    type=subscript,
    name={\ensuremath{BC}},
    description={Indicates a heat pump}
}
\newglossaryentry{CarnotCycle}{
    type=subscript,
    name={\ensuremath{C}},
    description={Indicates a property of the cycle of Carnot}
}
\newglossaryentry{base}{
    type=hidden,
    name={\ensuremath{b}},
    description={Indicates a property at a reference arbitrary base}
}

\newglossaryentry{vaporQuality}{
    type=latin,
    name={\ensuremath{x}},
    description={%
        Vapor quality or vapor mass fraction of a saturated liquid-vapor mixture
    },
    unit={\si{\pascal}}
}
\newglossaryentry{gasConstant}{
    type=latin,
    name={\ensuremath{R}},
    description={%
        Constant of a specific gas
    },
    unit={\si{\kilo\joule\per\kilo\gram\per\kelvin}}
}
\newglossaryentry{universalGasConstant}{
    type=latin,
    name={\ensuremath{\molar{R}}},
    description={%
        Universal constant of gases
    },
    unit={\si{\kilo\joule\per\kilo\mol\per\kelvin}}
}
\newglossaryentry{compressibilityFactor}{
    type=latin,
    name={\ensuremath{Z}},
    description={Compressibiility factor},
    unit={-}
}
\newglossaryentry{PitzerAcentricFactor}{
    type=greek,
    name={\ensuremath{\omega}},
    description={Pitzer's acentric factor},
    unit={-}
}
\newglossaryentry{WuPolarFactor}{
    type=latin,
    name={\ensuremath{Y}},
    description={Wu-Stiel's polarity factor},
    unit={-}
}
\newglossaryentry{massFlowRate}{
    type=latin,
    name={\ensuremath{\dot{m}}},
    description={Mass flow rate},
    unit={\si{\kilo\gram\per\second}}
}

\newglossaryentry{speciesA}{
    type=hidden,
    name={\ensuremath{A}},
    description={Species A},
    unit={\si{-}}
}
\newglossaryentry{speciesB}{
    type=hidden,
    name={\ensuremath{B}},
    description={Species B},
    unit={\si{-}}
}
\newglossaryentry{numberMoles}{
    type=latin,
    name={\ensuremath{n}},
    description={Number of moles of a species $i$ in a system},
    unit={\si{\mol}}
}
\newglossaryentry{numberSpecies}{
    type=latin,
    name={\ensuremath{N}},
    description={Number of moles of a species $i$ in a system},
    unit={-}
}
\newglossaryentry{numberPhases}{
    type=latin,
    name={\ensuremath{p}},
    description={Number of phases in a heterogeneous mixture},
    unit={-}
}
\newglossaryentry{specificVolume}{
    type=latin,
    name={\ensuremath{v}},
    description={Specific volume -- volume per mass},
    unit={\si{\cubic\meter\per\kilogram}}
}
\newglossaryentry{modifiedReducedVolume}{
    type=hidden,
    name={\ensuremath{\gls{specificVolume}^{'}_{\gls{reduced}}}},
    description={Modified specific volume used for the reduced definition},
    unit={\si{-}}
}
\newglossaryentry{totalEnergy}{
    type=latin,
    name={\ensuremath{E}},
    description={Total energy of a system},
    unit={\si{\joule}}
}
\newglossaryentry{kineticEnergy}{
    type=latin,
    name={\ensuremath{E_c}},
    description={Kinetic energy of a system},
    unit={\si{\joule}}
}
\newglossaryentry{potentialEnergy}{
    type=latin,
    name={\ensuremath{E_p}},
    description={Potential energy of a system},
    unit={\si{\joule}}
}
\newglossaryentry{internalEnergy}{
    type=latin,
    name={\ensuremath{U}},
    description={Internal energy of a system},
    unit={\si{\joule}}
}
\newglossaryentry{intInternalEnergy}{
    type=latin,
    name={\ensuremath{u}},
    description={Intensive internal energy of a system},
    unit={\si{\joule\per\kilo\gram}}
}
\newglossaryentry{intTotalEnergy}{
    type=latin,
    name={\ensuremath{e}},
    description={Intensive total energy of a system},
    unit={\si{\joule\per\kilo\gram}}
}
\newglossaryentry{enthalpy}{
    type=latin,
    name={\ensuremath{H}},
    description={Enthalpy of a system},
    unit={\si{\joule}}
}
\newglossaryentry{intEnthalpy}{
    type=latin,
    name={\ensuremath{h}},
    description={Intensive enthalpy of a system},
    unit={\si{\joule\per\kilo\gram}}
}

\newglossaryentry{entropy}{
    type=latin,
    name={\ensuremath{S}},
    description={Entropy of a system},
    unit={\si{\joule\per\kelvin}}
}
\newglossaryentry{intEntropy}{
    type=latin,
    name={\ensuremath{s}},
    description={Intensive entropy of a system},
    unit={\si{\joule\per\kilo\gram\per\kelvin}}
}

\newglossaryentry{HelmholtzFreeEnergy}{
    type=latin,
    name={\ensuremath{A}},
    description={Helmholtz free-energy of a system},
    unit={\si{\joule}}
}
\newglossaryentry{intHelmholtzFreeEnergy}{
    type=latin,
    name={\ensuremath{a}},
    description={Intensive Helmholtz free-energy of a system},
    unit={\si{\joule\per\kilo\gram}}
}

\newglossaryentry{GibbsFreeEnergy}{
    type=latin,
    name={\ensuremath{G}},
    description={Gibbs free-energy of a system},
    unit={\si{\joule}}
}
\newglossaryentry{intGibbsFreeEnergy}{
    type=latin,
    name={\ensuremath{g}},
    description={Intensive Gibbs free-energy of a system},
    unit={\si{\joule\per\kilo\gram}}
}

\newglossaryentry{entropyCreated}{
    type=latin,
    name={\ensuremath{S_{gen}}},
    description={%
        Entropy created or generated in a process undergone by a system
    },
    unit={\si{\joule\per\kelvin}}
}
\newglossaryentry{intEntropyCreated}{
    type=latin,
    name={\ensuremath{s_{gen}}},
    description={%
        Intensive entropy created or generated in a process undergone by a
        system
    },
    unit={\si{\joule\per\kelvin}}
}
\newglossaryentry{entropyCreatedRate}{
    type=latin,
    name={\ensuremath{{\transferRate{entropy}}_{gen}}},
    description={%
        Entropy created or generated in a process undergone by a system
    },
    unit={\si{\watt\per\kelvin}}
}
\newglossaryentry{intEntropyCreatedRate}{
    type=latin,
    name={\ensuremath{\dot{s}_{gen}}},
    description={%
        Intensive entropy created or generated in a process undergone by a
        system
    },
    unit={\si{\watt\per\kelvin}}
}

\newglossaryentry{chemicalPotential}{
    type=greek,
    name={\ensuremath{\mu}},
    description={%
        Chemical potential of a species crossing the boundary of a system
    },
    unit={\si{\joule\per\kelvin}}
}
\newglossaryentry{irreversibility}{
    type=latin,
    name={\ensuremath{I}},
    description={Irreversibility of a thermodynamical process},
    unit={\si{\joule}}
}
\newglossaryentry{intIrreversibility}{
    type=latin,
    name={\ensuremath{i}},
    description={Intensive irreversibility},
    unit={\si{\joule\per\kilo\gram}}
}
\newglossaryentry{irreversibilityRate}{
    type=latin,
    name={\ensuremath{\dot{I}}},
    description={Irreversibility rate of a thermodynamical process},
    unit={\si{\watt}}
}

\newglossaryentry{thermodynamicProperty}{
    type=hidden,
    name={\ensuremath{X}},
    description={%
        Generic extensive thermodynamic property of a system
    },
    unit={-}
}
\newglossaryentry{intThermodynamicProperty}{
    type=hidden,
    name={\ensuremath{x}},
    description={%
        Generic intensive thermodynamic property of a system
    },
    unit={-}
}
\newglossaryentry{thermodynamicSubstate}{
    type=greek,
    name={\ensuremath{\upsilon}},
    description={%
        Generic thermodynamic sub-state of a system, identified as the
        intensive counterpart of the generalized displacement producing work
        transfers
    },
    unit={-}
}
\newglossaryentry{generalizedForce}{
    type=greek,
    name={\ensuremath{\tau}},
    description={Generalized force applied to a system},
    unit={-}
}
\newglossaryentry{generalizedDisplacement}{
    type=greek,
    name={\ensuremath{\Upsilon}},
    description={Generalized displacement applied to a system},
    unit={-}
}
\newglossaryentry{heatTransfer}{
    type=latin,
    name={\ensuremath{Q}},
    description={Heat transfer interaction on the boundary of a system},
    unit={\si{\joule}}
}
\newglossaryentry{thermalConductivity}{
    type=latin,
    name={\ensuremath{k}},
    description={%
        Thermal conductivity of a material
    },
    unit={\si{\watt\per\meter\per\kelvin}}
}
\newglossaryentry{thermalDiffusivity}{
    type=greek,
    name={\ensuremath{\alpha}},
    description={%
        Thermal diffusivity of a material
    },
    unit={\si{\square\meter\per\second}}
}
\newglossaryentry{specificHeatTransfer}{
    type=latin,
    name={\ensuremath{q}},
    description={%
        Heat transfer interaction on the boundary of a system per unit
        mass
    },
    unit={\si{\joule\per\kilo\gram}}
}
\newglossaryentry{heatFlux}{
    type=latin,
    name={\ensuremath{q^{"}}},
    description={%
        Unidirection heat flux
    },
    unit={\si{\watt\per\square\meter}}
}
\newglossaryentry{heatFluxVector}{
    type=latin,
    name={\ensuremath{\vec{q}^{"}}},
    description={%
        Heat flux vector
    },
    unit={\si{\watt\per\square\meter}}
}
\newglossaryentry{logMeanTemperatureDifference}{
    type=hidden,
    name={\ensuremath{\Delta T_{\text{ml}}}},
    description={%
        Log mean tempratura difference
    },
    unit={\si{\kelvin}}
}
\newglossaryentry{specificHeatGeneration}{
    type=latin,
    name={\ensuremath{\dot{q}}},
    description={%
        Rate of heat generated per unit volume in the interior of a medium
    },
    unit={\si{\watt\per\cubic\meter}}
}
\newglossaryentry{heatFlowVector}{
    type=latin,
    name={\ensuremath{\vec{q}}},
    description={%
        Heat flux or heat flow vector, defined as the heat transfer per unit
        surface area
    },
    unit={\si{\watt\per\square\meter}}
}
\newglossaryentry{molarFlux}{
    type=latin,
    name={\ensuremath{N^{"}}},
    description={%
        Molar flux of a species
    },
    unit={\si{\mol\per\second}}
}
\newglossaryentry{massFlux}{
    type=latin,
    name={\ensuremath{n^{"}}},
    description={%
        Mass flux of a species
    },
    unit={\si{\kilo\gram\per\second\per\square\meter}}
}
\newglossaryentry{molarTransferRate}{
    type=latin,
    name={\ensuremath{N}},
    description={%
        Molar transfer rate of a species
    },
    unit={\si{\mol\per\second}}
}
\newglossaryentry{molarConcentration}{
    type=greek,
    name={\ensuremath{C}},
    description={Molar concentration},
    unit={\si{\kilo\mol\per\cubic\meter}}
}
\newglossaryentry{binaryDiffusionCoefficient}{
    type=latin,
    name={\ensuremath{D}},
    description={%
        Binary diffusion coefficent
    },
    unit={\si{\square\meter\per\second}}
}

\newglossaryentry{workTransfer}{
    type=latin,
    name={\ensuremath{W}},
    description={Generic work transfer interaction on the boundary of a system},
    unit={\si{\joule}}
}
\newglossaryentry{heatTransferRate}{
    type=latin,
    name={\ensuremath{\transferRate{heatTransfer}}},
    description={Heat transfer rate},
    unit={\si{\kilo\joule\per\second}}
}
\newglossaryentry{workTransferRate}{
    type=latin,
    name={\ensuremath{\transferRate{workTransfer}}},
    description={Work transfer rate},
    unit={\si{\kilo\joule\per\second}}
}
\newglossaryentry{specifcWorkRate}{
    type=latin,
    name={\ensuremath{w}},
    description={Work transfer rate},
    unit={\si{\kilo\joule\per\kilo\gram}}
}
\newglossaryentry{internalMechanicalPower}{
    type=latin,
    name={\ensuremath{\dot{w}_{int}}},
    description={%
        Internal mechanical power in a system, caused by the stress and
        deformation
    },
    unit={\si{\watt}}
}
\newglossaryentry{efficiency}{
    type=greek,
    name={\ensuremath{\eta}},
    description={Generic efficiency of a thermal machine or other system},
    unit={-}
}
\newglossaryentry{performanceCoeff}{
    type=greek,
    name={\ensuremath{\beta}},
    description={%
        Generic coefficient of performance of a refrigerator or heat pump
    },
    unit={-}
}
\newglossaryentry{secondLawEfficiency}{
    type=greek,
    name={\ensuremath{\gls{efficiency}_{\textsc{ii}}}},
    description={Efficiency of a thermal system in terms of the Second Law},
    unit={-}
}
\newglossaryentry{numberWorkSources}{
    type=latin,
    name={\ensuremath{m}},
    description={Number of work sources to a system},
    unit={}
}

\newglossaryentry{bulkModulus}{
    type=greek,
    name={\ensuremath{\kappa}},
    description={Bulk modulus},
    unit={\si{\pascal}}
}

\newglossaryentry{satTemperature}{
    type=latin,
    name={\ensuremath{\gls{temperature}_{SAT}}},
    description={Saturation temperature of a pure substance},
    unit={\si{\kelvin}}
}
\newglossaryentry{satPressure}{
    type=latin,
    name={\ensuremath{\gls{pressure}_{SAT}}},
    description={Saturation pressure of a pure substance},
    unit={\si{\pascal}}
}

\newglossaryentry{exergy}{
    type=latin,
    name={\ensuremath{E}},
    description={Generic symbol representing exergy},
    unit={\si{\joule}}
}
\newglossaryentry{exergyRate}{
    type=latin,
    name={\ensuremath{\dot{E}}},
    description={Generic symbol representing exergy rate},
    unit={\si{\watt}}
}
\newglossaryentry{flowExergy}{
    type=latin,
    name={\ensuremath{\Psi}},
    description={Flow exergy},
    unit={\si{\joule}}
}
\newglossaryentry{nonFlowExergy}{
    type=latin,
    name={\ensuremath{\Phi}},
    description={Non-flow exergy},
    unit={\si{\joule}}
}
\newglossaryentry{intFlowExergy}{
    type=latin,
    name={\ensuremath{\psi}},
    description={Intensive flow exergy},
    unit={\si{\joule\per\kilo\gram}}
}
\newglossaryentry{intNonFlowExergy}{
    type=latin,
    name={\ensuremath{\varphi}},
    description={Intensive non-flow exergy},
    unit={\si{\joule\per\kilo\gram}}
}
\newglossaryentry{specificHeat}{
    type=latin,
    name={\ensuremath{c}},
    description={Specific heat of a pure substance},
    unit={\si{\joule\per\kilo\gram\per\kelvin}}
}
\newglossaryentry{constVolumeSpecificHeat}{
    type=latin,
    name={\ensuremath{\gls{specificHeat}_{\gls{volume}}}},
    description={Constant-volume specific heat of a pure substance},
    unit={\si{\joule\per\kilo\gram\per\kelvin}}
}
\newglossaryentry{constPressureSpecificHeat}{
    type=latin,
    name={\ensuremath{\gls{specificHeat}_{\gls{pressure}}}},
    description={Constant-pressure specific heat of a pure substance},
    unit={\si{\joule\per\kilo\gram\per\kelvin}}
}
\newglossaryentry{specificHeatRatio}{
    type=latin,
    name={\ensuremath{k}},
    description={Specific heat ratio},
    unit={\si{-}}
}
\newglossaryentry{fugacity}{
    type=latin,
    name={\ensuremath{f}},
    description={Fugacity of a system},
    unit={\si{\pascal}}
}
\newglossaryentry{activity}{
    type=latin,
    name={\ensuremath{a}},
    description={Activity of a component in a mixture},
    unit={-}
}
\newglossaryentry{volumeExpansivity}{
    type=greek,
    name={\ensuremath{\alpha_{\gls{pressure}}}},
    description={Volume expansivity of a system},
    unit={\si{\per\kelvin}}
}
\newglossaryentry{isothermalCompressibility}{
    type=greek,
    name={\ensuremath{\beta_{\gls{temperature}}}},
    description={Isothermal compressibility of a system},
    unit={\si{\per\pascal}}
}

\newglossaryentry{volumetricThermalExpansionCoeff}{
    type=greek,
    name={\ensuremath{\beta}},
    description={Volumetric thermal expansion coefficient of a fluid},
    unit={\si{\per\kelvin}}
}
\newglossaryentry{latentHeat}{
    type=latin,
    name={\ensuremath{\gls{intEnthalpy}_{fg}}},
    description={Latent heat},
    unit={\si{\kilo\joule\per\kilo\gram}}
}
\newglossaryentry{adiabaticCompressibility}{
    type=greek,
    name={\ensuremath{\beta_{\gls{entropy}}}},
    description={Adiabatic compressibility of a system},
    unit={\si{\per\pascal}}
}
\newglossaryentry{moleFraction}{
    type=latin,
    name={\ensuremath{y}},
    description={%
        Mole fraction of a component in a mixture
    },
    unit={-}
}
\newglossaryentry{mixture}{
    type=subscript,
    name={\ensuremath{m}},
    description={%
        Indicates a proprty of a homogeneous mixture.
    }
}
\newglossaryentry{reactionExtent}{
    type=greek,
    name={\ensuremath{\varepsilon}},
    description={Extent of a chemical reaction (or degree of advancement)},
    unit={\si{\mol}}
}
\newglossaryentry{formationEnthalpy}{
    type=greek,
    name={\ensuremath{\molar{\gls{intEnthalpy}}^{\mathrm{0}}_{f}}},
    description={Enthalpy of formation of a component},
    unit={\si{\kilo\joule\per\kilo\mol}}
}
\newglossaryentry{formationGibbsFreeEnergy}{
    type=greek,
    name={\ensuremath{\molar{\gls{intGibbsFreeEnergy}}^{\mathrm{0}}_{f}}},
    description={Gibbs free energy of formation of a component},
    unit={\si{\kilo\joule\per\kilo\mol}}
}
\newglossaryentry{absoluteEntropy}{
    type=greek,
    name={\ensuremath{\molar{\gls{intEntropy}}^{\mathrm{0}}}},
    description={Absolute entropy of a substance},
    unit={\si{\kilo\joule\per\kilo\mol\per\kelvin}}
}
\newglossaryentry{standardState}{
    type=superscript,
    name={\ensuremath{\mathrm{0}}},
    description={Indicates the standard state}
}
\newglossaryentry{standardTemperature}{
    type=latin,
    name={\ensuremath{\gsuper{temperature}{standardState}}},
    description={Standard temperature, \SI{298.15}{\kelvin}},
    unit={\si{\kelvin}}
}
\newglossaryentry{standardPressure}{
    type=latin,
    name={\ensuremath{\gsuper{pressure}{standardState}}},
    description={Standard pressure, \SI{0.1}{\mega\pascal}},
    unit={\si{\mega\pascal}}
}
\newglossaryentry{carbonQuantity}{
    type=hidden,
    name={\ensuremath{\alpha}},
    description={%
        Quantity of carbon atoms in the generic
        $\mathrm{C}_{\alpha}\mathrm{H}_{\beta}\mathrm{O}_{\gamma}\mathrm{N}_{\delta}$
    },
    unit={-}
}
\newglossaryentry{hidrogenQuantity}{
    type=hidden,
    name={\ensuremath{\beta}},
    description={%
        Quantity of hidrogen atoms in the generic
        $\mathrm{C}_{\alpha}\mathrm{H}_{\beta}\mathrm{O}_{\gamma}\mathrm{N}_{\delta}$
    },
    unit={-}
}
\newglossaryentry{oxygenQuantity}{
    type=hidden,
    name={\ensuremath{\gamma}},
    description={%
        Quantity of oxigen atoms in the generic
        $\mathrm{C}_{\alpha}\mathrm{H}_{\beta}\mathrm{O}_{\gamma}\mathrm{N}_{\delta}$
    },
    unit={-}
}
\newglossaryentry{nitrogenQuantity}{
    type=hidden,
    name={\ensuremath{\delta}},
    description={%
        Quantity of nitrogen atoms in the generic
        $\mathrm{C}_{\alpha}\mathrm{H}_{\beta}\mathrm{O}_{\gamma}\mathrm{N}_{\delta}$
    },
    unit={-}
}
\newglossaryentry{chemicalGenericFuel}{
    type=hidden,
    name={\ensuremath{%
        \mathrm{C}_{\alpha}\mathrm{H}_{\beta}\mathrm{O}_{\gamma}\mathrm{N}_{\delta}%
    }},
    description={%
        Chemical formula for a generic fuel
    },
    unit={-}
}
\newglossaryentry{airFuelRatio}{
    type=latin,
    name={\ensuremath{\mathit{AC}}},
    description={Air-fuel ratio},
    unit={-}
}
\newglossaryentry{stoichiometricAirFuelRatio}{
    type=latin,
    name={\ensuremath{\mathit{AC_s}}},
    description={Stoichiometric or theoretical air-fuel ratio},
    unit={-}
}
\newglossaryentry{molarAirFuelRatio}{
    type=latin,
    name={\ensuremath{a_s}},
    description={Stoichiometric molar air-fuel ratio},
    unit={-}
}
\newglossaryentry{fuelAirRatio}{
    type=latin,
    name={\ensuremath{\mathit{CA}}},
    description={Fuel-air ratio},
    unit={-}
}
\newglossaryentry{stoichiometricFuelAirRatio}{
    type=latin,
    name={\ensuremath{\mathit{CA_s}}},
    description={Stoichiometric or theoretical fuel-air ratio},
    unit={-}
}
\newglossaryentry{equivalenceRatio}{
    type=latin,
    name={\ensuremath{\phi}},
    description={Air-fuel equivalent ratio},
    unit={-}
}
\newglossaryentry{rEquivalenceRatio}{
    type=latin,
    name={\ensuremath{\lambda}},
    description={Reciprocal of the air-fuel equivalent ratio},
    unit={-}
}
\newglossaryentry{chProducts}{
    type=subscript,
    name={\ensuremath{P}},
    description={Indicates the products of a chemical reaction}
}
\newglossaryentry{chReactants}{
    type=subscript,
    name={\ensuremath{R}},
    description={Indicates the reactants of a chemical reaction}
}
\newglossaryentry{flameAdiabaticTemperature}{
    type=latin,
    name={\ensuremath{\gls{temperature}_{ad}}},
    description={Flame adiabatic temperature},
    unit={\si{\kelvin}}
}
\newglossaryentry{combustionEnthalpy}{
    type=greek,
    name={\ensuremath{%
        \molar{\gls{intEnthalpy}}^{\mathrm{0}}_{\gls{chReactants}\gls{chProducts}}}%
    },
    description={Enthalpy of combustion of a fuel},
    unit={\si{\kilo\joule\per\kilo\mol}}
}
\newglossaryentry{LagrangeMultiplyer}{
    type=greek,
    name={\ensuremath{\lambda}},
    description={Lagrange multiplyer},
    unit={-}
}
\newglossaryentry{chEquilibriumConstant}{
    type=latin,
    name={\ensuremath{K}},
    description={Constant of equilibrium of a chemical reaction},
    unit={-}
}
\newglossaryentry{PoyntingFactor}{
    type=latin,
    name={\ensuremath{F}},
    description={Poynting factor},
    unit={-}
}

\newglossaryentry{molecularMass}{
    type=latin,
    name={\ensuremath{M}},
    description={Molecular mass},
    unit={\si{\kilo\mol}}
}

\newglossaryentry{convectionHeatTransferCoeff}{
    type=latin,
    name={\ensuremath{h}},
    description={Convection heat transfer coefficient},
    unit={\si{\watt\per\square\meter\kelvin}}
}
\newglossaryentry{meanConvectionHeatTransferCoeff}{
    type=hidden,
    name={\ensuremath{\avg{h}}},
    description={Mean convection heat transfer coefficient},
    unit={\si{\watt\per\square\meter\kelvin}}
}
\newglossaryentry{convectionMassTransferCoeff}{
    type=latin,
    name={\ensuremath{h_m}},
    description={Convection mass transfer coefficient},
    unit={\si{\meter\per\second}}
}
\newglossaryentry{meanConvectionMassTransferCoeff}{
    type=hidden,
    name={\ensuremath{\avg{h}_m}},
    description={Mean convection mass transfer coefficient},
    unit={\si{\meter\per\second}}
}
\newglossaryentry{conductionShapeFactor}{
    type=latin,
    name={\ensuremath{S}},
    description={Conduction shape factor},
    unit={\si{\meter}}
}
\newglossaryentry{freeStream}{
    type=subscript,
    name={\ensuremath{\infty}},
    description={Indicates free-stream properties}
}
\newglossaryentry{velocityLayerThickness}{
    type=latin,
    name={\ensuremath{\delta}},
    description={Thickness of the velocity boundary layer},
    unit={\si{\meter}}
}
\newglossaryentry{thermalLayerThickness}{
    type=latin,
    name={\ensuremath{\delta_t}},
    description={Thickness of the thermal boundary layer},
    unit={\si{\meter}}
}
\newglossaryentry{totalHeatTransferRate}{
    type=latin,
    name={\ensuremath{q}},
    description={%
        Total heat transfer rate
    },
    unit={\si{\watt}}
}
\newglossaryentry{thermalResistance}{
    type=latin,
    name={\ensuremath{R_t}},
    description={%
        Thermal resistance
    },
    unit={\si{\kelvin\per\watt}}
}
\newglossaryentry{convectiveThermalResistance}{
    type=hidden,
    name={\ensuremath{R_{t,\mathrm{conv}}}},
    description={%
        Convective thermal resistance
    },
    unit={\si{\kelvin\per\watt}}
}
\newglossaryentry{conductionThermalResistance}{
    type=hidden,
    name={\ensuremath{R_{t,\mathrm{cond}}}},
    description={%
        Thermal resistance for condution
    },
    unit={\si{\kelvin\per\watt}}
}
\newglossaryentry{radiationThermalResistance}{
    type=hidden,
    name={\ensuremath{R_{t,\mathrm{rad}}}},
    description={%
        Thermal resistance for radiation
    },
    unit={\si{\kelvin\per\watt}}
}
\newglossaryentry{totalThermalResistance}{
    type=hidden,
    name={\ensuremath{R_{\mathrm{tot}}}},
    description={%
        Total thermal resistance of a thermal circuit
    },
    unit={\si{\kelvin\per\watt}}
}
\newglossaryentry{globalHeatTransferCoeff}{
    type=hidden,
    name={\ensuremath{U}},
    description={%
        Global heat transfer coefficient
    },
    unit={\si{\watt\per\square\meter\kelvin}}
}

\newglossaryentry{electricalResistance}{
    type=hidden,
    name={\ensuremath{R_e}},
    description={%
        Electrical resistance
    },
    unit={\si{\ohm}}
}
\newglossaryentry{electricalCurrent}{
    type=hidden,
    name={\ensuremath{I}},
    description={%
        Electrical current
    },
    unit={\si{\ampere}}
}

\newglossaryentry{radiation}{
    type=subscript,
    name={\ensuremath{\mathrm{rad}}},
    description={Indicates radiation heat exchange}
}
\newglossaryentry{surroundings}{
    type=subscript,
    name={\ensuremath{\mathrm{viz}}},
    description={Indicates the properties of the surroundings surfaces}
}
\newglossaryentry{emissivePower}{
    type=hidden,
    name={\ensuremath{E}},
    description={%
        Emissive power of a surface
    },
    unit={\si{\watt\per\square\meter}}
}
\newglossaryentry{irradiation}{
    type=hidden,
    name={\ensuremath{G}},
    description={%
        Irradiation on a surface
    },
    unit={\si{\watt\per\square\meter}}
}
\newglossaryentry{StefanBoltzmannConstant}{
    type=hidden,
    name={\ensuremath{\sigma}},
    description={%
        Stefan-Boltazmann constant ($ =
        \SI{5.67e-8}{\watt\per\square\meter\kelvin^4}$)
    },
    unit={\si{\watt\per\square\meter\kelvin^4}}
}
\newglossaryentry{emissivity}{
    type=hidden,
    name={\ensuremath{\varepsilon}},
    description={%
        Emissivity of a surface
    },
    unit={\si{-}}
}
\newglossaryentry{absorptivity}{
    type=hidden,
    name={\ensuremath{\alpha}},
    description={%
        Absorptivity of a surface
    },
    unit={\si{-}}
}

\newglossaryentry{cvNumber}{
    type=latin,
    name={\ensuremath{N_{CV}}},
    description={Number of control volumes in a finite volume mesh},
    unit={\si{-}}
}

\newglossaryentry{centroid}{
    type=subscript,
    name={\ensuremath{C}},
    description={%
        Indicates variables evaluated at the centroid of a control volume
    }
}
\newglossaryentry{nbCentroid}{
    type=subscript,
    name={\ensuremath{N}},
    description={%
        Indicates variables evaluated at the centroid of a neighbor control
        volume to cell $V_C$
    }
}
\newglossaryentry{owner}{
    type=subscript,
    name={\ensuremath{\gls{centroid}}},
    description={%
        Indicates variables evaluated at the centroid of a the face owner
        control volume
    }
}
\newglossaryentry{neighbor}{
    type=subscript,
    name={\ensuremath{\gls{nbCentroid}}},
    description={%
        Indicates variables evaluated at the centroid of a face neighbor
        control volume
    }
}
\newglossaryentry{face}{
    type=subscript,
    name={\ensuremath{f}},
    description={%
        Indicate a face of a polyhedral control volume and a property evaluated
        at it
    }
}
\newglossaryentry{skewFace}{
    type=subscript,
    name={\ensuremath{\gls{face}^{\prime}}},
    description={%
        Indicates the point where the vector joining two face adjacent cell's
        centroids intercepts the face
    }
}

\newglossaryentry{fvCellVolume}{
    type=latin,
    name={\ensuremath{\gls{volume}_{\gls{centroid}}}},
    description={Volume of a cell with centroid $C$ in a finite volume mesh},
    unit={\si{\cubic\meter}}
}
\newglossaryentry{fvCellSurface}{
    type=latin,
    name={\ensuremath{\gls{surface}_{\gls{centroid}}}},
    description={Surface of a cell with centroid $C$ in a finite volume mesh},
    unit={\si{\square\meter}}
}
\newglossaryentry{fvFaceNormal}{
    type=latin,
    name={\ensuremath{\vec{S}_{\gls{face}}}},
    description={Normal vector to the face $f$ of a polyhedral cell},
    unit={\si{\square\meter}}
}
\newglossaryentry{joinCentroidVector}{
    type=latin,
    name={\ensuremath{\vec{d}_{\gls{centroid}\gls{nbCentroid}}}},
    description={%
        Vector joining the centroids of adjacent cells in a finite volume mesh
    },
    unit={\si{\meter}}
}
\newglossaryentry{skewnessVector}{
    type=latin,
    name={\ensuremath{\vec{m}}},
    description={%
        Vector joining the face centroid with the interception point between
        the face and the line between the owner and neighbour cells.
    },
    unit={\si{\meter}}
}
\newglossaryentry{facePyramidHeigth}{
    type=latin,
    name={\ensuremath{\vec{h}_{\gls{face}\gls{centroid}}}},
    description={%
        Heigth of a cell face pyramid -- is the opposite of the vector joining
        the cell centroid and the face centroid
    },
    unit={\si{\meter}}
}
\newglossaryentry{centroidFaceVector}{
    type=latin,
    name={\ensuremath{\vec{d}_{\gls{centroid}\gls{face}}}},
    description={%
        Vector joining the centroids of a cell \gls{fvCellVolume} and the
        centroid of a face \gls{face}
    },
    unit={\si{\meter}}
}
\newglossaryentry{ownerFaceVector}{
    type=latin,
    name={\ensuremath{\vec{d}_{\gls{owner}\gls{face}}}},
    description={%
        Vector joining the centroid of a face owner cell and the centroid of
        the face \gls{face}
    },
    unit={\si{\meter}}
}
\newglossaryentry{neighFaceVector}{
    type=latin,
    name={\ensuremath{\vec{d}_{\gls{neighbor}\gls{face}}}},
    description={%
        Vector joining the centroid of a face neighbor cell and the centroid of
        the face \gls{face}
    },
    unit={\si{\meter}}
}

\newglossaryentry{projCentroidNormalVec}{
    type=hidden,
    name={\ensuremath{\vec{d}_{fn}}},
    description={%
        Vector formed by the projection of \gls{joinCentroidVector}
        on \gls{sNormalVector}
    },
    unit={\si{\meter}}
}
\newglossaryentry{projCentroidNormalNorm}{
    type=latin,
    name={\ensuremath{d_{fn}}},
    description={%
        Projection of the vector \gls{joinCentroidVector} on the face normal
        \gls{sNormalVector}
    },
    unit={\si{\meter}}
}
\newglossaryentry{advWeightFactor}{
    type=hidden,
    name={\ensuremath{w}},
    description={%
        Weight of cell \gls{centroid} on the advective interpolation profile
    },
    unit={\si{-}}
}
\newglossaryentry{faceVolumeSwept}{
    type=latin,
    name={\ensuremath{\dot{V}_f}},
    description={%
        Volume swept by face \gls{face} of control volume \gls{fvCellVolume}
    },
    unit={\si{\cubic\meter\per\second}}
}
\newglossaryentry{normalDecomposition1}{
    type=latin,
    name={\ensuremath{\vec{\beta}}},
    description={%
        Vector joining the centroid of face \gls{face} and the neighbor cell
        centroid \gls{nbCentroid}
    },
    unit={\si{-}}
}
\newglossaryentry{normalDecomposition1Norm}{
    type=hidden,
    name={\ensuremath{\beta}},
    description={%
        Modulus of the vector joining the centroid of face \gls{face} and the
        neighbor cell centroid \gls{nbCentroid}
    },
    unit={\si{-}}
}
\newglossaryentry{normalDecomposition2}{
    type=greek,
    name={\ensuremath{\vec{\kappa}}},
        description={Complement vector of the decomposition of
        \gls{fvFaceNormal} in \gls{normalDecomposition1}},
        unit={\si{-}}
}
\newglossaryentry{fvInterpolationWeight}{
    type=hidden,
    name={\ensuremath{g}},
    description={%
        Generic symbol for the interpolation weights of the finite volume
        method
    },
    unit={\si{-}}
}
\newglossaryentry{joinCentroidCellFace}{
    type=latin,
    name={\ensuremath{\vec{d}_{\gls{face}\gls{centroid}}}},
    description={%
        Vector ligando os centroides \gls{centroid} da célula e de uma face
        \gls{face}
    },
    unit={\si{\meter}}
}
\newglossaryentry{joinNbCentroidCellFace}{
    type=latin,
    name={\ensuremath{\vec{d}_{\gls{face}\gls{nbCentroid}}}},
    description={%
        Vector ligando os centroides da célula vizinha \gls{nbCentroid} da
        célula e de uma face \gls{face}
    },
    unit={\si{\meter}}
}

\newglossaryentry{transportVariable}{
    type=greek,
    name={\ensuremath{\phi}},
    description={%
        Generic intensive (per unit mass) variable transported by a fluid flow
    },
    unit={$[\phi]/$\si{\kilo\gram}}
}
\newglossaryentry{integralTransportVariable}{
    type=greek,
    name={\ensuremath{\Phi}},
    description={%
        Generic intensive variable transported by a fluid flow integrated in a
        finite control volume
    },
    unit={$[\Phi]$}
}
\newglossaryentry{transportVariablePerVolume}{
    type=greek,
    name={\ensuremath{\varphi}},
    description={%
        Generic intensive (per unit volume) variable transported by a fluid
        flow
    },
    unit={$[\Phi]/$\si{\cubic\meter}}
}
\newglossaryentry{transportDiffusivity}{
    type=greek,
    name={\ensuremath{\Gamma^{\phi}}},
    description={Diffusivity of \gls{transportVariable}},
    unit={\si{\square\meter\per\second}}
}
\newglossaryentry{transportSource}{
    type=latin,
    name={\ensuremath{S^{\phi}}},
    description={%
        Source and/or sink terms of the differential transport equation of
        \gls{transportVariable}
    },
    unit={$[\phi]/$\si{\cubic\meter\per\second}}
}
\newglossaryentry{linearTransportSource}{
    type=latin,
    name={\ensuremath{\bar{S}^{\phi}}},
    description={%
        Mean source and/or sink terms of the integral transport equation of
        \gls{transportVariable}
    },
    unit={$[\phi]/$\si{\cubic\meter\per\second}}
}
\newglossaryentry{fvCoeffs}{
    type=latin,
    name={\ensuremath{a}},
    description={%
        Generic coefficients of the linearized transport equation for the
        volume $V_C$
    },
    unit={\si{\kilo\gram\per\second}}
}
\newglossaryentry{fvDiagCoeffs}{
    type=latin,
    name={\ensuremath{\gls{fvCoeffs}_{\gls{centroid}}}},
    description={%
        Diagonal coefficients of the linearized transport equation for the
        volume $V_C$
    },
    unit={\si{\kilo\gram\per\second}}
}
\newglossaryentry{fvOffDiagCoeffs}{
    type=latin,
    name={\ensuremath{\gls{fvCoeffs}_{\gls{nbCentroid}}}},
    description={%
        Off-diagonal coefficients of the linearized transport equation for the
        volume $V_C$
    },
    unit={\si{\kilo\gram\per\second}}
}
\newglossaryentry{fvMatrix}{
    type=latin,
    name={\ensuremath{\matrix{A}}},
    description={%
        Coefficients matrix of a system of equations assembled for a finite
        volume mesh
    },
    unit={$[\phi]$}
}
\newglossaryentry{sourceScalar}{
    type=hidden,
    name={\ensuremath{b}},
    description={%
        Source term of the linearized transport equation for a control volume
        $V_C$
    },
    unit={$[\phi]$\si{\kilo\gram\per\second}}
}
\newglossaryentry{fvSource}{
    type=hidden,
    name={\ensuremath{\cvec{b}}},
    description={%
        Source term of the linearized system equation for a finite volume mesh
    },
    unit={\si{-}}
}
\newglossaryentry{relaxationFactor}{
    type=hidden,
    name={\ensuremath{\alpha}},
    description={%
        Relaxation factor
    },
    unit={\si{-}}
}
\newglossaryentry{sourceVector}{
    type=latin,
    name={\ensuremath{\vec{b}}},
    description={%
        Source vector of the linearized momentum equations for a control volume
    },
    unit={$[\phi]$\si{\kilo\gram\per\second}}
}
\newglossaryentry{pressureCorrection}{
    type=latin,
    name={\ensuremath{p'}},
    description={%
        Pressure correction field in the pressure-velocity algorithms to solve
        the coupled Navier-Stokes equations
    },
    unit={\si{\pascal}}
}
\newglossaryentry{transportOperator}{
    type=hidden,
    name={\ensuremath{\matrix{H}}},
    description={Transport operator},
    unit={\si{-}}
}
\newglossaryentry{sourceOperator}{
    type=hidden,
    name={\ensuremath{\matrix{B}}},
    description={Source operator},
    unit={\si{-}}
}
\newglossaryentry{pressureDiffOperator}{
    type=hidden,
    name={\ensuremath{\matrix{D}}},
    description={Pressure diffusive operator},
    unit={\si{-}}
}

\newglossaryentry{meshNonOrthog}{
    type=latin,
    name={\ensuremath{mesh_{nonOrtho}}},
    description={Mesh non-orthogonality}
}

\newglossaryentry{meshSkewness}{
    type=latin,
    name={\ensuremath{mesh_{skew}}},
    description={Mesh skewness}
}

\newglossaryentry{faceTwist}{
    type=latin,
    name={\ensuremath{mesh_{faceTwist}}},
    description={Mesh face twist}
}

\newglossaryentry{faceWeight}{
    type=latin,
    name={\ensuremath{mesh_{faceWeight}}},
    description={Mesh face weight}
}

\newglossaryentry{faceVolumeRatio}{
    type=latin,
    name={\ensuremath{mesh_{faceVolRatio}}},
    description={Mesh face volume ratio}
}

\newglossaryentry{faceTriangTwist}{
    type=latin,
    name={\ensuremath{mesh_{faceTriTwist}}},
    description={Mesh face twist}
}

\newglossaryentry{cellConvexity}{
    type=latin,
    name={\ensuremath{mesh_{cellConcave}}},
    description={Mesh cells convexity: it measures if a cell is concave.}
}

\newglossaryentry{pyramidVolume}{
    type=latin,
    name={\ensuremath{mesh_{cellPyrVol}}},
    description={Mesh cells pyramids volume}
}

\newglossaryentry{cellTetQuality}{
    type=latin,
    name={\ensuremath{mesh_{tetQuality}}},
    description={Mesh cells tetrahedral decomposition quality}
}
\newglossaryentry{polyCellTensor}{
    type=latin,
    name={\ensuremath{\tensor{T}}},
    description={%
        Second-order tensor formed by the outer product of the face vector of a
        generic polyhedral cell
    }
}
\newglossaryentry{cellDeterminant}{
    type=latin,
    name={\ensuremath{mesh_{cellDeterminant}}},
    description={Mesh face twist}
}

\newglossaryentry{dynamicViscosity}{
    type=greek,
    name={\ensuremath{\mu}},
    description={Dynamic viscosity of a fluid},
    unit={\si{\pascal\second}}
}
\newglossaryentry{surfaceTension}{
    type=greek,
    name={\ensuremath{\sigma}},
    description={Surface tension},
    unit={\si{\newton\per\meter}}
}
\newglossaryentry{secondViscosity}{
    type=greek,
    name={\ensuremath{\lambda}},
    description={%
        Second viscosity of a fluid (sometimes referred to as
        \emph{bulk viscosity})
    },
    unit={\si{\pascal\second}}
}
\newglossaryentry{kinematicViscosity}{
    type=greek,
    name={\ensuremath{\nu}},
    description={Kinematic viscosity of a fluid},
    unit={\si{\square\meter\per\second}}
}
\newglossaryentry{pressure}{
    type=latin,
    name={\ensuremath{p}},
    description={Pressure field of a fluid flow},
    unit={\si{\pascal}}
}
\newglossaryentry{viscousDissipationRateFunction}{
    type=greek,
    name={\ensuremath{\Phi}},
    description={Viscous dissipation rate function},
    unit={\si{\watt\per\cubic\meter}}
}
\newglossaryentry{meshDiffusivity}{
    type=greek,
    name={\ensuremath{\gamma}},
    description={%
        Mesh diffusivity for the Laplace equation governing the mesh motion
    },
    unit={\si{-}}
}
\newglossaryentry{centroidToBoundaryDistance}{
    type=latin,
    name={\ensuremath{l}},
    description={Cell centroid distance to a selected boundary},
    unit={\si{\meter}}
}
\newglossaryentry{viscousCauchyStress}{
    type=greek,
    name={\ensuremath{\gvec{\tau}}},
    description={Viscous part of the Cauchy stress tensor of a fluid flow},
    unit={\si{\pascal}}
}
\newglossaryentry{viscousCauchyStressComp}{
    type=hidden,
    name={\ensuremath{\tau}},
    description={Viscous part of the Cauchy stress tensor of a fluid flow},
    unit={\si{\pascal}}
}
\newglossaryentry{vorticityTensor}{
    type=greek,
    name={\ensuremath{\tensor{\Omega}}},
    description={Vorticity or spin tensor},
    unit={\si{\per\second}}
}
\newglossaryentry{vorticityTensorComp}{
    type=hidden,
    name={\ensuremath{\Omega}},
    description={Vorticity or spin tensor component},
    unit={\si{\per\second}}
}
\newglossaryentry{vorticity}{
    type=greek,
    name={\ensuremath{\gvec{\zeta}}},
    description={Vorticity vector},
    unit={\si{\per\second}}
}
\newglossaryentry{vorticityComp}{
    type=hidden,
    name={\ensuremath{\zeta}},
    description={Vorticity vector component},
    unit={\si{\per\second}}
}
\newglossaryentry{circulation}{
    type=hidden,
    name={\ensuremath{\Gamma}},
    description={Circulation},
    unit={\si{\square\meter\per\second}}
}
\newglossaryentry{ReynoldsNumber}{
    type=latin,
    name={\ensuremath{Re}},
    description={Reynolds number},
    unit={\si{-}}
}
\newglossaryentry{PecletNumber}{
    type=latin,
    name={\ensuremath{Pe}},
    description={Péclet number},
    unit={\si{-}}
}
\newglossaryentry{frictionCoefficient}{
    type=latin,
    name={\ensuremath{C_f}},
    description={Friction coefficient},
    unit={\si{-}}
}
\newglossaryentry{frictionFactor}{
    type=latin,
    name={\ensuremath{f}},
    description={Friction factor for head loss computation in internal flow},
    unit={\si{-}}
}
\newglossaryentry{MachNumber}{
    type=latin,
    name={\ensuremath{Ma}},
    description={Mach number},
    unit={\si{-}}
}
\newglossaryentry{soundSpeed}{
    type=latin,
    name={\ensuremath{c}},
    description={Sound speed},
    unit={\si{\meter\per\second}}
}
\newglossaryentry{WeberNumber}{
    type=latin,
    name={\ensuremath{We}},
    description={Weber number},
    unit={\si{-}}
}
\newglossaryentry{FroudeNumber}{
    type=latin,
    name={\ensuremath{Fr}},
    description={Froude number},
    unit={\si{-}}
}
\newglossaryentry{EulerNumber}{
    type=latin,
    name={\ensuremath{Eu}},
    description={Euler number},
    unit={\si{-}}
}
\newglossaryentry{CavitationNumber}{
    type=latin,
    name={\ensuremath{Ca}},
    description={Cavitation number},
    unit={\si{-}}
}
\newglossaryentry{CourantNumber}{
    type=latin,
    name={\ensuremath{Co}},
    description={Courant number},
    unit={\si{-}}
}
\newglossaryentry{PrandtlNumber}{
    type=latin,
    name={\ensuremath{Pr}},
    description={%
        Prandtl number
    },
    unit={\si{-}}
}
\newglossaryentry{SchmidtNumber}{
    type=latin,
    name={\ensuremath{Sc}},
    description={%
        Schmidt number
    },
    unit={\si{-}}
}
\newglossaryentry{NusseltNumber}{
    type=latin,
    name={\ensuremath{Nu}},
    description={%
        Nusselt number
    },
    unit={\si{-}}
}
\newglossaryentry{SherwoodNumber}{
    type=latin,
    name={\ensuremath{Sh}},
    description={%
        Sherwood number
    },
    unit={\si{-}}
}
\newglossaryentry{BiotNumber}{
    type=latin,
    name={\ensuremath{Bi}},
    description={Biot number},
    unit={\si{-}}
}
\newglossaryentry{GrashofNumber}{
    type=latin,
    name={\ensuremath{Gr}},
    description={Grashof number},
    unit={\si{-}}
}
\newglossaryentry{RayleighNumber}{
    type=latin,
    name={\ensuremath{Ra}},
    description={Rayleigh number},
    unit={\si{-}}
}
\newglossaryentry{FourierNumber}{
    type=latin,
    name={\ensuremath{Fo}},
    description={Fourier number},
    unit={\si{-}}
}
\newglossaryentry{shearRate}{
    type=latin,
    name={\ensuremath{\dot{\gamma}}},
    description={%
        Shear rate components in a fluid flow
    },
    unit={\si{\per\second}}
}
\newglossaryentry{yieldStress}{
    type=hidden,
    name={\ensuremath{\tau_C}},
    description={Yield stress of a Herschel-Bulkley fluid},
    unit={\si{\pascal}}
}
\newglossaryentry{consistency}{
    type=hidden,
    name={\ensuremath{k_n}},
    description={Consistency of a Herschel-Bulkley fluid},
    unit={\si{\pascal\second^n}}
}
\newglossaryentry{indexFlow}{
    type=hidden,
    name={\ensuremath{n}},
    description={Index flow of a Herschel-Bulkley fluid},
    unit={\si{-}}
}
\newglossaryentry{apparentViscosity}{
    type=greek,
    name={\ensuremath{\eta^{\gls{fluid}}}},
    description={Apparent viscosity of a non-Newtonian fluid},
    unit={\si{\pascal\second}}
}
\newglossaryentry{CassonCoefficient}{
    type=latin,
    name={\ensuremath{k}},
    description={Consistency coefficient of Casson's model},
    unit={\si{\pascal\second}}
}

\newglossaryentry{firstInvariant}{
    type=latin,
    name={\ensuremath{I_1}},
    description={First principal invariant of the deformation tensors},
    unit={\si{-}}
}
\newglossaryentry{secondInvariant}{
    type=latin,
    name={\ensuremath{I_2}},
    description={Second principal invariant of the deformation tensors},
    unit={\si{-}}
}
\newglossaryentry{thirdInvariant}{
    type=latin,
    name={\ensuremath{I_3}},
    description={Third principal invariant of the deformation tensors},
    unit={\si{-}}
}
\newglossaryentry{modifiedSecondInvariant}{
    type=latin,
    name={\ensuremath{J_2}},
    description={Modified second invariant of the rate of deformation tensor},
    unit={\si{-}}
}

\newglossaryentry{elasticModuliTensor}{
    type=latin,
    name={\ensuremath{\fourthTensor{C}}},
    description={Elastic moduli tensor},
    unit={\si{\pascal}}
}
\newglossaryentry{elasticModuliTensorComp}{
    type=hidden,
    name={\ensuremath{C}},
    description={Elastic moduli tensor component},
    unit={\si{\pascal}}
}
\newglossaryentry{YoungModulus}{
    type=latin,
    name={\ensuremath{E}},
    description={Young's modulus of a solid},
    unit={\si{\pascal}}
}
\newglossaryentry{shearModulus}{
    type=latin,
    name={\ensuremath{G}},
    description={Shear modulus of a solid},
    unit={\si{\pascal}}
}
\newglossaryentry{OgdenCoeffs}{
    type=greek,
    name={\ensuremath{\mu}},
    description={Generic representation of Ogden's law material constants},
    unit={\si{\pascal}}
}
\newglossaryentry{PoissonRatio}{
    type=greek,
    name={\ensuremath{\nu^{\gls{solid}}}},
    description={Poisson's ratio of a solid},
    unit={\si{-}}
}
\newglossaryentry{strainEnergyFunction}{
    type=greek,
    name={\ensuremath{\Psi}},
    description={%
        Deformation work function or strain energy function
    },
    unit={\si{\joule\per\cubic\meter}}
}
\newglossaryentry{deformationWorkFunction}{
    type=latin,
    name={\ensuremath{\Psi}},
    description={%
        Deformation work function or strain energy function -- function of
        \gls{rightCauchyGreenDeformation}. Energy from deformation work per
        unit volume
    },
    unit={\si{\joule\per\cubic\meter}}
}
\newglossaryentry{lameConst1}{
    type=greek,
    name={\ensuremath{\mu^{\gls{solid}}}},
    description={First Lamé's constant of a solid},
    unit={\si{\pascal}}
}
\newglossaryentry{lameConst2}{
    type=greek,
    name={\ensuremath{\lambda^{\gls{solid}}}},
    description={Second Lamé's constant of a solid},
    unit={\si{\pascal}}
}
\newglossaryentry{nominalStress}{
    type=latin,
    name={\ensuremath{\tensor{P}}},
    description={Nominal stress tensor},
    unit={\si{\pascal}}
}
\newglossaryentry{nominalStressComp}{
    type=hidden,
    name={\ensuremath{P}},
    description={Nominal stress tensor component},
    unit={\si{\pascal}}
}
\newglossaryentry{strainRateTensor}{
    type=latin,
    name={\ensuremath{\tensor{D}}},
    description={Rate of deformation tensor},
    unit={\si{\per\second   }}
}
\newglossaryentry{strainRateTensorComp}{
    type=hidden,
    name={\ensuremath{D}},
    description={Rate of deformation tensor component},
    unit={\si{\per\second}}
}
\newglossaryentry{deformationGradient}{
    type=latin,
    name={\ensuremath{\tensor{F}}},
    description={Deformation gradient tensor},
    unit={\si{-}}
}
\newglossaryentry{deformationGradientComp}{
    type=hidden,
    name={\ensuremath{F}},
    description={Deformation gradient tensor component},
    unit={\si{-}}
}
\newglossaryentry{rotationTensor}{
    type=latin,
    name={\ensuremath{\tensor{R}}},
    description={Rotation tensor},
    unit={\si{-}}
}

\newglossaryentry{stretch}{
    type=greek,
    name={\ensuremath{\lambda}},
    description={%
        Uniaxial stretch of a solid deforming
    },
    unit={\si{-}}
}
\newglossaryentry{deformationGradComp}{
    type=hidden,
    name={\ensuremath{F}},
    description={Deformation gradient tensor component},
    unit={\si{-}}
}
\newglossaryentry{solidDisplacement}{
    type=latin,
    name={\ensuremath{\vec{u}}},
    description={Solid displacement vector},
    unit={m}
}
\newglossaryentry{solidDisplacementComp}{
    type=hidden,
    name={\ensuremath{u}},
    description={Solid displacement vector component},
    unit={\si{-}}
}
\newglossaryentry{xSolidDisplacement}{
    type=hidden,
    name={\ensuremath{u_{\gls{xAxis}}}},
    description={%
        Solid displacement vector component in the x direction of Cartesian
        coordinates
    },
    unit={\si{-}}
}
\newglossaryentry{ySolidDisplacement}{
    type=hidden,
    name={\ensuremath{u_{\gls{yAxis}}}},
    description={%
        Solid displacement vector component in the y direction of Cartesian
        coordinates
    },
    unit={\si{-}}
}
\newglossaryentry{zSolidDisplacement}{
    type=hidden,
    name={\ensuremath{u_{\gls{zAxis}}}},
    description={%
        Solid displacement vector component in the z direction of Cartesian
        coordinates
    },
    unit={\si{-}}
}
\newglossaryentry{secondPiolaKirchhoffStress}{
    type=latin,
    name={\ensuremath{\tensor{S}}},
    description={Second Piola-Kirchhoff stress tensor},
    unit={\si{\pascal}}
}
\newglossaryentry{secondPiolaKirchhoffStressComp}{
    type=hidden,
    name={\ensuremath{S}},
    description={Second Piola-Kirchhoff stress tensor component},
    unit={\si{\pascal}}
}
\newglossaryentry{KirchhoffStress}{
    type=greek,
    name={\ensuremath{\tensor{\sigma}^{\mathrm{K}}}},
    description={Kirchhoff stress tensor},
    unit={\si{\pascal}}
}
\newglossaryentry{KirchhoffStressComp}{
    type=hidden,
    name={\ensuremath{\sigma^{\mathrm{K}}}},
    description={Kirchhoff stress tensor component},
    unit={\si{\pascal}}
}
\newglossaryentry{GLstrain}{
    type=latin,
    name={\ensuremath{\tensor{E}}},
    description={Green-Lagrange strain tensor},
    unit={\si{-}}
}
\newglossaryentry{leftCauchyGreenDeformation}{
    type=latin,
    name={\ensuremath{\tensor{B}}},
    description={Left Cauchy-Green deformation tensor},
    unit={\si{-}}
}
\newglossaryentry{rightCauchyGreenDeformation}{
    type=latin,
    name={\ensuremath{\tensor{C}}},
    description={Right Cauchy-Green deformation tensor},
    unit={\si{-}}
}
\newglossaryentry{2PKstressComp}{
    type=hidden,
    name={\ensuremath{S}},
    description={Second Piola-Kirchhoff stress tensor component},
    unit={\si{\pascal}}
}
\newglossaryentry{GLstrainComp}{
    type=hidden,
    name={\ensuremath{E}},
    description={Green-Lagrange strain tensor component},
    unit={\si{-}}
}
\newglossaryentry{leftCauchyGreenDeformationComp}{
    type=hidden,
    name={\ensuremath{B}},
    description={Left Green-Cauchy deformation tensor component},
    unit={\si{-}}
}
\newglossaryentry{rightCauchyGreenDeformationComp}{
    type=hidden,
    name={\ensuremath{C}},
    description={Right Green-Cauchy deformation tensor component},
    unit={\si{-}}
}
\newglossaryentry{linearStrain}{
    type=greek,
    name={\ensuremath{\tensor{\varepsilon}}},
    description={Linear or engineering strain tensor},
    unit={\si{-}}
}
\newglossaryentry{linearStrainComp}{
    type=hidden,
    name={\ensuremath{\varepsilon}},
    description={Linear or engineering strain tensor},
    unit={\si{-}}
}

\newglossaryentry{trueLinearStrain}{
    type=greek,
    name={\ensuremath{\tensor{e}}},
    description={True or logarithmic strain tensor},
    unit={\si{-}}
}

\newglossaryentry{tensorInvariant}{
    type=hidden,
    name={\ensuremath{I}},
    description={},
    unit={\si{-}}
}
\newglossaryentry{fourthInvariant}{
    type=latin,
    name={\ensuremath{I_4}},
    description={%
        Fourth invariant of the right Cauchy-Green deformation tensor with
        direction information
    },
    unit={\si{-}}
}
\newglossaryentry{fifthInvariant}{
    type=latin,
    name={\ensuremath{I_5}},
    description={%
        Third invariant of the right Cauchy-Green deformation tensor with
        directional information
    },
    unit={\si{-}}
}
\newglossaryentry{sixthInvariant}{
    type=latin,
    name={\ensuremath{I_6}},
    description={%
        Sixth invariant of the right Cauchy-Green deformation tensor with
        directional information
    },
    unit={\si{-}}
}
\newglossaryentry{seventhInvariant}{
    type=latin,
    name={\ensuremath{I_7}},
    description={%
        Seventh invariant of the right Cauchy-Green deformation tensor with
        directional information
    },
    unit={\si{-}}
}
\newglossaryentry{eighthInvariant}{
    type=latin,
    name={\ensuremath{I_8}},
    description={%
        Eighth invariant of the right Cauchy-Green deformation tensor with
        directional information
    },
    unit={\si{-}}
}

\newglossaryentry{MooneyCoeffs}{
    type=hidden,
    name={\ensuremath{c}},
    description={%
        Coefficients of Rivlin series expansion of the strain-energy function
    },
    unit={\si{\kilo\joule\per\cubic\meter}}
}
\newglossaryentry{OgdenExponents}{
    type=hidden,
    name={\ensuremath{\alpha}},
    description={Exponents of Ogden's strain-energy function},
    unit={\si{-}}
}
\newglossaryentry{FungArgumentFunction}{
    type=latin,
    name={\ensuremath{Q}},
    description={%
        Function of the strain tensor components used in the generic Fung-type
        models
    },
    unit={\si{-}}
}
\newglossaryentry{fiberDirection}{
    type=latin,
    name={\ensuremath{\vec{m}}},
    description={%
        Vector of the preferred direction of a fiber reinforced material
    },
    unit={\si{-}}
}
\newglossaryentry{orientationTensor}{
    type=latin,
    name={\ensuremath{\tensor{M}}},
    description={Orientation tensor},
    unit={\si{-}}
}
\newglossaryentry{orientationTensorComp}{
    type=hidden,
    name={\ensuremath{M}},
    description={Orientation tensor component},
    unit={\si{-}}
}
\newglossaryentry{fiberDirectionComp}{
    type=hidden,
    name={\ensuremath{m}},
    description={%
        Vector component of the preferred direction of a fiber reinforced
        material
    },
    unit={\si{-}}
}
\newglossaryentry{groundSubstance}{
    type=subscript,
    name={\ensuremath{g}},
    description={Ground substance}
}
\newglossaryentry{collagen}{
    type=subscript,
    name={\ensuremath{c}},
    description={Collagen fibers}
}
\newglossaryentry{wallLumenRatio}{
    type=latin,
    name={\ensuremath{\mathit{WLR}}},
    description={Wall to lumen ratio},
    unit={\si{-}}
}

\newglossaryentry{fsiInterface}{
    type=greek,
    name={\ensuremath{\gls{domainBoundary}^{\gls{fs}}}},
    description={Fluid-solid interface  = $\Lambda^f \cap \Lambda^s$},
    unit={\si{-}}
}
\newglossaryentry{fsiInterfaceFluid}{
    type=greek,
    name={\ensuremath{\gls{domainBoundary}_{fluid}^{\gls{fs}}}},
    description={%
        Fluid mesh boundary corresponding to the fluid-solid interface
    },
    unit={\si{-}}
}
\newglossaryentry{fsiInterfaceSolid}{
    type=greek,
    name={\ensuremath{\gls{domainBoundary}_{solid}^{\gls{fs}}}},
    description={%
        Solid mesh boundary corresponding to the fluid-solid interface
    },
    unit={\si{-}}
}
\newglossaryentry{fsiTolerance}{
    type=greek,
    name={\ensuremath{{\epsilon}_0}},
    description={Tolerance of the FSI coupling iterations},
    unit={\si{-}}
}
\newglossaryentry{fluidSolver}{
    type=latin,
    name={\ensuremath{\bm{\mathcal{F}}}},
    description={Symbolic representation of the fluid solver},
    unit={\si{-}}
}
\newglossaryentry{solidSolver}{
    type=latin,
    name={\ensuremath{\bm{\mathcal{S}}}},
    description={Symbolic representation of the solid solver},
    unit={\si{-}}
}
\newglossaryentry{addedMassOperator}{
    type=latin,
    name={\ensuremath{\bm{\mathcal{M}_{\mathrm{a}}}}},
    description={Symbolic representation of the added-mass operator},
    unit={\si{-}}
}
\newglossaryentry{addedMassEigenvalue}{
    type=greek,
    name={\ensuremath{\mu_\mathrm{a}}},
    description={Eigenvalue of the added-mass operator},
    unit={\si{-}}
}
\newglossaryentry{fsiTraction}{
    type=latin,
    name={\ensuremath{\cvec{y}}},
    description={%
        Solid solver input: traction exerted by the flow on the FSI interface
    },
    unit={\si{\pascal}}
}
\newglossaryentry{fsiInterfaceDispl}{
    type=latin,
    name={\ensuremath{\cvec{d}}},
    description={Position of the FSI interface},
    unit={\si{\meter}}
}
\newglossaryentry{fsDisplOutput}{
    type=latin,
    name={\ensuremath{\cvec{\tilde{d}}}},
    description={FSI interface position after a coupling iteration},
    unit={\si{\meter}}
}
\newglossaryentry{fsiResidualOperator}{
    type=latin,
    name={\ensuremath{\bm{\mathcal{R}}}},
    description={%
        Residual operator resulted from the fixed-point formulation of the FSI
        problem
    },
    unit={\si{\meter}}
}
\newglossaryentry{fsiInterfaceDoF}{
    type=latin,
    name={\ensuremath{N^{\gls{fs}}}},
    description={Number of degrees of freedom on the FSI interface},
    unit={\si{-}}
}
\newglossaryentry{fsiResidual}{
    type=latin,
    name={\ensuremath{\cvec{r}}},
    description={Discrete residual vector of the coupling iterations},
    unit={\si{\meter}}
}
\newglossaryentry{fsiReuse}{
    type=latin,
    name={\ensuremath{q}},
    description={Number of time-steps reused in the IQN-ILS algorithm},
    unit={\si{-}}
}

\newglossaryentry{peakSystole}{
    type=subscript,
    name={\ensuremath{ps}},
    description={%
        Indicates variables defined at the peak systole instant
    }
}
\newglossaryentry{lowDiastole}{
    type=subscript,
    name={\ensuremath{ld}},
    description={%
        Indicates variables defined at the low diastole instant
    }
}
\newglossaryentry{aneurysm}{
    type=subscript,
    name={\ensuremath{ia}},
    description={%
        Indicates variables defined or integrated on an aneurysm surface
    }
}
\newglossaryentry{convexHull}{
    type=subscript,
    name={\ensuremath{ch}},
    description={%
        Indicates variables defined or integrated on an aneurysm surface
        convex hull
    }
}
\newglossaryentry{ostium}{
    type=subscript,
    name={\ensuremath{o}},
    description={%
        Indicates variables defined or integrated on an aneurysm ostium
        surface
    }
}

\newglossaryentry{domeHeight}{
    type=latin,
    name={\ensuremath{h_d}},
    description={%
        Maximum height of an aneurysm's dome perpendicular to ostium
        surface
    },
    unit={\si{\meter}}
}
\newglossaryentry{maximumDomeHeight}{
    type=latin,
    name={\ensuremath{h_{max}}},
    description={%
        Maximum height of an aneurysm's dome, measured from the ostium center
    },
    unit={\si{\meter}}
}
\newglossaryentry{bulgeHeight}{
    type=latin,
    name={\ensuremath{h_b}},
    description={%
        Bulge height of an aneurysm's dome
    },
    unit={\si{\meter}}
}
\newglossaryentry{neckDiameter}{
    type=latin,
    name={\ensuremath{d_n}},
    description={Maximum diameter of an aneurysm's neck},
    unit={\si{\meter}}
}
\newglossaryentry{domeDiameter}{
    type=latin,
    name={\ensuremath{d_d}},
    description={Maximum diameter of an aneurysm's dome},
    unit={\si{\meter}}
}
\newglossaryentry{aspectRatio}{
    type=latin,
    name={\ensuremath{\mathit{AR}}},
    description={%
        Aspect ratio of an aneurysm, defined as the ratio between dome height
        and neck diameter
    },
    unit={\si{-}}
}
\newglossaryentry{meanBloodFlowRate}{
    type=latin,
    name={\ensuremath{\bar{q}_a}},
    description={Mean blood flow rate through an artery section},
    unit={\si{\milli\litre\per\second}}
}
\newglossaryentry{timeBloodFlowRate}{
    name={\ensuremath{q_a}},
    description={%
        Blood flow rate over time through an artery section
    },
    unit={\si{\milli\litre\per\second}}
}
\newglossaryentry{normBloodFlowRate}{
    name={\ensuremath{q^{*}}},
    description={%
        Normalized blood flow rate profile through the ICA section measured
        experimentally for a population.
    },
    unit={\si{-}}
}
\newglossaryentry{arteryDiameter}{
    type=latin,
    name={\ensuremath{d_{pa}}},
    description={Diameter of an aneurysm's parent artery section},
    unit={\si{\meter}}
}

\newglossaryentry{bottleneckFactor}{
    type=latin,
    name={\ensuremath{\mathit{BF}}},
    description={Bottleneck factor},
    unit={\si{-}}
}
\newglossaryentry{bulgeLocation}{
    type=latin,
    name={\ensuremath{\mathit{BL}}},
    description={Bulge location},
    unit={\si{-}}
}
\newglossaryentry{sizeRatio}{
    type=latin,
    name={\ensuremath{\mathit{SR}}},
    description={Size ratio},
    unit={\si{-}}
}
\newglossaryentry{convexityRatio}{
    type=latin,
    name={\ensuremath{\mathit{CR}}},
    description={Convexity ratio},
    unit={\si{-}}
}
\newglossaryentry{isoperimetricRatio}{
    type=latin,
    name={\ensuremath{\mathit{IPR}}},
    description={Isoperimetric ratio},
    unit={\si{-}}
}
\newglossaryentry{undulationIndex}{
    type=latin,
    name={\ensuremath{\mathit{UI}}},
    description={Undulation index},
    unit={\si{-}}
}
\newglossaryentry{ellipticityIndex}{
    type=latin,
    name={\ensuremath{\mathit{EI}}},
    description={Ellipticity index},
    unit={\si{-}}
}
\newglossaryentry{nonsphericityIndex}{
    type=latin,
    name={\ensuremath{\mathit{NSI}}},
    description={Nonsphericity index},
    unit={\si{-}}
}
\newglossaryentry{conicityParameter}{
    type=latin,
    name={\ensuremath{\mathit{CP}}},
    description={Conicity Parameter},
    unit={\si{-}}
}
\newglossaryentry{firstPrincipalCurvature}{
    type=greek,
    name={\ensuremath{\kappa_{1}}},
    description={First principal curvature of a surface},
    unit={\si{\per\meter}}
}
\newglossaryentry{secondPrincipalCurvature}{
    type=greek,
    name={\ensuremath{\kappa_{2}}},
    description={Second principal curvature of a surface},
    unit={\si{\per\meter}}
}
\newglossaryentry{meanCurvature}{
    type=latin,
    name={\ensuremath{H}},
    description={Mean curvature of a surface},
    unit={\si{\per\meter}}
}
\newglossaryentry{GaussianCurvature}{
    type=latin,
    name={\ensuremath{K}},
    description={Gaussian curvature of a surface},
    unit={\si{\per\square\meter}}
}
\newglossaryentry{areaAvgGaussianCurvature}{
    type=latin,
    name={\ensuremath{\mathit{GAA}}},
    description={Area-averaged Gaussian curvature of a surface},
    unit={\si{\per\square\milli\meter}}
}
\newglossaryentry{areaAvgMeanCurvature}{
    type=latin,
    name={\ensuremath{\mathit{MAA}}},
    description={Area-averaged mean curvature of a surface},
    unit={\si{\per\milli\meter}}
}
\newglossaryentry{l2NormGaussianCurvature}{
    type=latin,
    name={\ensuremath{\mathit{GLN}}},
    description={L2-norm of the Gaussian curvature field of a surface},
    unit={\si{-}}
}
\newglossaryentry{l2NormMeanCurvature}{
    type=latin,
    name={\ensuremath{\mathit{MLN}}},
    description={L2-norm of the mean curvature of a surface},
    unit={\si{-}}
}

\newglossaryentry{wallShearStressVector}{
    type=greek,
    name={\ensuremath{ \vec{\tau}_w }},
    description={%
        Wall shear stress vector defined on a vessel and aneurysm surface
    },
    unit={\si{\pascal}}
}
\newglossaryentry{wallShearStressMag}{
    type=hidden,
    name={\ensuremath{{\tau}_w}},
    description={%
        Wall shear stress magnitude defined on a vessel and aneurysm surface
    },
    unit={\si{\pascal}}
}
\newglossaryentry{timeAvgWallShearStress}{
    type=latin,
    name={\ensuremath{\mathit{TAWSS}}},
    description={%
        Time-average WSS field
    },
    unit={\si{\pascal\per\meter}}
}

\newglossaryentry{wallShearStressGradient}{
    type=latin,
    name={\ensuremath{\vec{G}}},
    description={%
        Wall shear stress gradient vector, defined on the surface
        $pq$-coordinate system
    },
    unit={\si{\pascal\per\meter}}
}
\newglossaryentry{wallShearStressSpatialGradient}{
    type=latin,
    name={\ensuremath{\avg{G}}},
    description={%
        Temporal-average of the wall shear stress (spatial) gradient vector
        magnitude, defined on the surface $pq$-coordinate system,
        \gls{wallShearStressGradient}
    },
    unit={\si{\pascal\per\meter}}
}
\newglossaryentry{timeAvgWallShearStressGradient}{
    type=latin,
    name={\ensuremath{\mathit{TAWSSG}}},
    description={%
        Surface gradient of the time-averaged wall shear stress gradient
        vector, defined on the surface $p$-coordinate direction (the flow
        direction)
    },
    unit={\si{\pascal\per\meter}}
}
\newglossaryentry{avgShearDirectionVector}{
    type=latin,
    name={\ensuremath{\vec{p}}},
    description={Mean flow direction adjacent to an artery's lumen},
    unit={\si{-}}
}
\newglossaryentry{transShearDirectionVector}{
    type=latin,
    name={\ensuremath{\vec{q}}},
    description={%
        Transversal direction to the mean flow direction (\vec{p})
    },
    unit={\si{-}}
}
\newglossaryentry{oscillatoryShearIndex}{
    type=latin,
    name={\ensuremath{\mathit{OSI}}},
    description={The oscillatory shear index},
    unit={\si{-}}
}
\newglossaryentry{relativeResidenceTime}{
    type=latin,
    name={\ensuremath{\mathit{RRT}}},
    description={The relative residence time indicator},
    unit={\si{-}}
}
\newglossaryentry{gradientOscillatoryNumber}{
    type=latin,
    name={\ensuremath{\mathit{GON}}},
    description={The gradient oscillatory number},
    unit={\si{-}}
}
\newglossaryentry{aneurysmFormationIndicator}{
    type=latin,
    name={\ensuremath{\mathit{AFI}}},
    description={The \textit{potential} aneurysm formation indicator},
    unit={\si{-}}
}
\newglossaryentry{cardiacPeriod}{
    type=latin,
    name={\ensuremath{T_c}},
    description={Cardiac period},
    unit={\si{\second}}
}
\newglossaryentry{lowShearArea}{
    type=latin,
    name={\ensuremath{\mathit{lsa}}},
    description={Normalized area of low WSS defined on the aneurysm surface},
    unit={\si{-}}
}
\newglossaryentry{avgParentArteryWSS}{
    type=latin,
    name={\ensuremath{\avg{\tau}_{w,pa}}},
    description={%
        Averaged parent artery WSS, defined in the anterior portion of the
        parent artery
    },
    unit={\si{\pascal}}
}
\newglossaryentry{lowWSSReference}{
    type=latin,
    name={\ensuremath{(\gls{wallShearStressMag})_{low}}},
    description={Low WSS reference},
    unit={\si{-}}
}
\newglossaryentry{wssPulsatilityIndex}{
    type=latin,
    name={\ensuremath{\mathit{WSSPI}}},
    description={Pulsatility index of the wall shear stress magnitude},
    unit={\si{-}}
}
\newglossaryentry{wssTemporalGradient}{
    type=latin,
    name={\ensuremath{\mathit{WSSTG}}},
    description={%
        Maximal temporal derivative of the wall shear stress magnitude
    },
    unit={\si{\pascal\per\second}}
}
\newglossaryentry{transWallShearStress}{
    type=latin,
    name={\ensuremath{\mathit{transWSS}}},
    description={%
        Temporal average of the wall shear stress vector along its average
        transversal direction
    },
    unit={\si{\pascal}}
}
\newglossaryentry{speedup}{
    type=latin,
    name={\ensuremath{s}},
    description={Speedup},
    unit={\si{-}}
}
\newglossaryentry{residualVector}{
    type=latin,
    name={\ensuremath{\cvec{r}}},
    description={%
        Residual vector of a linear system $\matrix{A} \cvec{x} = \cvec{b}$
    },
    unit={\si{-}}
}
\newglossaryentry{residualNorm}{
    type=hidden,
    name={\ensuremath{R_s}},
    description={%
        Scaled normalized residual of a linear system in an iterative solver
    },
    unit={\si{-}}
}
\newglossaryentry{scaleFactor}{
    type=latin,
    name={\ensuremath{F_s}},
    description={Scale factor used for scaling the residual norm},
    unit={\si{-}}
}
\newglossaryentry{identityMatrix}{
    type=hidden,
    name={\ensuremath{\matrix{I}}},
    description={Identity matrix of $\realnumbers^{N \times N}$},
    unit={\si{-}}
}

\newglossaryentry{turbIntegralLength}{
    type=latin,
    name={\ensuremath{\mathcal{L}}},
    description={Length integral scale of turbulent flow},
    unit={\si{\meter}}
}
\newglossaryentry{turbIntegralTime}{
    type=latin,
    name={\ensuremath{\mathcal{T}}},
    description={Time integral scale of turbulent flow},
    unit={\si{\second}}
}
\newglossaryentry{turbIntegralVelocity}{
    type=latin,
    name={\ensuremath{\mathcal{V}}},
    description={Velocity integral scale of turbulent flow},
    unit={\si{\meter\per\second}}
}
\newglossaryentry{KolmogorovLength}{
    type=latin,
    name={\ensuremath{\eta}},
    description={Kolmogorov length scale of turbulent flow},
    unit={\si{\meter}}
}
\newglossaryentry{KolmogorovTime}{
    type=latin,
    name={\ensuremath{\tau}},
    description={KOlmogorov time scale of turbulent flow},
    unit={\si{\second}}
}
\newglossaryentry{KolmogorovVelocity}{
    type=latin,
    name={\ensuremath{\upsilon}},
    description={Kolmogorov velocity scale of turbulent flow},
    unit={\si{\meter\per\second}}
}
\newglossaryentry{fluctuationVelocity}{
    type=latin,
    name={\ensuremath{\vec{u^{'}}}},
    description={Vetor de velocidade flutuante},
    unit={\si{\meter\per\second}}
}
\newglossaryentry{meanVelocity}{
	type=latin,	
    name={\ensuremath{\overline{\gls{velocityComp}}}},				
	description={Vetor de velocidade média},
	unit={\si{\meter\per\second}}
}
\newglossaryentry{yPlus}{
    type=latin,
    name={\ensuremath{y^+}},
    description={Dimensionless wall distance},
    unit={-}
}
\newglossaryentry{turbulenceIntensity}{
    type=latin,
    name={\ensuremath{I_t}},
    description={Turbulence intensity for inlet boundaries},
    unit={\si{-}}
}

\newglossaryentry{firstCellSize}{
    type=latin,
    name={\ensuremath{h}},
    description={The distance nearest to the wall of the first cell},
    unit={\si{\meter}}
}

\newglossaryentry{boundaryLayerThickness}{
    type=latin,
    name={\ensuremath{\delta}},
    description={Boundary-layer thickness},
    unit={\si{\meter}}
}
\newglossaryentry{velocityPlus}{
    type=latin,
    name={\ensuremath{u^{+}}},
    description={Dimensionless velocity in the turbulent boundary layer},
    unit={\si{\meter\per\second}}
}
\newglossaryentry{frictionVelocity}{
    type=latin,
    name={\ensuremath{u_{\tau}}},
    description={Friction velocity},
    unit={\si{\meter\per\second}}
}
\newglossaryentry{turbKineticEnergy}{
    type=latin,
    name={\ensuremath{k}},
    description={Energia cinética turbulenta por unidade de massa},
    unit={\si{\square\meter\per\square\second}}
}

\newglossaryentry{turbDissipationRate}{
    type=greek,
    name={\ensuremath{\varepsilon}},
    description={Taxa de dissipação da energia cinética turbulenta},
    unit={\si{\square\meter\per\cubic\second}}
}

\newglossaryentry{turbDynamicViscosity}{
    type=greek,
    name={\ensuremath{\mu_{\text{t}}}},
    description={Viscosidade din\^amica turbulenta},
    unit={\si{\pascal\second}}
}
\newglossaryentry{turbViscosity}{
    type=greek,
    name={\ensuremath{\nu_{\text{t}}}},
    description={Viscosidade cinemática turbulenta},
    unit={\si{\square\meter\per\second}}
}
\newglossaryentry{eddyViscosityParameter}{
    type=greek,
    name={\ensuremath{\tilde{\nu}}},
    description={%
        Kinematic eddy viscosity parameter of the Spalart-Allmaras models%
    },
    unit={\si{\square\meter\per\second}}
}
\newglossaryentry{wallDampingFunction}{
    type=greek,
    name={\ensuremath{f_{\nu 1}}},
    description={%
        Wall-damping function of the Spalart-Allmaras model%
    },
    unit={\si{-}}
}

\newglossaryentry{effTurbViscosity}{
    type=greek,
    name={\ensuremath{\nu_{\text{ef}}}},
    description={Viscosidade cinemática efetiva},
    unit={\si{\square\meter\per\second}}
}

\newglossaryentry{effkDiffusivity}{
    type=latin,
    name={\ensuremath{d_{\gls{turbKineticEnergy},\text{ef}}}},
    description={Difusividade efetiva de $k$},
    unit={\si{\square\meter\per\second}}
}

\newglossaryentry{effEpsilonDiffusivity}{
    type=latin,
    name={\ensuremath{d_{\gls{turbDissipationRate},\text{ef}}}},
    description={Difusividade efetiva de $\epsilon$},
    unit={\si{\square\meter\per\second}}
}

\newglossaryentry{effOmegaDiffusivity}{
    type=latin,
    name={\ensuremath{d_{\omega,\text{ef}}}},
    description={Difusividade efetiva de $\omega$},
    unit={\si{\square\meter\per\second}}
}

\newglossaryentry{effPhiDiffusivity}{
    type=latin,
    name={\ensuremath{d_{\phi,\text{ef}}}},
    description={Difusividade efetiva de $\phi$},
    unit={\si{\square\meter\per\second}}
}

\newglossaryentry{turbKineticEnergyProduction}{
    type=latin,
    name={\ensuremath{P_{\gls{turbKineticEnergy}}}},
    description={Produção de energia cinética turbulenta},
    unit={\si{\square\meter\per\cubic\second}}
}

\newglossaryentry{limitedTurbProduction}{
    type=latin,
    name={\ensuremath{P_{k}^{'}}},
    description={Produção de energia cinética turbulenta limitado},
    unit={\si{\square\meter\per\cubic\second}}
}

\newglossaryentry{limitedOmegaTurbProduction}{
    type=latin,
    name={\ensuremath{P_{\omega}^{'}}},
    description={Produção de energia cinética turbulenta limitado},
    unit={\si{\square\meter\per\cubic\second}}
}

\newglossaryentry{specificDissipationRate}{
    type=greek,
    name={\ensuremath{\omega}},
    description={Turbulence frequency},
    unit={\si{\per\second}}
}

\newglossaryentry{F1}{
    type=latin,
    name={\ensuremath{F_{1}}},
    description={Função do modelo k-$\omega$ SST},
    unit={-}
}

\newglossaryentry{F2}{
    type=latin,
    name={\ensuremath{F_{2}}},
    description={Função do modelo k-$\omega$ SST},
    unit={-}
}

\newglossaryentry{nearestDistWall}{
    type=latin,
    name={\ensuremath{y_{\perp}}},
    description={Distância mais próxima da parede},
    unit={\si{\meter}}
}

\newglossaryentry{cdkw}{
    type=latin,
    name={\ensuremath{cd_{k\omega}}},
    description={Parte positiva do termo cruzado de difusão da equação de transporte de $\omega$},
    unit={\si{1\per\second^{4}}}
}

\newglossaryentry{psi}{
    type=latin,
    name={\ensuremath{\psi}},
    description={Coeficientes do modelo de turbulência k-$omega$ SST},
    unit={-}
}

\newglossaryentry{cmu}{
    type=latin,
    name={\ensuremath{C_{\mu}}},
    description={Constante do modelo de turbulência},
    unit={-}
}

\newglossaryentry{kPrandtlNumber}{
    type=greek,
    name={\ensuremath{\sigma_{k}}},
    description={Número de Prandtl turbulento para $k$},
    unit={-}
}

\newglossaryentry{epsilonPrandtlNumber}{
    type=greek,
    name={\ensuremath{\sigma_{\epsilon}}},
    description={Número de Prandtl turbulento para $\epsilon$},
    unit={-}
}
\newglossaryentry{turbFrequencyPrandtlNumber}{
    type=greek,
    name={\ensuremath{\sigma_{\omega}}},
    description={Turbulence frequency Prandtl number},
    unit={-}
}

\newglossaryentry{alphaPhi}{
    type=greek,
    name={\ensuremath{\alpha_{\phi}}},
    description={Inverso do número de Prandtl turbulento},
    unit={-}
}

\newglossaryentry{a1}{
    type=latin,
    name={\ensuremath{a_{1}}},
    description={Constante do modelo de turbulência},
    unit={-}
}

\newglossaryentry{b1}{
    type=latin,
    name={\ensuremath{b_{1}}},
    description={Constante do modelo de turbulência},
    unit={-}
}

\newglossaryentry{c1}{
    type=latin,
    name={\ensuremath{c_{1}}},
    description={Constante do modelo de turbulência},
    unit={-}
}

\newglossaryentry{C1epsilon}{
    type=latin,
    name={\ensuremath{C_{1\gls{turbDissipationRate}}}},
    description={Constante do modelo de turbulência},
    unit={-}
}

\newglossaryentry{C2epsilon}{
    type=latin,
    name={\ensuremath{C_{2\gls{turbDissipationRate}}}},
    description={Constante do modelo de turbulência},
    unit={-}
}

\newglossaryentry{beta}{
    type=greek,
    name={\ensuremath{\beta}},
    description={Constante do modelo de turbulência},
    unit={-}
}

\newglossaryentry{betaStar}{
    type=greek,
    name={\ensuremath{\beta^{*}}},
    description={Constante do modelo de turbulência},
    unit={-}
}

\newglossaryentry{gamma}{
    type=greek,
    name={\ensuremath{\gamma}},
    description={Constante do modelo de turbulência},
    unit={-}
}

\newglossaryentry{sigmaOmega1}{
    type=greek,
    name={\ensuremath{\sigma_{\omega 1}}},
    description={Constante do modelo de turbulência},
    unit={-}
}

\newglossaryentry{sigmaOmega2}{
    type=greek,
    name={\ensuremath{\sigma_{\omega 2}}},
    description={Constante do modelo de turbulência},
    unit={-}
}

\newglossaryentry{alphak1}{
    type=greek,
    name={\ensuremath{\alpha_{k 1}}},
    description={Constante do modelo de turbulência},
    unit={-}
}

\newglossaryentry{alphak2}{
    type=greek,
    name={\ensuremath{\alpha_{k 2}}},
    description={Constante do modelo de turbulência},
    unit={-}
}

\newglossaryentry{alphaOmega1}{
    type=greek,
    name={\ensuremath{\alpha_{\omega 1}}},
    description={Constante do modelo de turbulência},
    unit={-}
}

\newglossaryentry{alphaOmega2}{
    type=greek,
    name={\ensuremath{\alpha_{\omega 2}}},
    description={Constante do modelo de turbulência},
    unit={-}
}

\newglossaryentry{beta1}{
    type=greek,
    name={\ensuremath{\beta_{1}}},
    description={Constante do modelo de turbulência},
    unit={-}
}

\newglossaryentry{beta2}{
    type=greek,
    name={\ensuremath{\beta_{2}}},
    description={Constante do modelo de turbulência},
    unit={-}
}

\newglossaryentry{gamma1}{
    type=greek,
    name={\ensuremath{\gamma_{1}}},
    description={Constante do modelo de turbulência},
    unit={-}
}

\newglossaryentry{gamma2}{
    type=greek,
    name={\ensuremath{\gamma_{2}}},
    description={Constante do modelo de turbulência},
    unit={-}
}
\newglossaryentry{ReynoldsTensor}{
    type=greek,
    name={\ensuremath{\tensor{\tau^{R}}}},
    description={Tensor das tensões de Reynolds},
    unit={\si{\per\second}}
}
\newglossaryentry{ReynoldsTensorComp}{
    type=hidden,
    name={\ensuremath{\tau^{R}}},
    description={Tensor das tensões de Reynolds},
    unit={\si{\per\second}}
}

\newglossaryentry{meanStrainRateTensor}{
    type=latin,
    name={\ensuremath{\overline{\gls{strainRateTensor}}}},
    description={Tensor taxa de deformação média},
    unit={\si{\per\second}}
}

\newglossaryentry{mesh}{
    type=hidden,
    name={\ensuremath{M}},
    description={%
        Símbolo genérico representando uma malha computacional
    },
    unit={\si{-}}
}
\newglossaryentry{coarse}{
    type=subscript,
    name={\ensuremath{c}},
    description={%
        Indica propriedades da malha mais grosseira (\textit{coarse})
    },
    unit={\si{-}}
}
\newglossaryentry{intermediate}{
    type=subscript,
    name={\ensuremath{i}},
    description={%
        Indica propriedades da malha intermediária
    },
    unit={\si{-}}
}
\newglossaryentry{fine}{
    type=subscript,
    name={\ensuremath{f}},
    description={%
        Indica propriedades da malha mais refinada (\textit{finer})
    },
    unit={\si{-}}
}
\newglossaryentry{meshCellsNumber}{
    type=hidden,
    name={\ensuremath{N_M}},
    description={%
        Number of cells in a computational mesh
    },
    unit={\si{-}}
}
\newglossaryentry{meshSurfaceCellDensity}{
    type=hidden,
    name={\ensuremath{\rho_{S}}},
    description={%
        Cells surface density in a computational mesh
    },
    unit={\si{cells\per\milli\square\meter}}
}
\newglossaryentry{meshVolumeCellDensity}{
    type=hidden,
    name={\ensuremath{\rho_{V}}},
    description={%
        Cells volumetric density in a computational mesh
    },
    unit={\si{cells\per\milli\cubic\meter}}
}
\newglossaryentry{maxEdgeSize}{
    type=hidden,
    name={\ensuremath{l_e}},
    description={%
        Maximum edge length size in a computational mesh
    },
    unit={\si{\milli\meter}}
}
\newglossaryentry{meshSize}{
    type=hidden,
    name={\ensuremath{h}},
    description={%
        Edge length size of a generic structured computational mesh
    },
    unit={\si{\milli\meter}}
}
\newglossaryentry{meshRefinementRatio}{
    type=hidden,
    name={\ensuremath{r}},
    description={%
        Ratio between element size of systematically refined meshes
    },
    unit={\si{-}}
}
\newglossaryentry{fvmOrder}{
    type=hidden,
    name={\ensuremath{p}},
    description={%
        Formal order of accuracy of the Finite Volume Method
    },
    unit={\si{-}}
}
\newglossaryentry{realFvmOrder}{
    type=hidden,
    name={\ensuremath{\tilde{p}}},
    description={%
        Real or practical order of accuracy of the Finite Volume Method
    },
    unit={\si{-}}
}
\newglossaryentry{solutionDifference}{
    type=latin,
    name={\ensuremath{\epsilon}},
    description={%
        Diferença (absoluta ou relativa) entre as soluções de duas malhas
        sistematicamente refinadas
    },
    unit={\si{-}}
}
\newglossaryentry{discretizationError}{
    type=latin,
    name={\ensuremath{E^{\mathrm{R}}}},
    description={%
        Discretization error of a particular mesh computed using Richardson's
        extrapolation
    },
    unit={\si{-}}
}
\newglossaryentry{truncationError}{
    type=latin,
    name={\ensuremath{e^{\mathrm{t}}}},
    description={%
        Truncation error of a discretization scheme.
    },
    unit={\si{-}}
}
\newglossaryentry{gridConvergenceIndex}{
    type=latin,
    name={\ensuremath{\mathit{GCI}}},
    description={%
        Grid convergence index
    },
    unit={\si{-}}
}
\newglossaryentry{gciSafetyFactor}{
    type=latin,
    name={\ensuremath{F_s}},
    description={%
        Safety factor defined with the grid convergence index
    },
    unit={\si{-}}
}

\newacronym[category=standard,type=hidden]{1d}{1D}{one-dimensional}
\newacronym[category=standard,type=hidden]{2d}{2D}{two-dimensional}
\newacronym[category=standard,type=hidden]{3d}{3D}{three-dimensional}
\newacronym[category=standard,type=hidden]{4d}{4D}{four-dimensional}

\newacronym{cfd}{CFD}{Computational Fluid Dynamics}
\newacronym{csd}{CSD}{Computational Solid Dynamics}
\newacronym{cae}{CAE}{Computer-Aided Engineering}
\newacronym{fea}{FEA}{Finite Element Analysis}
\newacronym{cem}{CEM}{Computational Electromagnetic}
\newacronym{mbd}{MBD}{Multi-body Dynamics}
\newacronym{fsi}{FSI}{Fluid-Solid Interaction}

\newacronym{pde}{PDE}{partial differential equations}
\newacronym{fvm}{FVM}{Finite Volume Method}
\newacronym{fdm}{FDM}{Finite Difference Method}
\newacronym{fem}{FEM}{Finite Element Method}
\newacronym{ale}{ALE}{arbitrary Lagrangian-Eulerian}
\newacronym{iqn-ils}{IQN-ILS}{Interface Quasi-Newton-Implicit Jacobian Least-Squares}
\newacronym[category=only-short,type=hidden]{simple}{SIMPLE}{Semi-Implicit Method for Pressure Linked Equations}
\newacronym[category=only-short,type=hidden]{piso}{PISO}{Pressure Implicit with Splitting Operators}

\newacronym{bc}{BC}{boundary condition}
\newacronym{am}{AM}{added-mass}
\newacronym{dnf}{DNF}{Dirichlet-Neumann formulation}
\newacronym{dof}{DOF}{degrees of freedom}

\newacronym{ia}{IA}{intracranial aneurysm}
\newacronym{aaa}{AAA}{abdominal aortic aneurysm}
\newacronym{voi}{VOI}{volume of interest}
\newacronym{dsa}{DSA}{digital subtraction angiography}
\newacronym{cta}{CTA}{computational tomography angiography}
\newacronym{wlr}{WLR}{wall-to-lumen ratio}
\newacronym{pae}{PAE}{peri-aneurysmal environment}
\newacronym{sah}{SAH}{subarachnoid hemorrhage}

\newacronym[category=standard,type=hidden]{vmtk}{VMTK}{Vascular Modeling Toolkit}
\newacronym[category=standard,type=hidden]{vtk}{VTK}{Visualization Toolkit}

\newacronym{wss}{WSS}{wall shear stress}
\newacronym{pswss}{PSWSS}{peak systole wall shear stress}
\newacronym{tawss}{TAWSS}{time-averaged wall shear stress}
\newacronym{tawssg}{TAWSSG}{time-averaged WSS gradient}
\newacronym{wsstg}{WSSTG}{maximum wall shear stress temporal gradient}
\newacronym{wsssg}{WSSSG}{wall shear stress surface gradient}
\newacronym{wsspi}{WSSPI}{wall shear stress pulsatility index}
\newacronym{transwss}{transWSS}{transversal wall shear stress}
\newacronym{osi}{OSI}{oscillatory shear index}
\newacronym{rrt}{RRT}{relative residence time}
\newacronym{afi}{AFI}{aneurysm formation indicator}
\newacronym{gon}{GON}{gradient oscillatory number}
\newacronym{lsa}{LSA}{low shear area}

\newacronym[category=never-short,type=hidden]{mca}{MCA}{middle cerebral artery}
\newacronym[category=never-short,type=hidden]{ica}{ICA}{internal carotid artery}
\newacronym[category=never-short,type=hidden]{aca}{ACA}{anterior cerebral artery}
\newacronym[category=never-short,type=hidden]{acoa}{ACoA}{anterior communicating artery}
\newacronym{pica}{PICA}{posterior inferior cerebellar artery}
\newacronym{pca}{PCA}{posterior cerebral artery}
\newacronym{ba}{BA}{basilar artery}
\newacronym{va}{VA}{vertebral arteries}
\newacronym{oa}{OA}{ophthalmic artery}

\newacronym[category=standard,type=hidden]{fapesp}{FAPESP}{São Paulo Research Foundation}
\newacronym[category=standard,type=hidden]{unesp}{UNESP}{São Paulo State University}
\newacronym[category=never-short,type=hidden]{hiae}{HIAE}{Albert Einstein Israeli Hospital}
\newacronym[category=never-short,type=hidden]{hbp}{HBP}{Beneficiência Portuguesa Hospital}
\newacronym{feb}{FEB}{School of Engineering of Bauru}
\newacronym{feis}{FEIS}{School of Engineering of Ilha Solteira}
\newacronym{rdidp}{RDIDP}{Regime de Dedicação Integral à Docência e à Pesquisa}
\newacronym{lncc}{LNCC}{Laboratório Nacional de Computação Científica}
\newacronym{ic}{IC}{scientific initiation}

\newacronym{cep}{CEP}{Committee of Research Ethics}
\newacronym{hpc}{HPC}{high-performance computing}

\newacronym{ui}{UI}{undulation index}

\newacronym{web}{WEB}{Woven EndoBridge}
\newacronym[category=never-short,type=hidden]{wnba}{WNBA}{wide-necked bifurcation aneurysm}
\newacronym[category=standard,type=hidden]{mri}{MRI}{magnetic resonance imaging}
\newacronym[category=never-short,type=hidden]{1wfsi}{1WFSI}{one-way fluid-solid interaction}

\newacronym[category=standard,type=hidden]{uiats}{UIATS}{unruptured \gls{ia} treatment score}
\newacronym[category=standard,type=hidden]{rrs}{RRS}{rupture resamblance score}
\newacronym[category=never-short,type=hidden]{uwm}{UWM}{uniform-wall model}
\newacronym[category=never-short,type=hidden]{awm}{AWM}{abnormal-wall model}
\newacronym{sd}{SD}{standard deviation}

\newglossaryentry{vasculature}{
    type=subscript,
    name={\ensuremath{v}},
    description={%
        Indicates variables defined or integrated on the vasculature surface
    }
}
\newglossaryentry{parentArtery}{
    type=subscript,
    name={\ensuremath{pa}},
    description={%
        Indicates variables defined or integrated on the parent artery surface
        of an aneurysm
    }
}
\newglossaryentry{branch}{
    type=subscript,
    name={\ensuremath{b}},
    description={%
        Indicates variables defined on the branches of a vasculature
    }
}
\newglossaryentry{populationAvgDiameter}{
    type=latin,
    name={\ensuremath{d_{avg}}},
    description={%
        Populational-average diameter of the ICA section
    },
    unit={\si{\milli\meter}}
}
\newglossaryentry{aneurysmPulsatility}{
    type=greek,
    name={\ensuremath{\delta_{v}}},
    description={Pulsatility of an intracranial aneurysm},
    unit={\si{-}}
}
\newglossaryentry{distanceToNeckLine}{
    type=latin,
    name={\ensuremath{g_n}},
    description={%
        Local geodesic distance to the aneurysm neck contour
    },
    unit={\si{\milli\meter}}
}
\newglossaryentry{lumenDiameter}{
    type=latin,
    name={\ensuremath{d_l}},
    description={%
        Vasculature lumen diameter at a specified section
    },
    unit={\si{\milli\meter}}
}
\newglossaryentry{intracranialPressure}{
    type=latin,
    name={\ensuremath{\gls{pressure}_{o}}},
    description={Intracranial pressure},
    unit={\si{\pascal}}
}
\newglossaryentry{patientBloodFlowRate}{
    name={\ensuremath{q_p}},
    description={%
        Blood flow rate over time through an artery section of a particular
        patient
    },
    unit={\si{\milli\litre\per\second}}
}
\newglossaryentry{ellipticPatch}{
    type=subscript,
    name={\ensuremath{E}},
    description={%
        Elliptic patches of a surface
    }
}
\newglossaryentry{hyperbolicPatch}{
    type=subscript,
    name={\ensuremath{H}},
    description={%
        Hyperbolic patches of a surface
    }
}
\newglossaryentry{rupturedGroup}{
    type=latin,
    name={\ensuremath{\mathcal{R}}},
    description={%
        The group of ruptured aneurysms
    },
    unit={\si{-}}
}
\newglossaryentry{unrupturedGroup}{
    type=latin,
    name={\ensuremath{\mathcal{U}}},
    description={%
        The group of unruptured aneurysms
    },
    unit={\si{-}}
}
\newglossaryentry{pValue}{
    type=hidden,
    name={\ensuremath{p}},
    description={P-value of statistical tests},
    unit={\si{-}}
}
\newglossaryentry{significanceLevel}{
    type=hidden,
    name={\ensuremath{\alpha}},
    description={Significance level for statistical tests},
    unit={\si{-}}
}
\newglossaryentry{standardDeviation}{
    type=hidden,
    name={\ensuremath{\mathit{SD}}},
    description={Standard-deviation of a sample's distribution},
    unit={\si{-}}
}
\newglossaryentry{SpearmanCoeff}{
    type=greek,
    name={\ensuremath{\rho_{\mathrm{S}}}},
    description={Spearman's correlation coefficient},
    unit={\si{-}}
}
\newglossaryentry{PearsonCoeff}{
    type=greek,
    name={\ensuremath{\rho_{\mathrm{P}}}},
    description={Pearson's correlation coefficient},
    unit={\si{-}}
}
\newglossaryentry{surfMeshEdgeLength}{
    type=latin,
    name={\ensuremath{l_e}},
    description={Average edge length of a surface mesh},
    unit={\si{\milli\meter}}
}
\newglossaryentry{meshEdgeSubdivisions}{
    type=latin,
    name={\ensuremath{N_s}},
    description={Number of subdivisions of the triangles of a surface mesh},
    unit={\si{-}}
}
\newglossaryentry{deploymentTolerance}{
    type=latin,
    name={\ensuremath{\epsilon}},
    description={Deployment tolerance for the device deployment algorithm},
    unit={\si{-}}
}
\newglossaryentry{deploymentMaxIterations}{
    type=latin,
    name={\ensuremath{N_{\mathrm{iter}}}},
    description={Maximum number of iterations for the device deployment algorithm},
    unit={\si{-}}
}
\newglossaryentry{deploymentAlpha}{
    type=greek,
    name={\ensuremath{\alpha}},
    description={Alpha parameter for the device deployment algorithm},
    unit={\si{-}}
}
\newglossaryentry{deploymentTotalTime}{
    type=greek,
    name={\ensuremath{T_{\mathrm{tot}}}},
    description={Total deployment time for the device deployment algorithm},
    unit={\si{\second}}
}
\newglossaryentry{signedDistanceToSurface}{
    type=greek,
    name={\ensuremath{d_{\mathrm{web}}}},
    description={Signed distance to the deployed device surface},
    unit={\si{\milli\meter}}
}
\newglossaryentry{signedDistanceTargetToSurface}{
    type=greek,
    name={\ensuremath{d_{\mathrm{web_{target}}}}},
    description={Signed distance to the target device surface},
    unit={\si{\milli\meter}}
}
\newglossaryentry{avgSignedDistance}{
    type=greek,
    name={\ensuremath{\bar{d}_{\mathrm{web}}}},
    description={Surface-average of the signed distance to the deployed device surface},
    unit={\si{\milli\meter}}
}
\newglossaryentry{avgSignedDistanceNB}{
    type=greek,
    name={\ensuremath{\savg{\gls{signedDistanceToSurface}}_{\mathrm{NB}}}},
    description={Surface-average of the signed distance in the neck and body regions},
    unit={\si{\milli\meter}}
}
\newglossaryentry{avgSignedDistanceDome}{
    type=greek,
    name={\ensuremath{\savg{\gls{signedDistanceToSurface}}_{\mathrm{D}}}},
    description={Surface-average of the signed distance in the aneurysm dome region},
    unit={\si{\milli\meter}}
}
\newglossaryentry{crimpRatio}{
    type=greek,
    name={\ensuremath{c_{\mathrm{r}}}},
    description={Crimp ratio used in deployment},
}
\newglossaryentry{strutWireRadius}{
    type=greek,
    name={\ensuremath{r_{\mathrm{w}}}},
    description={Strut wire radius of the WEB deployed},
    unit={\si{\milli\meter}}
}
\newglossaryentry{nThreadPairs}{
    type=greek,
    name={\ensuremath{n_{\mathrm{t}}}},
    description={Number of thread pairs},
}
\newglossaryentry{targetOrientationControl}{
    type=greek,
    name={\ensuremath{\lambda_{tg}}},
    description={Control parameter for the target
     orientation of the WEB during deployment},
}

    \usepackage{import}
    \usepackage[capitalize]{cleveref}
    \usepackage{soul}

\begin{document}

        \title{%
        Adaptation of a Virtual Physics-based Stent Deployment Algorithm for
        WEBs: A Parametric and Morphological Study%
    }

    \author[1]{Marcella P. A. Dallavanzi\thanks{Corresponding author. Email: marcella.avanzi@unesp.br}}

    \author[2]{José L. Gasche}

    \author[1]{Iago L. Oliveira}

    \affil[1]{%
    	São Paulo State University (UNESP), School of Engineering, Bauru, %
    	Department of Mechanical Engineering. %
    	Av. Engenheiro Luiz Edmundo Carrijo Coube, 14-01, 17033-360, Bauru, SP,
    	Brazil %
    }
    
    \affil[2]{%
    	São Paulo State University (UNESP), School of Engineering, Ilha
    	Solteira, %
    	Mechanical Engineering Department
    }
    
    \maketitle

    \begin{abstract}
    	
        \noindent\textbf{Background and Objective:} Selecting the optimal \gls{web} device size for a patient-specific
        \gls{ia} case is important to avoid protrusion or recanalization. This
        study adapts a physics-based stent-deployment algorithm to predict the
        final configuration of \gls{web} SLS devices in patient-specific
        \glsxtrlongpl{ia}. \\
        \textbf{Methods:} A virtual deployment algorithm, originally designed
        for stents, was adapted for the WEB SLS geometry. The device shape was
        generated using helical parametric curves, and ten cases of wide-neck
        bifurcation aneurysms from the \emph{Aneurisk} repository were selected
        for evaluation. The surface-average signed distance between the
        deployed \gls{web} and the aneurysm wall was analyzed to assess
        apposition quality. \\
        \textbf{Results:} An average surface edge length of
        \SI{0.03}{\milli\meter} and a tolerance of \num{1e-4} provided the
        minimum accuracy-to-efficiency balance. Wall apposition was
        successfully computed in all cases selected by adjusting the algorithm
        parameters. The algorithm was able to predict that oversized devices
        significantly improved dome filling and neck-body apposition for three
        of four cases tested without causing parent artery protrusion. \\
        \textbf{Conclusions:} This
        physics-based approach captures device to wall adjustment more
        realistically than geometric methods. It serves as a tool for geometry
        prediction in \gls{cfd} investigations and it may, potentially, serve
        as a basis algorithm for pre-procedural planning.

	\begin{keyword}
		\Glsxtrlongpl{ia}, %
		Woven EndoBridge (WEB), %
		Virtual Deployment, %
		Surgical Planning, %
		Computational Fluid Dynamics (CFD)
	\end{keyword}
	
	\end{abstract}

    \glsresetall
    \section{Introduction}
\label{sec:introduction}

    \Glspl{ia} pose a severe risk to patients if their walls rupture, causing
    an intracranial haemorrhage, typically presenting as a subarachnoid
    haemorrhage. Reports indicate that \SI{85}{\percent} of subarachnoid
    haemorrhages are caused by an \gls{ia} \citep{VanGijn2007}. Rupture-related
    mortality rates range between \num{40} and \SI{60}{\percent}
    \citep{Vlak2013,Saqr2019}, with morbidity affecting approximately one-third
    of survivors \citep{Malmivaara2012}. Classified as either lateral or
    bifurcation based on their location, bifurcation aneurysms are more
    prevalent and are associated with a higher rupture risk \citep{Liu2019b},
    particularly among small aneurysms \citep{Feng2017}.

    Endovascular procedures are the preferred treatment for both ruptured and
    unruptured aneurysms due to superior surgical outcomes compared to clipping
    \citep{Liu2015}. However, \glspl{wnba} --- neck larger than
    \SI{4}{\milli\meter} and a dome-to-neck ratio between \num{1} and \num{2}
    \citep{DeLeacy2019,pierot_ten_2021,chung_comparative_2024} --- remain a
    challenge. In these cases, standalone coiling carries a high risk of coil
    migration and protrusion into the parent artery
    \citep{Armoiry2016,maroufi_comparison_2024}, with
    recanalization and retreatment rates of \SI{20}{\percent} and
    \SI{10}{\percent} at \SI{12} months, respectively \citep{Armoiry2016}.
    While stent- and balloon-assisted coiling are alternative options
    \citep{chung_comparative_2024}, complex geometries often necessitate
    intricate \enquote{X} or \enquote{Y} stent configurations
    \citep{Bartolini2014,Armoiry2016}. These procedures carry the
    risk of obstructing branching artery flow or causing other complications,
    such as thromboembolic events and in-stent stenosis
    \citep{Armoiry2016,gherasim_endovascular_2015,spiotta_mid-term_2011}.
    Given that \glspl{wnba} constitute \SIrange{26}{36}{\percent} of all
    \glspl{ia} \citep{goyal_how_2020}, they represent a significant clinical
    concern.

    Intrasaccular devices, such as the \gls{web} (Terumo Neuro)
    \citep{Ding2011}, LUNA-AED (Medtronic) \citep{Piotin2018}, and Contour
    (Cerus Endovascular) \citep{Akhunbay-Fudge2020}, have been developed to
    address these limitations. Particularly, the \gls{web} device, the most
    widely used to date among these, consists of a self-expanding braided mesh
    implanted within the sac to promote blood stasis and subsequent thrombosis
    without affecting the parent artery flow
    \citep{Aghli2021,pierot_ten_2021,chung_comparative_2024}. Clinical reports
    demonstrate adequate occlusion rates of \SI{90}{\percent}
    \citep{hecker_comparison_2023}, with lower complication and retreatment
    rates compared to standalone or stent-assisted coiling
    \citep{maroufi_comparison_2024,rodriguez-calienes_woven_2024}, offering a
    safer and more effective alternative \citep{arthur_safety_2019}.

    Successful \gls{web} implantation requires precise wall apposition, defined
    as the degree of physical contact between the device wires and the
    aneurysm sac wall. Once selected, the device diameter must exceed the
    aneurysm's average diameter to achieve lateral compression, necessitating a
    corresponding reduction in height \citep{Cagnazzo2023}. Improper
    sizing risks either \gls{web} protrusion into the parent artery if too
    large, or recanalization, if insufficiently small \citep{pierot_ten_2021}.
    As a retrievable device with standardized sizes, the final deployed
    geometry of the initial selection may lack stability, requiring
    intra-procedural replacement which prolongs surgery and increases
    complication risks
    \citep{Cagnazzo2023,goyal_how_2020,munoz_simulation_2022}. Consequently,
    predictive tools for pre-procedural testing play an important role in
    surgical planning. \citet{mut_imagebased_2019} modeled post-treatment
    hemodynamics via angiographic matching, although they did not simulate the
    deployment process itself. Subsequent efforts by
    \citet{dazeo_intra-saccular_2021} and \citet{munoz_simulation_2022}
    utilized morphological parameters via the ID-fit method. However, these
    approaches may not capture the fundamental physics and structural forces of
    deployment.

    Tools to predict a \gls{web} deployed geometry would play an important role
    to assess the device's effect on the aneurysm hemodynamics using numerical
    methods, such as \gls{cfd}. Since those tools are still limited, there are
    few studies that have numerically solved the flow with the device geometry
    within the aneurysm sac
    \citep{mut_imagebased_2019,sharifi_numerical_2020,dazeo_intra-saccular_2021}.
    Moreover, in these studies, the \gls{web} was either modeled based on an
    idealized aneurysm geometry and an idealized model of the \gls{web} as a
    porous media \citep{sharifi_numerical_2020}, thus failing to represent the
    actual wires and their complexity, or did not simulate the \gls{web}
    deformed by the aneurysm sac, which can produce gaps between the device and
    aneurysm lumen surfaces that may not exist \citep{Aghli2021}. Such
    approaches may not capture the real blood flow behavior within the
    aneurysm, given that treatment outcomes depend on the details of the
    interaction between blood flow and the aneurysm wall
    \citep{Berod2023}. Therefore, a geometry that captures the
    morphological features and closely resembles the deployed device would
    yield more reliable results in \gls{cfd} simulations. This study addresses
    this gap by adapting a method of virtual deployment developed for stents
    \citep{Larrabide2011,Spranger2014,Nainggolan2021} to \gls{web} geometries
    generated via parametric equations \citep{Frank2024} to produce more
    realistic geometries for \gls{cfd} analyzes. Our goal is to evaluate the
    robustness of the algorithm across a sample of representative
    patient-specific aneurysms. To achieve this, we identified parameter
    combinations that optimize virtual wall apposition while minimizing
    computational cost. Finally, we applied the algorithm to assess the impact
    of device oversizing, thus, an example of selection that may serve as an
    evaluation to the algorithm's potential effectiveness for surgical
    planning.

    \section{Methods}
\label{sec:methods}

\subsection{Dataset Selection}
\label{subsec:geometriesTested}

    We utilized patient-specific aneurysm geometries  (as discretized surfaces
    in the STL format) from the open-access \emph{Aneurisk} repository to
    evaluate the deployment algorithm. The size of the aneurysm sac was
    computed based on the definition of the aneurysm neck contour, defined as a
    \gls{3d} contour that separates the aneurysm sac from the bifurcating
    artery. Then, the neck ostium surface is defined as the minimum surface
    enclosing the neck contour. The aneurysm neck diameter, \gls{neckDiameter},
    was calculated by using the hydraulic diameter definition, following its
    introduction by \citet{Ma2004} and \citet{Raghavan2005}:
    \begin{equation} \label{eq:neckDiameterDefinition}
        \gls{neckDiameter}
        =
        \frac{
            4\gsub{surfaceArea}{ostium}
        }{
            \gsub{perimeter}{ostium}
        }\,,
    \end{equation}
    where \gsub{surfaceArea}{ostium} and \gsub{perimeter}{ostium} are the
    surface area and perimeter of the ostium surface, respectively. Based on
    the ostium surface definition, the aneurysm height, \gls{domeHeight}, was
    computed as the maximum perpendicular distance from the ostium surface
    centroid to the points on the aneurysm sac \citep{Ma2004}. The maximum
    diameter of the \gls{ia} sac, \gls{domeDiameter}, was defined as the
    hydraulic diameter of the planar section of the sac surface along the
    normal direction of the ostium surface with largest area. Finally, the
    dome-to-neck ratio was computed as
    $\dfrac{\gls{domeDiameter}}{\gls{neckDiameter}}$.

    Inclusion criteria first required cases to satisfy wide-neck definition by
    \citet{pierot_ten_2021} --- neck diameter $\gls{neckDiameter} \geq$
    \SI{4}{\milli\meter} and dome-to-neck ratio of \SIrange[range-phrase =
    --]{1}{2}{}. Additionally, we accounted for the standardized Terumo Neuro
    WEB SLS dimensions, featuring diameters of \SIrange[range-phrase =
    --]{4}{11}{\milli\meter} and heights of \SIrange[range-phrase =
    --]{2.6}{9.6}{\milli\meter} \citep{goyal_how_2020,gajera_woven_2022}.
    Consequently, only aneurysms compatible with these sizes --- specifically,
    those with maximum diameters between \num{3} and \SI{10}{\milli\meter} and
    heights between \num{3.6} and \SI{10.6}{\milli\meter} --- were eligible for
    selection. A sample of \num{12} bifurcation aneurysms was initially
    selected.

    Following this, a visual inspection of the remaining geometries was
    conducted to assess morphology and microcatheter accessibility. Two cases
    were excluded due to extreme morphological features as measured by their
    degree of surface irregularity identified with a relatively large
    undulation index ($>0.225$) \citep{Raghavan2005}. Hence, ten cases
    ultimately constituted the selected sample: five in the \gls{mca} (cases
    C0020, C0045, C0051, C0071 and C0078; for reference, we used the same
    labels given in the \textit{Aneurisk} database), three in the \gls{aca}
    (cases C0033, C0052 and C0060), and two in the \gls{ica} (cases C0046 and
    C0066). For each case, the most suitable WEB SLS device was chosen based on
    the dimensions provided in the Terumo Neuro selection table
    \citep{goyal_how_2020}.

\subsection{WEB Geometry Modeling}
\label{subsec:webGeometry}

    The \gls{web} SLS device was modeled based on the parametric equations of
    helical curves presented by \citet{Frank2024}. The wires composing the
    structure were represented in pairs, where each pair consists of one thread
    moving clockwise and one counter-clockwise, totalizing $n_t$ wire pairs.
    The curves are parameterized by $\theta \in [0,1]$, such that each wire
    extends from the \enquote{south} pole ($\theta = 0$) to the \enquote{north}
    pole ($\theta = 1$) of the device. To ensure uniform positioning, the
    angular distribution of the wire pairs around the central axis of the
    device was determined by the variable $\phi_i$, as follows:
    \begin{equation} \label{eq:phi}
        \phi_i
        =
        \frac{
            2 \pi (i-1)
        }{
            n_t
        },
        \quad
        i = 1,2, \dots, n_t\,.
    \end{equation}

    The three-dimensional position of each point on a wire is defined by the
    following parametric equations, $\vec{\gamma}_{i,c/cc}(\theta)$:
    \begin{equation} \label{eq:gamma}
        \vec{\gamma}_{i,c/cc}
        (\theta)
        =
        \begin{bmatrix}
            r_{d,max}r_d(\theta)\cos(\pm 2\pi n_w \theta + \phi_i) \\
            r_{d,max}r_d(\theta)\sin(\pm 2\pi n_w \theta + \phi_i) \\
            h_{d,max}z_d(\theta)
        \end{bmatrix}
    \end{equation}
    where $r_{d,max}$ and $h_{d,max}$ represent the maximum radius and height
    of the device, respectively; $n_w$ is the number of turns the wire makes
    around its own axis (assumed to be one in this case); and the sign
    \enquote{$\pm$} distinguishes clockwise (subscript \enquote{c}, positive
    orientation) and counter-clockwise (subscript \enquote{cc}, negative
    orientation) directions. The function $r_d(\theta)$ describes the radial
    profile in the $x$-$y$ plane, which reproduces the characteristic spherical
    geometry of the \gls{web} SLS device and ensures the closure of the poles,
    and is given as:
    \begin{equation} \label{eq:rd}
        r_d(\theta) = \sin(\pi \theta)\,.
    \end{equation}

    The height profile is defined by the function $z_d(\theta)$:
    \begin{equation} \label{eq:zd}
        z_d(\theta)
        =
        \left\{
            \frac{1}{2}
            \tanh{
                \left[
                    b
                    \left(
                        \theta
                        -
                        \frac{1}{2}
                    \right)
                \right]
            }
            +
            \frac{1}{2}
        \right\}
        +
        \left\{
            l
            \text{sign}
            \left(
                \theta-\frac{1}{2}
            \right)
            \left[
                m(\theta; w)
                - 1
            \right]
        \right\}\,,
    \end{equation}
    which incorporates a smoothing function, or \emph{mollifier},
    $m(\theta;w)$, enabling the creation of recessed tips at the device poles.
    The complete description of the mollifier function can be found in
    \citet{Frank2024}. The parameters $b$, $l$, and $w$ represent, respectively,
    the \gls{web} device \enquote{\textit{boxiness}} and the depth and width of
    the poles tips. In all tests executed for this work, we used a boxiness of
    3 and pole width and depth of \num{0.15} and \num{0.20}, respectively.

    The device was generated using an in-house Python script that calculated
    the point coordinates and subsequently merged the wire points into a single
    mesh using \gls{vtk}\R. Consequently, in contrast to the actual physical
    structure of the \gls{web} --- where wires are braided in an overlapping
    pattern that allows for relative
    displacement --- this model assumes a fused junction at each wire
    intersection, referred to herein as device nodes.

\subsection{Virtual Deployment Algorithm}
\label{suibsec:deploymentAlgorith}

    Like stents, \gls{web} devices are delivered via microcatheters. This
    process involves radial compression of the device to fit within the
    catheter and subsequent expansion upon deployment as the device releases
    its stored deformation energy \citep{Ding2011}. To simulate this process,
    we adapted the algorithm presented by \citet{Nainggolan2021} (based on
    \citet{Spranger2014}), originally designed for stent deployment, to
    accommodate the specific geometry of the \gls{web} SLS.

\begin{figure}[!t]
        \centering
        \caption{%
            Examples for the geometry of case C0020 with a representative
            \gls{web} model with 20 wires: (a) a crimped \gls{web} geometry
            (white) with a crimping ratio $\gls{crimpRatio} = \num{0.4}$ along
            the sac centerline (green line; the blue line marks the aneurysm
            neck contour) and its target geometry (gray); (b) comparison of the
            sac centerline path (green) and the centerline following from the
            vessel inlet to the aneurysm top dome point (red); (c) example of
            extreme \gls{web} target orientation control, where the target
            \gls{web} is aligned with the tangent at the neck centroid (blue,
            $\gls{targetOrientationControl} = 0.0$) and with the average
            tangents to the sac centerline (gray,
            $\gls{targetOrientationControl} = 1.0$).
        }

        \includegraphics[%
           width=\textwidth
        ]   {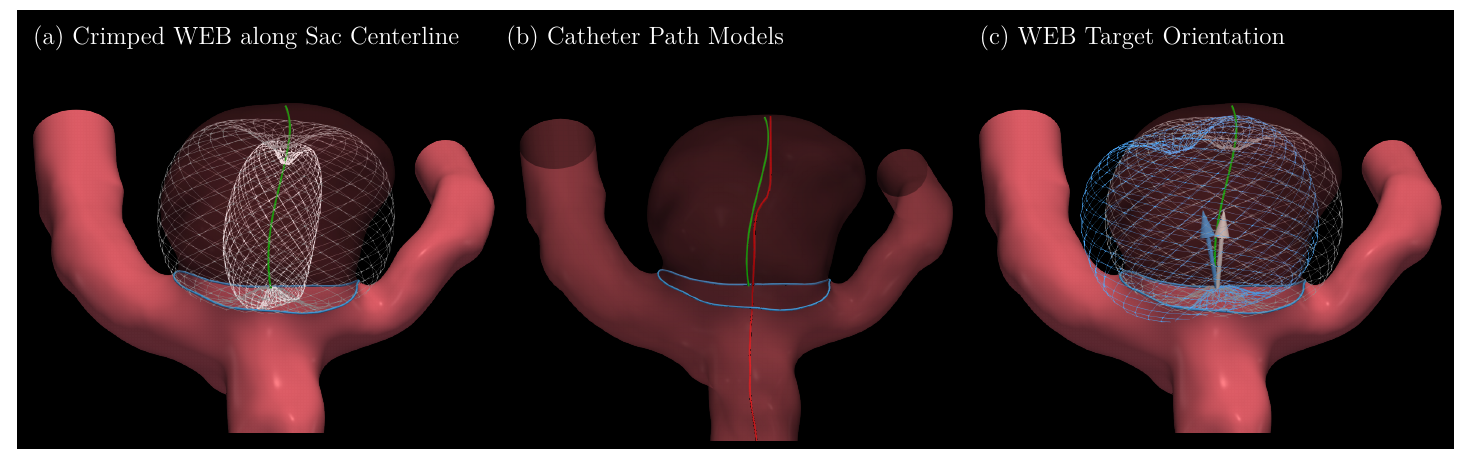}

        \label{fig:predeploymentConfigurationC0020}
    \end{figure}

    The \gls{web} geometry generated via parametric equations represents its
    fully relaxed state. The deployment procedure begins with this geometry
    being virtually crimped; subsequently, it iteratively expands toward its
    relaxed state—referred to herein as the \emph{target geometry}. The
    crimping phase entails applying an inward radial displacement to the nodes
    according to a crimping ratio, \gls{crimpRatio}, ranging from $0$ to $1$.
    To better replicate catheter positioning and the clinical variability of
    the deployment procedure, the crimping path was defined using two distinct
    approaches. The first, designated here as the \textit{sac-centerline
    catheter model}, utilizes a \gls{3d} contour defined by the aneurysm sac
    isolation algorithm proposed by \citet{Piccinelli2012}. In this approach,
    the crimping path forms a \gls{3d} contour traversing the center of the sac
    (see \cref{fig:predeploymentConfigurationC0020}a for a representation of
    the crimped \gls{web} and the sac centerline). The second strategy,
    referred to as the \textit{parent-artery catheter model}, constructs the
    centerline along the direction of the parent artery entering the aneurysm,
    thereby simulating the device constrained within the microcatheter prior to
    deployment.

    To create the target geometry, the device was translated and rotated to be
    positioned relative to the aneurysm sac surface. The position of the target
    geometry was defined such that its south pole coincided with the aneurysm
    neck centroid, similar to the crimped geometry (see
    \cref{fig:predeploymentConfigurationC0020}a). This position, for both the
    crimped and target virtual geometries, can be shifted along the two
    catheter models via the south pole location. This allows to model a
    \gls{web} device deployed either fully within the aneurysm sac or partially
    protruding from it. This is implemented by, first, determining the maximum
    available length along the catheter model path that prevents the device
    from extending beyond the dome. The exact position is then specified as a
    fraction of this distance, ranging from $0.0$ (\gls{web} south pole at the
    neck centroid) to $1.0$ (south pole at the maximum available distance along
    the catheter model line). The device orientation was computed based on the
    tangent vectors to the assumed catheter model path --- i.e., the
    sac-centerline or parent-artery catheter models --- mimicking the device
    placement during an intervention (see
    \cref{fig:predeploymentConfigurationC0020}b; here, the curvature in the red
    centerline arises from \gls{vmtk}\R default behavior that connects the end
    point of the centerline entering the aneurysm with the selected point on
    the aneurysm surface necessary in the procedure to create this centerline,
    in this work defined as the point on the dome furthest from the neck
    centroid, otherwise the centerline would stop inside the aneurysm). To
    allow a better control of this orientation, a target orientation control
    parameter, labeled as \gls{targetOrientationControl}, was introduced and
    defined as the percentage of the centerline length used for this tangent
    averaging. For example, a value of \num{1.0} aligns the target \gls{web}
    with the average tangent of the entire catheter model path within the
    aneurysm sac, whereas a value of \num{0.0} aligns it with the tangent at
    its point closest to the neck centroid (see
    \cref{fig:predeploymentConfigurationC0020}c for these two examples of the
    target geometry).

    Subsequently, the deployment algorithm \citep{Spranger2014,Nainggolan2021}
    was applied, where the position of each node of the crimped geometry was
    iteratively displaced, towards the target geometry, according to an
    elastic-damping connection model, as follows:
    \begin{equation} \label{eq:elasticDampingModel}
        \gls{mass}_i
        \frac{\mathrm{d}^2\gsub{positionVector}{i}}{dt^2}
        +
        b_i
        \DDt{\gls{positionVector}_i}
        =
        \vec{F}_i
    \end{equation}
    where \gsub{positionVector}{i} is the position vector of the $i$-th node,
    $\gls{mass}_i$ is its mass, $b_i$ is the damping coefficient, and
    $\vec{F}_i$ is the resulting spring force acting on the node. In this
    model, the device's elastic behavior is represented by the forces that
    drive the nodes towards their equilibrium positions, while the damping term
    prevents oscillations during expansion. Each node is assumed to be a
    zero-mass point to avoid artificial oscillatory motion caused by rapid
    expansion, which would occur if the inertial term were included. Based on
    the discrete solution of \cref{eq:elasticDampingModel}, the coordinates of
    the \gls{web} nodes are updated as follows:
    \begin{equation} \label{eq:webPointUpdate}
        \gls{positionVector}_i^{k+1}
        =
        \gls{positionVector}_i^{k}
        +
        \gls{deploymentAlpha}
        \gls{force}_i^k
    \end{equation}
    where the superscript $k$ denotes an iteration, and \gls{deploymentAlpha}
    represents a lumped parameter combining the numerical time-step, the spring
    constant, and the damping coefficient. Following the original algorithm, we
    initially employed $\gls{deploymentAlpha} = 1$. The adoption of this value
    does not alter the final shape of the device, as its only consequence is to
    not simulate the real-time dynamic
    expansion of the device from the catheter, i.e. the node updates during
    expansion do not follow the actual physical path.

\begin{figure}[!htp]
        \centering
        \caption{%
            Distances for node $\gls{positionVector}_{i}^{k}$ and its neighbors $\gls{positionVector}_{j}^{k}$ relative to their relaxed states $\gls{positionVector}_{i}^{*}$ and $\gls{positionVector}_{j}^{*}$.
        }

        \includegraphics[%
           width=0.40\textwidth
        ]   {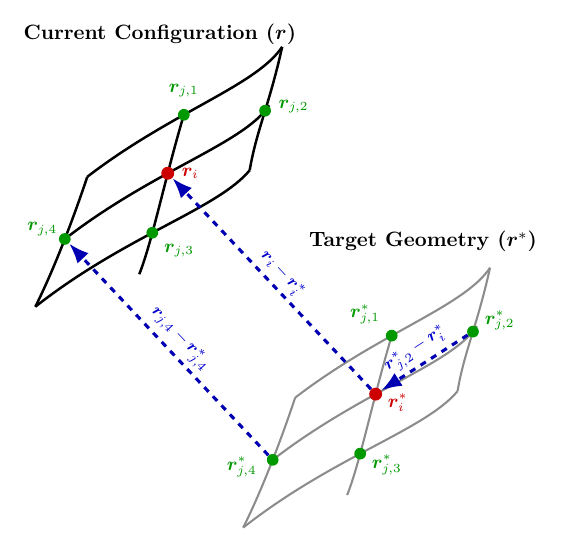}

        \label{fig:distancenodesdiagram}
    \end{figure}

    The term $\gls{force}_i^k$ represents a normalized resultant force moving
    nodes toward their equilibrium configuration \citep{Nainggolan2021} and is
    computed by the difference between a node's displacement from its relaxed
    state and the displacement of its neighbors relative to their own relaxed
    positions, as follows:
    \begin{equation} \label{eq:forceComp}
        \gls{force}_i^k
         =
         \left(
            \sum_{j \in N_{i}}
                \norm{
                    \gls{positionVector}_{j}^{*}
                    -
                    \gls{positionVector}_{i}^{*}
                }^{-1}
        \right)^{-1}
        \sum_{j\in N_{i}}
            \frac{
                (
                    \gls{positionVector}_{j}^{k}
                    -
                    \gls{positionVector}_{j}^{*}
                )
                -
                (
                    \gls{positionVector}_{i}^{k}
                    -
                    \gls{positionVector}_{i}^{*}
                )
            }{
                \norm{
                    \gls{positionVector}_{j}^{*}
                    -
                    \gls{positionVector}_{i}^{*}
                }
            }
    \end{equation}
    where $j \in N_i$ denotes the set of neighboring nodes connected to node
    $i$ by different wires and $\gls{positionVector}_i^{*}$ represents the
    position of node $i$ in the target geometry (see
    \cref{fig:distancenodesdiagram} for a schematic representation of a group
    of nodes of a device mesh where these vector distances are represented).
    This formulation ensures the \gls{web} expands in a manner compatible with
    its elastic behavior, allowing the structure to gradually recover its
    target shape through these local nodal interactions.

    Expansion ceases when the device nodes satisfy two complementary conditions.
    First, a physical constraint where the wires are not allowed to cross or
    touch the aneurysm and vessel boundary surface. This is guaranteed by
    updating a node's position only if its distance to the vessel boundary
    exceeds the sum of the wire radius and a user-specified tolerance distance
    --- selected as zero, by default, in this work. It is important to clarify,
    however, that the algorithm does not account for the physical interaction
    between the wires and the aneurysm wall, as the walls are assumed to be
    rigid. Second, an equilibrium
    criterion determines the end of the deployment when the displacement of all
    points between successive iterations falls below an increment deployment
    tolerance, \gls{deploymentTolerance}. This approach ensures that the device
    either conforms to the aneurysm wall or achieves its fully relaxed state,
    mimicking its natural expansion within the aneurysm sac.

    The algorithm implementation was adapted from the version made available by
    \citet{Nainggolan2021} and integrated with an in-house implementation of
    the \gls{web} geometry generation and virtual crimping. All code was
    implemented in Python and utilizes \gls{vtk}\R and PyVista libraries for
    geometry manipulation and distance calculations. The results presented were
    computed in a workstation with a 13th-generation Intel Core CPU i7-13700
    with \SI{32}{\giga\byte} of RAM.

\subsection{Data Analysis}
\label{subsec:parametricalAnalysis}

\subsubsection*{Parametric Analysis of Deployment Algorithm}

    We conducted a parametric analysis using case C0020 from the
    \emph{Aneurisk} repository to determine optimal algorithm arguments and
    identify factors influencing virtual deployment. The analyzed parameters
    were:

    \begin{enumerate}
        \item the input aneurysm sac surface average edge length
            (\gls{surfMeshEdgeLength}), varied between the original coarse mesh
            available in the database with, approximately,
            \SI{0.13}{\milli\meter} to a finer mesh with average edge length of
            \SI{0.02}{\milli\meter};

        \item deployment tolerance (\gls{deploymentTolerance}) which was set to
            four values (\num{1e-3}, \num{1e-4}, \num{1e-5}, and
            \SI{1e-6}{\milli\meter}).
    \end{enumerate}

    The deployment results for both catheter models were analyzed. We selected
    a \gls{web} SLS with nominal dimensions of \SI{6}{\milli\meter} $\times$
    \SI{4.6}{\milli\meter} for case C0020. The model consisted of
    $\gls{nThreadPairs} = \num{50}$ thread pairs, totaling \num{100} wires,
    approximating the actual geometry of 144 wires
    \citep{Armoiry2016,pierot_web_2015}. We fixed the crimp ratio
    at $\gls{crimpRatio} = \num{0.5}$ (lower crimp ratios can be applied
    without restriction, as the algorithm allows any value between 0 and 1,
    enabling the simulation of higher compression levels if required), the
    target orientation control at $\gls{targetOrientationControl} = 0.5$, and
    two wire diameter were employed, the reported diameter of
    \SI{0.0254}{\milli\meter} (\num{0.001} inch) \citep{Aghli2021} and twice
    this value, thus \SI{0.045}{\milli\meter}, applying a consistent neck
    contour across all tests.

\begin{figure}[!htp]
        \centering
        \caption{%
            Anatomical regions of the aneurysm sac used for surface-averaged
            signed distance calculations illustrated for case C0020: neck,
            body, and dome.
        }

        \includegraphics[%
           width=0.40\textwidth
        ]   {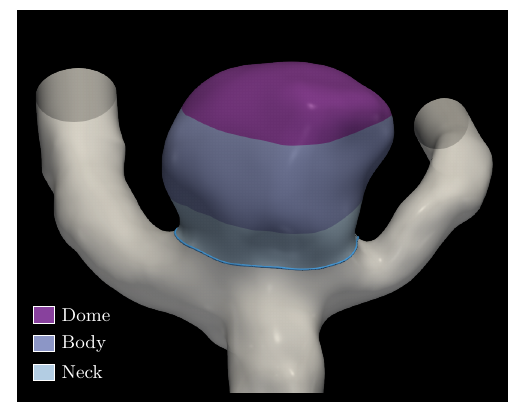}

        \label{fig:sacRegionsC0020}
    \end{figure}

    We measured the total computational time --- encompassing target
    generation, device crimping, and deployment ---, \gls{deploymentTotalTime},
    in minutes, and the distance between the \gls{web} and the aneurysm wall,
    in millimeters. To assess wall apposition with or without penetration of
    the wall by the \gls{web}, we calculated the minimum point-wise vector
    distance between the deployed \gls{web} surface and the aneurysm sac
    surface using \gls{vmtk}\R. This distance was, then, projected along the
    outward unit normals of the aneurysm surface. We labeled it here as
    \gls{signedDistanceToSurface}. Positive values indicate \gls{web} mesh
    portions that have penetrated and transpassed the wall, while negative
    values indicate that the device remains within the sac. The results were
    categorized into anatomical regions based on geodesic distance to the neck
    contour. Following the classification by \citet{SalimiAshkezari2021}, the
    sac was partitioned based on a proportion of the maximum geodesic distance
    from the neck contour: neck ($<$ \SI{20}{\percent}), body
    (\SIrange{20}{60}{\percent}), and dome ($>$ \SI{60}{\percent}). See
    \cref{fig:sacRegionsC0020} for an example of this partitioning. In this
    analysis, the body and neck were consolidated into a single region,
    separate from the dome, as the device-to-wall distance is inherently larger
    at the dome due to the proximal and distal recesses of the \gls{web} tips.

    For a quantitatively assessment, we also computed the surface-average of
    \gls{signedDistanceToSurface}, defined as:
    \begin{equation} \label{eq:surfaceAvg}
        \savg{\gls{signedDistanceToSurface}}_{\gls{surface}_i}
        =
        \surfaceavg{
            \gls{surface}_i
        }{
            \gls{signedDistanceToSurface}
        }\,,
    \end{equation}
    where $\gls{surface}_i$ represents the combined neck-body surfaces,
    resulting in the value labeled as \gls{avgSignedDistanceNB}, and the dome
    surface, indicated by \gls{avgSignedDistanceDome}. Thus, to quantitatively
    identify sac crossing, we computed, separately, the surface-average over
    regions of positive and negative \gls{signedDistanceToSurface} indicated as
    \enquote{Inside} and \enquote{Outside}, respectively. Ideally, the
    \enquote{Inside} \gls{avgSignedDistanceNB} values should be small and
    slightly negative with close-to-zero and positive \enquote{Outside}
    \gls{avgSignedDistanceNB} values, representing an optimal fit with minimal
    wall penetration.

\subsubsection*{Comparison of Deployments Across Cases}

    After determining the optimal parameters for the deployment algorithm, we
    applied it to the ten selected cases. We evaluated the total
    computational time and the surface-average of the signed distance to the
    aneurysm wall for each case. Additionally, we qualitatively assessed the
    wall apposition of the deployed devices by visually inspecting the 3D
    geometries of the deployed \gls{web} SLS and their proximity to the
    aneurysm sac walls. This visual inspection aimed to check the final shape
    and conformity of the structure, while verifying potential overshooting of
    the deployed \gls{web} into the aneurysm sac. This was particularly
    relevant in more complex and irregular surfaces.

    \section{Results}
\label{sec:results}

\subsection{Parametric Analysis}
\label{subsec:parametricAnalysisResults}

\begin{figure*}[!htb]
        \centering
        \caption{
            Signed distance between the aneurysm sac surface and \gls{web}
            device for case C0020, using a deployment tolerance of \num{1e-4},
            for edge lengths of (a) \num{0.131}, (b) \num{0.05}, (c)
            \num{0.03}, and (d) \SI{0.02}{\milli\meter} without (top) and with
            (bottom) transparency for better visualization of apposition to the
            neck and body regions. The thin line in the non-transparent figures
            indicates the divisions between the neck, body, and dome regions.
        }

            \includegraphics[%
                width=\textwidth
            ]   {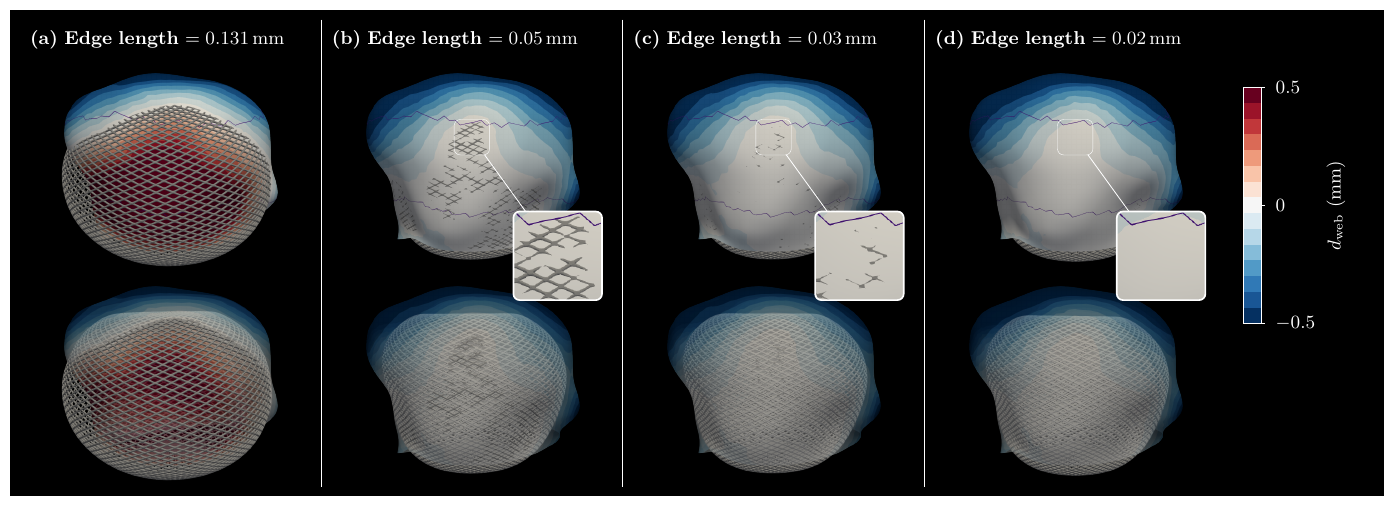}

        \label{fig:signedDistanceEdgeLength}
    \end{figure*}

    The field of the distance \gls{signedDistanceToSurface}, between the
    deployed \gls{web} device and the aneurysm sac surface, for case C0020 is
    shown in \cref{fig:signedDistanceEdgeLength} (fixed deployment tolerance of
    $\gls{deploymentTolerance}=\SI{1e-4}{\meter}$ and sac-centerline catheter
    model). For the original coarser discretized surface
    (\cref{fig:signedDistanceEdgeLength}a), the device overshoot the aneurysm
    sac, as indicated by the region on the neck and body patches with positive
    distance values (in dark red). However, the results for an edge length of
    \SI{0.05}{\milli\meter} (\cref{fig:signedDistanceEdgeLength}b) remain
    reasonable, with the device only slightly surpassing the surface, with an
    increasingly better wall apposition as the discretized sac surface is
    further refined (\cref{fig:signedDistanceEdgeLength}c and d). This behavior
    of the algorithm was already reported by \citet{Nainggolan2021}. Overall,
    the \gls{web} surface is well apposed to the body region as a portion of
    the device hits the aneurysm sac (large white region indicating a zero
    distance to the wall in \cref{fig:signedDistanceEdgeLength}b, c and d),
    although the device does not deform to accommodate the lateral bleb present
    in this case. Additionally, note that the signed distance is clearly
    negative on the aneurysm dome due to height undersizing.

\begin{figure*}[!htb]
        \centering
        \caption{
            \Gls{web} apposition inside (a) neck-body and (b) dome regions
            versus total preparation and deployment time under different
            deployment tolerances and catheter models for case C0020.
        }

        \includegraphics[%
            width=\textwidth
        ]   {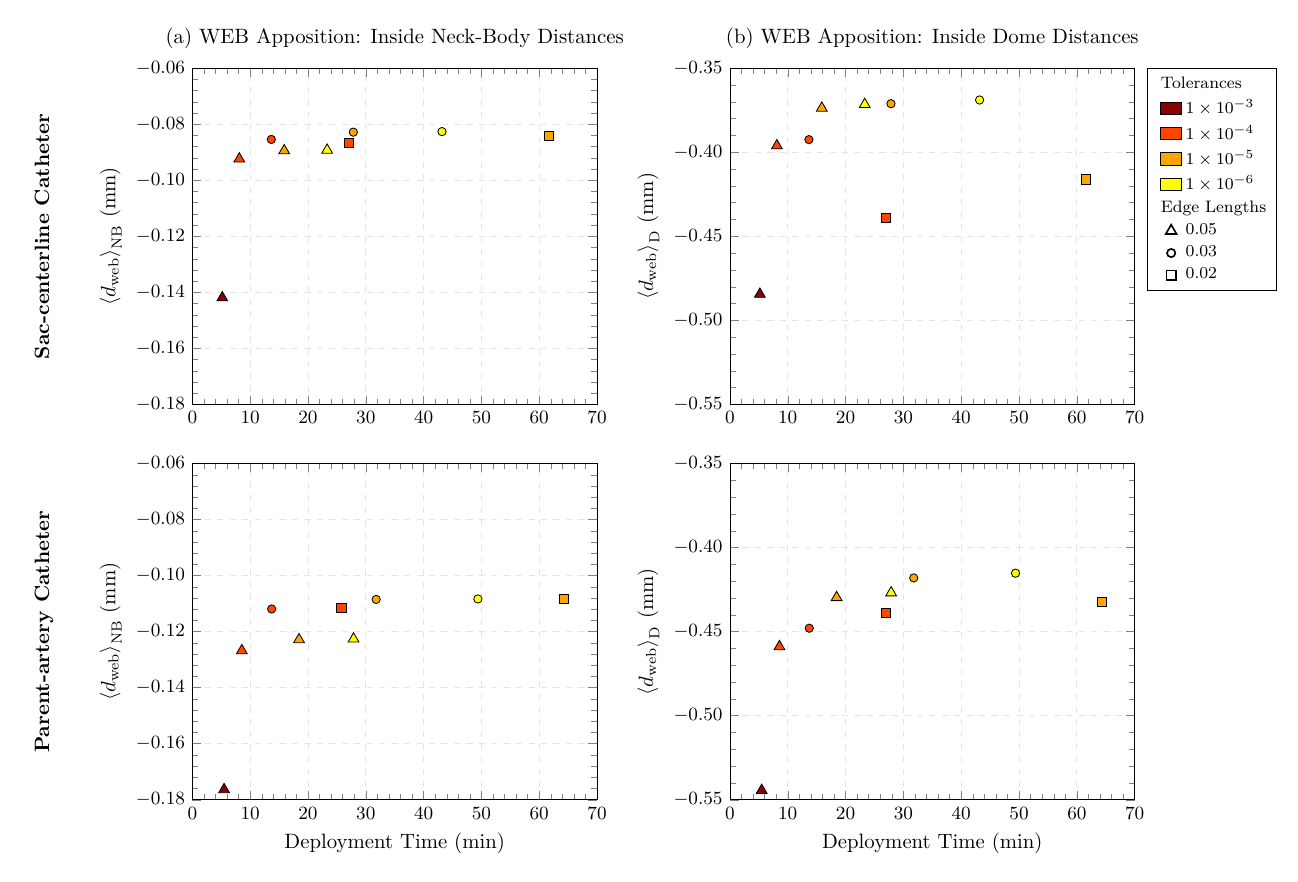}

        \label{fig:WebAppositionDistances}
    \end{figure*}

    Quantitatively, the effect of mesh refinement on the surface-averaged
    \gls{signedDistanceToSurface} is reflected by a slight increase in its mean
    value over the neck-body region (see \cref{fig:WebAppositionDistances}a;
    data for the mesh with an edge length of \SI{0.131}{\milli\meter} were
    excluded from this figure because the \gls{web} device clearly penetrated
    the aneurysm surface, as shown in \cref{fig:signedDistanceEdgeLength}a),
    accompanied by a substantial increase in computational time for both
    catheter models. Utilizing a very fine edge length of
    \SI{0.02}{\milli\meter} does not improve apposition, yet it increases the
    total computational time (at a tolerance of $\gls{deploymentTolerance} =
    \SI{1e-5}{\meter}$) by approximately \SI{32}{\minute} compared to
    $\gls{surfMeshEdgeLength} = \SI{0.03}{\milli\meter}$, and by
    \SI{46}{\minute} compared to $\gls{surfMeshEdgeLength} =
    \SI{0.05}{\milli\meter}$.

    Regarding the effect of deployment tolerance, which exhibited a consistent
    effect across both catheter models, a tolerance of
    $\gls{deploymentTolerance} = \SI{1e-3}{\milli\meter}$ yielded a relatively
    large \gls{avgSignedDistanceNB} at a surface edge length of
    \SI{0.05}{\milli\meter} (approximately $\gls{avgSignedDistanceNB} =
    \SI{-0.14}{\milli\meter}$ for the sac-centerline catheter). A similar
    outcome was observed for \gls{avgSignedDistanceDome}, which remained near
    \SI{-0.48}{\milli\meter}. Reducing the tolerance to \SI{1e-4}{\milli\meter}
    substantially improved apposition, decreasing \gls{avgSignedDistanceNB} and
    \gls{avgSignedDistanceDome} to approximately \SI{-0.09}{\milli\meter} and
    \SI{-0.40}{\milli\meter}, respectively. Further decreasing the tolerance to
    $\gls{deploymentTolerance} \le \SI{1e-5}{\milli\meter}$ led to an
    asymptotic behavior, with \gls{avgSignedDistanceNB} converging toward
    \SI{-0.08}{\milli\meter} (as evidenced by the yellow markers in
    \cref{fig:WebAppositionDistances} for a tolerance of
    \SI{1e-6}{\milli\meter} evaluated at edge lengths of
    \SI{0.03}{\milli\meter} and \SI{0.05}{\milli\meter}). For the parent-artery
    catheter model, the asymptotic \gls{avgSignedDistanceNB} settled at
    approximately \SI{-0.11}{\milli\meter} --- a value \SI{37.5}{\percent}
    higher than that of the sac-centerline model --- while incurring a
    comparable total computational time.

    Additionally, regarding the two tested wire diameters, the mean outside
    \gls{avgSignedDistanceNB} remained small for both values, showing only a
    minor increase from \SI{0.001}{\milli\meter} to \SI{0.003}{\milli\meter}
    when decreasing the wire diameter from \SI{0.045}{\milli\meter} to
    \SI{0.0254}{\milli\meter}. However, this diameter reduction resulted in an
    increased deployment time of approximately \SI{10}{\minute} (rising from
    \SI{14}{\minute} to \SI{24}{\minute}), while the remaining signed distance
    metrics exhibited comparable values.

\subsection{Assessment of Deployment in Patient-Specific Geometries}
\label{subsec:patientSpecificResults}

\begin{figure*}[!ht]
        \centering
        \caption{
            Views of the final WEB position after deployment for 4/8 cases
            selected (a) with a detail of the neck, body and dome regions and
            (b) a transparent sac and superior and lateral views of the WEB
            apposition.
        }

        \includegraphics[%
            width=0.9\textwidth
        ]   {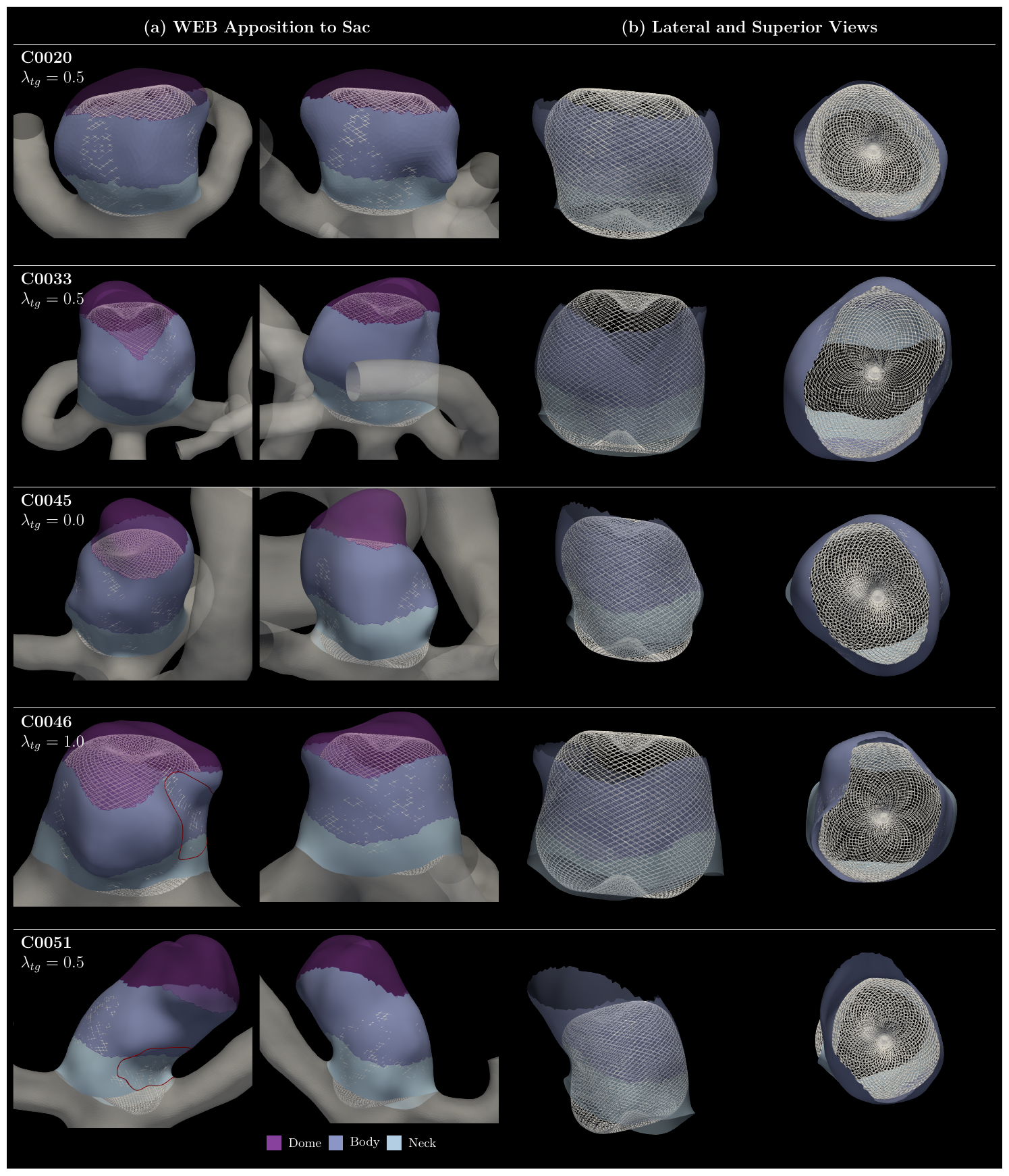}

        \label{fig:WebDeploymentsAllCases1}
    \end{figure*}

\begin{figure*}[!ht]
        \centering
        \caption{
            (continuation) Views of the final WEB position after deployment for
            4/8 cases selected (a) a detail of the neck, body and dome regions
            and (b) a transparent sac and superior and lateral views of the WEB
            apposition.
        }

        \includegraphics[%
            width=0.9\textwidth
        ]   {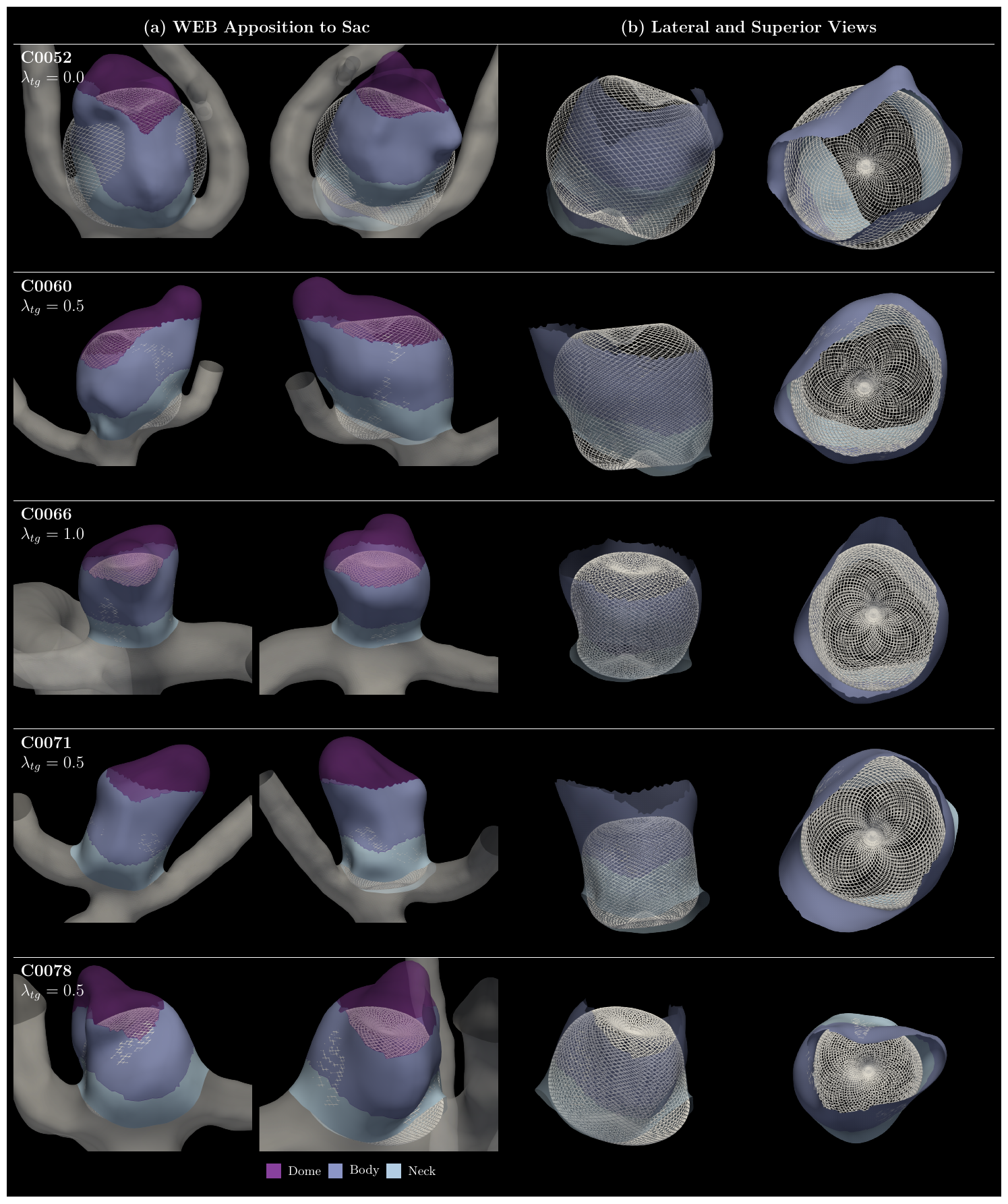}

        \label{fig:WebDeploymentsAllCases2}
    \end{figure*}

    For the complete dataset, we adopted only the reported diameter of
    \num{0.001} inch (\SI{0.0254}{\milli\meter}) \citep{Aghli2021}, with a
    crimping ratio of $\gls{crimpRatio} = \num{0.3}$, and fixed the
    sac-centerline catheter model while testing three values of target
    orientation --- \gls{targetOrientationControl} as \num{0.0}, \num{0.5}, and
    \num{1.0} --- to assess how different device alignments affect deployment
    and wall apposition success in cases with geometric complexity, such as
    those with lobulated and irregular sac surfaces. All tests employed a sac
    surface edge length of \num{0.03} and deployment tolerance of
    \SI{1e-4}{\milli\meter} as a result of the parametric analysis.
    \Cref{tab:ResultsPatientSpecificGeometries} summarizes the quantitative
    results, with information on the selected \gls{web} size, while views of
    the final apposition for the target orientation that promoted the best
    deployment (annotated in the figure) are shown in
    \cref{fig:WebDeploymentsAllCases1,fig:WebDeploymentsAllCases2}. Overall,
    the algorithm achieved effective apposition across most geometries. In nine
    of ten cases, the \gls{web} was successfully positioned without exceeding
    the wall with these parameters, as indicated by the \enquote{Outside}
    columns in \cref{tab:ResultsPatientSpecificGeometries} that remained near
    zero for the deployments showed in
    \cref{fig:WebDeploymentsAllCases1,fig:WebDeploymentsAllCases2}.

    \begin{table*}[!t]
        \centering
        \caption{%
            Mean signed distance --- split into \enquote{Inside} and
            \enquote{Outside} --- and total computation time,
            \gls{deploymentTotalTime}, for the patient-specific geometries (all
            tests assuming surface edge length of \num{0.03}, deployment
            tolerance of \SI{1e-4}{\milli\meter}, crimp ratio of \num{0.3}, and
            wire diameter of \SI{0.0254}{\milli\meter}).
        }
        \small

        \footnotesize
        \setlength{\tabcolsep}{1.8pt}
        \renewcommand{\arraystretch}{0.88}

        \begin{tabularx}{0.9\textwidth}{
            C
            c
            c
            S[table-format=1.3,round-mode=places,round-precision=3]
            S[table-format=1.3,round-mode=places,round-precision=3]
            S[table-format=1.3,round-mode=places,round-precision=3]
            S[table-format=1.3,round-mode=places,round-precision=3]
            C
        }
            \toprule
                \multirow{2}{*}{Case}
            &   \multirow{2}{*}{Selected \gls{web} Size}
            &   \multirow{2}{*}{%
                    \shortstack{%
                        Target \\ Orientation (\gls{targetOrientationControl})
                    }
                }
            &   \multicolumn{2}{c}{%
                    \gls{avgSignedDistanceNB} (\si{\milli\meter})%
                }
            &   \multicolumn{2}{c}{%
                    \gls{avgSignedDistanceDome} (\si{\milli\meter})%
                }
            &   \multirow{2}{*}{\shortstack{\gls{deploymentTotalTime} (\si{\minute})}}\\
            \cmidrule(lr){4-5}
            \cmidrule(lr){6-7}
            &
            &
            &   {\small Inside}
            &   {\small Outside}
            &   {\small Inside}
            &   {\small Outside}
            &   \\
            \midrule
                \multirow{3}{*}{C0020}
            &   \multirow{3}{*}{6.0 mm $\times$ 4.6 mm}
            &   0.0   & -0.130 & 0.002 & -0.328 & 0.002 & 43 \\
            & & 0.5 & -0.088 & 0.002 & -0.383 & 0.002 & 28 \\
            & & 1.0   & -0.088 & 0.001 & -0.392 & 0.002 & 25  \\
            \cmidrule(lr){1-8}
            \multirow{3}{*}{C0033}
            &   \multirow{3}{*}{10.0 mm $\times$ 8.6 mm}
            &   0.0   & -0.132 & 0.039 & -0.581 & 0.007 & 68 \\
            & & 0.5 & -0.127 & 0.002 & -0.632 & 0.002 & 48 \\
            & & 1.0   & -0.130 & 0.002 & -0.601 & 0.002 & 48 \\
            \cmidrule(lr){1-8}
            \multirow{3}{*}{C0045}
            &   \multirow{3}{*}{6.0 mm $\times$ 4.6 mm}
            &   0.0   & -0.247 & 0.001 & -1.417 & 0.000 & 122 \\
            & & 0.5 & -0.257 & 0.002 & -1.462 & 0.000 & 69 \\
            & & 1.0   & -0.249 & 0.002 & -1.426 & 0.000 & 98 \\
            \cmidrule(lr){1-8}
            \multirow{3}{*}{C0046}
            &   \multirow{3}{*}{8.0 mm $\times$ 6.6 mm}
            &   0.0   & -0.222 & 0.224 & -0.406 & 0.004 & 54 \\
            & & 0.5 & -0.169 & 0.002 & -0.396 & 0.002 & 39 \\
            & & 1.0   & -0.154 & 0.002 & -0.416 & 0.002 & 43 \\
            \cmidrule(lr){1-8}
            \multirow{3}{*}{C0051}
            &   \multirow{3}{*}{6.0 mm $\times$ 4.6 mm}
            &   0.0   & -0.167 & 0.269 & -1.599 & 0.000 & 69 \\
            & & 0.5 & -0.162 & 0.019 & -1.643 & 0.000 & 45 \\
            & & 1.0   & -0.163 & 0.024 & -1.603 & 0.000 & 59 \\
            \cmidrule(lr){1-8}
            \multirow{3}{*}{C0052}
            &   \multirow{3}{*}{9.0 mm $\times$ 7.6 mm}
            &   0.0   & -0.429 & 0.341 & -1.157 & 0.000 & 56  \\
            & & 0.5 & -0.385 & 0.338 & -1.172 & 0.000 & 47 \\
            & & 1.0   & -0.504 & 0.293 & -1.089 & 0.000 & 108 \\
            \cmidrule(lr){1-8}
            \multirow{3}{*}{C0060}
            &   \multirow{3}{*}{8.0 mm $\times$ 6.6 mm}
            &   0.0   & -0.154 & 0.036 & -0.597 & 0.002 & 47 \\
            & & 0.5 & -0.138 & 0.002 & -0.603 & 0.001 & 33 \\
            & & 1.0   & -0.141 & 0.002 & -0.614 & 0.001 & 36 \\
            \cmidrule(lr){1-8}
            \multirow{3}{*}{C0066}
            &   \multirow{3}{*}{6.0 mm $\times$ 4.6 mm}
            &   0.0   & -0.230 & 0.002 & -0.937 & 0.000 & 114 \\
            & & 0.5 & -0.153 & 0.002 & -1.025 & 0.000 & 71 \\
            & & 1.0   & -0.147 & 0.002 & -1.034 & 0.000 & 56 \\
            \cmidrule(lr){1-8}
            \multirow{3}{*}{C0071}
            &   \multirow{3}{*}{6.0 mm $\times$ 4.6 mm}
            &   0.0   & -0.225 & 0.002 & -1.962 & 0.000 & 89 \\
            & & 0.5 & -0.216 & 0.002 & -1.931 & 0.000 & 71 \\
            & & 1.0   & -0.222 & 0.060 & -1.850 & 0.000 & 91 \\
            \cmidrule(lr){1-8}
            \multirow{3}{*}{C0078}
            &   \multirow{3}{*}{5.0 mm $\times$ 3.6 mm}
            &   0.0   & -0.225 & 0.002 & -0.745 & 0.001 & 28 \\
            & & 0.5 & -0.227 & 0.004 & -0.766 & 0.007 & 25 \\
            & & 1.0   & -0.248 & 0.002 & -0.753 & 0.003 & 28 \\
            \bottomrule
        \end{tabularx}%
        \label{tab:ResultsPatientSpecificGeometries}
    \end{table*}

    Although the \enquote{Inside} \gls{avgSignedDistanceNB} values should be
    interpreted with caution --- as they depend on the specific anatomical
    partitioning of the sac --- qualitative inspection of
    \cref{fig:WebDeploymentsAllCases1,fig:WebDeploymentsAllCases2} confirms
    successful deployment and adequate apposition across nine cases. For seven
    of the ten cases (C0020, C0033, C0045, C0060, C0066, C0071, and C0078), all
    target orientations resulted in effective \gls{web} adjustment and
    conformity to the aneurysm sac surface, demonstrating the robustness of the
    algorithm (see \cref{tab:ResultsPatientSpecificGeometries}). Three cases
    (C0046, C0051, and C0052) exhibited unsuccessful deployment for at least
    one target orientation, wherein the device penetrated the aneurysm wall.
    However, the remaining target orientations for cases C0046 and C0051
    achieved good apposition. The sole exception across all cases where the
    device penetrated beyond the wall for all three target orientations was
    case C0052, which markedly penetrated the aneurysm surface near the concave
    regions of the sac (see the first row of
    \cref{fig:WebDeploymentsAllCases2}).

\begin{figure*}[ht]
        \centering
        \caption{
            Signed distance between the aneurysm sac surface and the \gls{web}
            device for case C0052, evaluated at deployment position fractions
            of \num{0.0} (a, b) and \num{0.15} (c) along the available
            sac-centerline length: (a) uniform surface mesh with an edge length
            of \SI{0.03}{\milli\meter}, and (b, c) non-uniform mesh refinement
            with a local edge length of \SI{0.01}{\milli\meter}.
        }

            \includegraphics[%
                width=\textwidth
            ]   {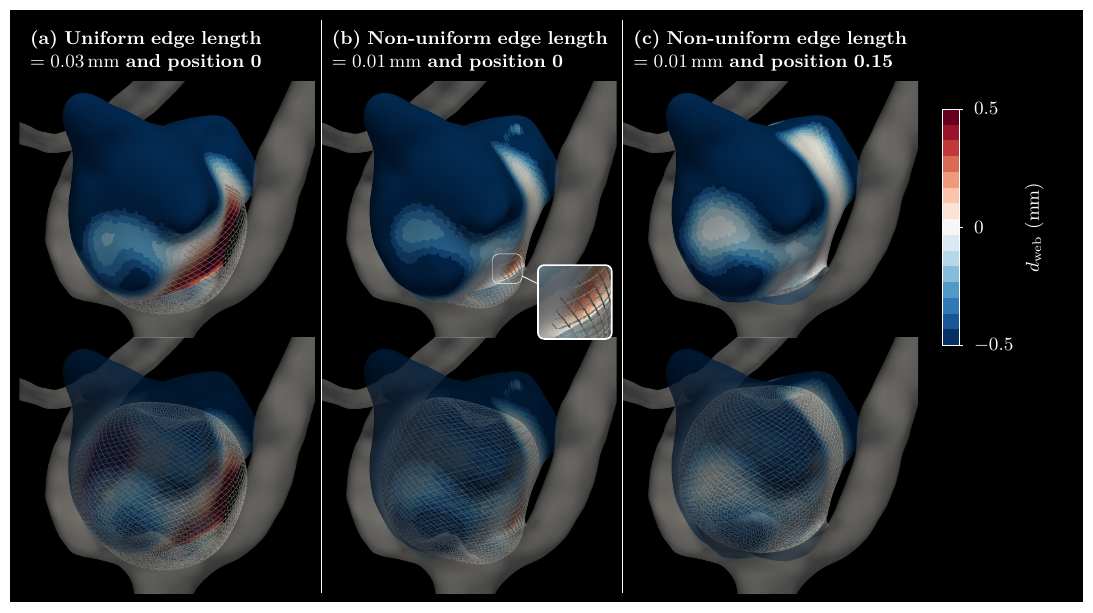}

        \label{fig:resolutionAndPosition52}
    \end{figure*}

    To assess the underlying causes and identify potential solutions for
    obtaining a correctly deployed geometry in the failed cases, we deployed
    the corresponding devices in cases C0046, C0051, and C0052 using finer sac
    surface meshes, given that finer surface meshes prevent the device from
    penetrating the aneurysm surface (see \cref{fig:signedDistanceEdgeLength}).
    For case C0051, a sac mesh with a uniform edge length of
    \SI{0.02}{\milli\meter} achieved complete apposition. For cases C0046 and
    C0052, we evaluated a targeted strategy using non-uniform mesh refinement,
    wherein only the specific surface patches where the \gls{web} penetrated
    the sac (highlighted by the red contours in
    \cref{fig:WebDeploymentsAllCases1} for cases C0046 and C0051, and the
    wall-penetrating device in case C0052)
    were refined to an edge length of \SI{0.01}{\milli\meter}, while
    maintaining an edge length of \SI{0.03}{\milli\meter} elsewhere on the
    surface. This approach yielded correct apposition across all three target
    orientations for case C0046, successfully resolving the initial deployment
    failure. For case C0052, the device deployed successfully in two of the
    tested target orientations; for a target orientation of \num{0.0}, slight
    wall penetration persisted near the neck region, though the improvement
    over the uniform refinement was substantial (see
    \cref{fig:resolutionAndPosition52}a,b). Case C0052 presents a notable
    challenge due to its lobulated surface, where initial \gls{web} wall
    penetration occurred primarily across highly concave surface patches.

    For case C0052, we also evaluated whether altering the device position
    along the sac centerline would impact apposition when using the
    non-uniformly refined mesh. A position fraction of \num{0.15} of the
    available deployment distance was tested for the fixed target orientation
    $\gls{targetOrientationControl} = 0.0$—the only configuration where the
    \gls{web} still penetrated the sac. The resulting
    \gls{signedDistanceToSurface} field between the deployed \gls{web} device
    and the aneurysm surface is shown in \cref{fig:resolutionAndPosition52}b
    and c. For visual comparison, the initial deployment configuration from
    \cref{fig:WebDeploymentsAllCases2} (i.e., with the south pole located at
    the neck centroid) is shown in gray in the first column. Translating the
    device by \num{0.15} of the available sac centerline length improved
    apposition at both the neck and dome, allowing the \gls{web} to occupy a
    larger proportion of the dome volume, as indicated by the expanded white
    region.

    Finally, regarding computational performance, an increase in computational
    time was observed for the patient-specific cases compared to case C0020
    evaluated in the parametric analysis. The total computational time peaked
    at \SI{122}{\minute} for case C0045, with an average across all cases of
    approximately \SI{51}{\minute}. However, it is important to note that these
    simulations employed a relatively fine mesh (edge length of
    \SI{0.03}{\milli\meter}) combined with the thinner wire diameter, both of
    which contributed to longer deployment times relative to the baseline
    values observed in the parametric study.

\subsubsection{Impact of Device Oversizing}
\label{subsubsec:impactOfDeviceOversizing}

    To further test the algorithm's robustness, we evaluated
    \enquote{oversized} devices for cases where the \gls{web} did not fully
    fill the sac's dome, i.e. cases C0045, C0051, C0066, and C0071 that
    presented the larger averaged web-to-dome distance after deployment (see
    \cref{tab:ResultsPatientSpecificGeometries}, column
    \gls{avgSignedDistanceDome} \enquote{Inside}). Regarding the remaining
    cases, as $\gls{avgSignedDistanceDome} < \SI{1}{\milli\meter}$ for the
    standard size, employing an \enquote{oversized} device would theoretically
    lead to protrusion into the parent artery, which is, clinically, an
    undesirable outcome that requires device replacement or the addition of a
    stent \citep{ansari_optimal_2021}.

    We defined \enquote{oversized} as the next available size in the
    standardized Terumo Neuro \gls{web} SLS dimensions. For instance, in case
    C0045, where the recommended size was \SI{6}{\milli\meter} $\times$
    \SI{4.6}{\milli\meter} (see \cref{tab:ResultsPatientSpecificGeometries}),
    the oversized configuration tested was \SI{7}{\milli\meter} $\times$
    \SI{5.6}{\milli\meter}. For all tests, we maintained the same deployment
    parameters as in the previous section.

\begin{figure*}[!htb]
        \centering
        \caption{
            Assessment of the deployment for \gls{web} oversized
            in the patient-specific geometries of cases C0045, C0051, C0066
            and C0071.
        }

        \includegraphics[%
            width=0.9\textwidth
        ]   {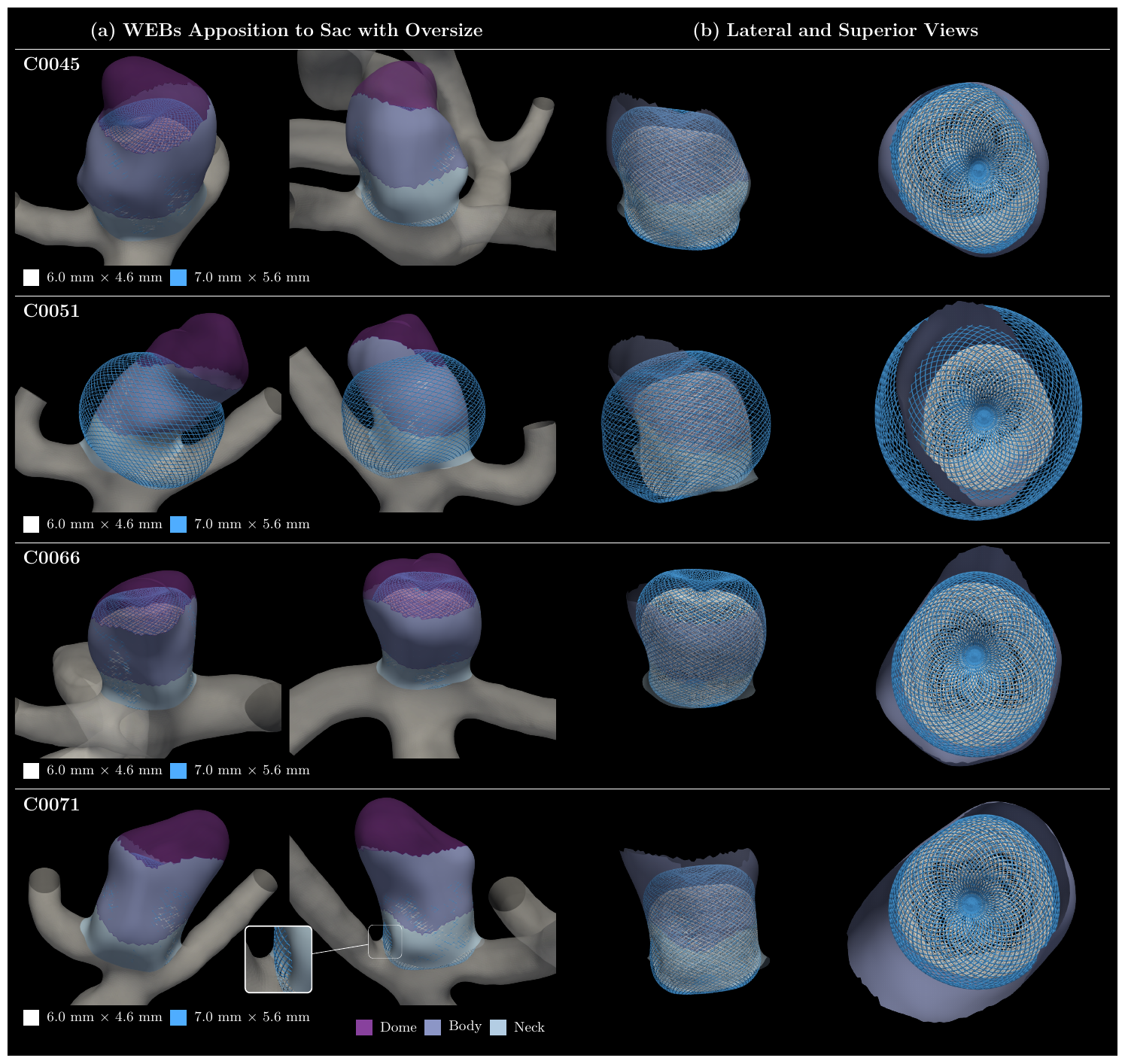}

        \label{fig:comparisonOversizedWEB}
    \end{figure*}

    Device oversizing improved apposition at the sac dome across all cases, as
    expected (see \cref{fig:comparisonOversizedWEB}; oversized device depicted
    in blue). Compared to the optimal target orientation values reported in
    \cref{tab:ResultsPatientSpecificGeometries}, the mean reduction in the
    surface-averaged distance to the dome, \gls{avgSignedDistanceDome}, was
    \SI{35.4}{\percent}. Furthermore, oversizing also yielded improved
    apposition in the neck-body region. With the exception of case C0051, the
    neck-body distance, \gls{avgSignedDistanceNB}, decreased by
    \SI{50}{\percent} for case C0071 and by over \SI{60}{\percent} for cases
    C0045 and C0066. In case C0071, a small region of the \gls{web} initially
    penetrated the wall, which was resolved using local mesh refinement
    (\SI{0.01}{\milli\meter} edge length) across that region of the sac
    surface. Conversely, the oversized device for case C0051 completely
    penetrated the sac surface. Case C0051 features the narrowest aneurysm
    geometry among those evaluated for oversizing; consequently, an oversized
    device begins further from the sac wall in the initial configuration,
    increasing the driving force per incremental displacement step and thereby
    precipitating wall penetration. Nevertheless, this artifact could be
    corrected through local surface mesh refinement in the penetration region,
    as demonstrated for cases C0052 and C0071.

    \section{Discussion}
\label{sec:discussion}

    This work presented the application of a physics-based algorithm—originally
    proposed for virtual stent deployment—to the virtual deployment of the
    intrasaccular \gls{web} device used in the endovascular treatment of
    \glspl{ia}. The algorithm employs a dynamic elastic-damped mechanical model
    that simulates device expansion iteratively. Our findings demonstrate that
    this approach effectively conforms a \gls{web} device to the
    patient-specific geometry of an \glsxtrlong{ia}. Furthermore, the algorithm
    exhibits robustness against variations in device size, initial orientation,
    and positioning along modeled catheter paths. Unlike purely geometric
    approximations, this physics-based formulation models the structural
    behavior of the device, wherein internal elastic forces—arising from the
    braid's tendency to recover its stress-free configuration—interact with the
    boundary constraints of the vascular wall. This mechanical interaction
    yields an equilibrium state that naturally reflects device apposition
    against the aneurysm sac.

    Adapting the algorithm to the \gls{web} device highlighted a known
    numerical sensitivity previously reported by \citet{Nainggolan2021}, where
    device nodes may penetrate the surface mesh if it is
    insufficiently refined. Nevertheless, our parametric analysis established
    optimal computational settings: a surface edge length of
    \SI{0.03}{\milli\meter} and a deployment tolerance of
    \SI{1e-4}{\milli\meter} provided an ideal compromise, substantially
    improving apposition relative to a \SI{0.05}{\milli\meter} mesh without
    incurring the high computational expense of finer discretizations.
    Consequently, an aneurysm sac refinement between \SI{0.03}{\milli\meter}
    and \SI{0.05}{\milli\meter} can be rationally selected based on target
    accuracy and available computational resources. Additionally, thinner braid
    wires exhibited a slightly higher propensity for wall penetration,
    indicating that smaller wire diameters necessitate finer sac
    discretizations to prevent sac penetration. Regarding catheter path models,
    both tested approaches yielded comparable overall deployments; subtle
    differences in neck-body surface distance arose because each path
    inherently defines a distinct target orientation. Thus, these differences
    reflect realistic variations across alternative clinical catheter
    trajectories.

    Regarding computational efficiency, while some cases may require longer
    computational times, the algorithm is able to provide relatively fast results
    for geometry predictions for \gls{cfd} simulations, while appearing high
    for a use within a clinical context. Nevertheless, there is significant
    potential for optimization. Substantial algorithmic improvements could
    still be achieved through parallelization, for example. The current main
    bottleneck is due to the large number of points necessary in the
    discretized aneurysm sac surface and \gls{web} geometry, which was
    responsible for a large increase in computational time, as shown by the
    parametric analysis. Our tests have shown that it is possible to obtain a
    correct deployment by using non-uniform discretizations of the sac surface,
    which allows to use a smaller edge length where the device penetrated the
    wall, which could improve the computational times. Moreover, implementing
    parallel processing for these point operations represents a viable
    alternative to further reduce the deployment times, potentially enabling
    real-time clinical application.

    Current literature presents few strategies for \gls{web} deployment
    modeling. For instance, \citet{dazeo_intra-saccular_2021} and
    \citet{munoz_simulation_2022} proposed the \enquote{ID-Fit} method, which
    relies on aneurysm morphology to estimate deployment. Their approach
    divides the deployment into three sections and compares the section radius
    to the nominal device radius. Such a method may not fully capture the
    device behavior in complex aneurysms with significant surface
    irregularities. In contrast, by employing a simpler mechanical model, the
    approach proposed in our work attempts to predict deformation based on the
    device approximate physical behavior and successfully handles relatively
    complex shapes. Furthermore, our work proposes different strategies to
    model the catheter path and the target \gls{web} orientation, which can be
    adjusted to better fit the specific anatomy of each case if necessary,
    emulating the manual adjustments an interventionist would make during the
    procedure.

    For hemodynamic investigations of treated aneurysms, this predictive
    deployment framework offers a robust procedure for generating realistic
    post-deployment \gls{web} surfaces for \gls{cfd} simulations. Unlike
    retrospective studies that rely on post-operative angiographic imaging to
    reconstruct approximate device contours \citep{mut_imagebased_2019}, our
    approach is entirely predictive. Consequently, these virtually deployed
    geometries could be integrated into \gls{cfd} workflows to evaluate
    pre-procedural blood flow changes, neck coverage, and treatment efficacy,
    serving as a potential tool for hemodynamic research.

    Despite these capabilities, patient-specific evaluations revealed
    challenges in highly irregular or lobulated geometries, as observed in case
    C0052. The aneurysm surface of case C0052 is highly irregular, leading to
    localized \gls{web} wall penetration across concave patches --- a behavior
    also noted in cases C0046 and C0051. These high-curvature regions are
    computationally prone to wall penetration because the local wall distance
    to the target nodes tends to be larger, as also indicated by the behavior
    of the oversized device tested for case C0051. To quantify this
    morphological effect, we evaluated the \gls{ui}, which measures surface
    irregularity as defined by \citet{Raghavan2005}:
    \begin{equation}
        \mathrm{UI} = 1 - \frac{V}{V_{\mathrm{ch}}}
    \end{equation}
    where $V$ and $V_{\mathrm{ch}}$ correspond to the volume of the aneurysm
    sac and its convex hull, respectively (see \cref{tab:undulationIndices}).
    Cases C0052 and C0051 exhibited the highest \gls{ui} values across the
    sample, indicating that increased sac irregularity may explain a higher
    susceptibility to numerical wall penetration. Solving this issue requires
    local surface mesh refinement across high-curvature patches to prevent
    boundary violations without unnecessarily increasing the overall
    computational cost.

    \begin{table*}[!htb]
        \centering
        \caption{Undulation indices for the patient-specific geometries.}
        \begin{tabular}{lcccccccccc}
            \toprule
                    Case
                &   C0020 & C0033 & C0045 & C0046 & C0051
                &   C0052 & C0060 & C0066 & C0071 & C0078 \\
            \cmidrule(lr){1-11}
                    \gls{ui}
                &   0.022 & 0.034 & 0.042 & 0.049 & 0.100
                &   0.098 & 0.036 & 0.063 & 0.079 & 0.081 \\
            \bottomrule
        \end{tabular}
        \label{tab:undulationIndices}
    \end{table*}

    These situations, however, may also arise from the choice of the target
    \gls{web} geometry, which in this study was defined as a relaxed
    configuration broadly oriented along the chosen catheter path model.
    Consequently, the final configuration is highly dependent on the initial
    orientation and positioning of the target \gls{web}. This is what explains
    the different asymptotic \gls{avgSignedDistanceNB} between the
    sac-centerline and parent-artery catheter models, as, if the target
    geometry is not aligned to encompass certain irregular features of the sac,
    the device may not deform sufficiently to occupy those spaces. This
    behavior is not a limitation of the numerical algorithm itself, but rather
    a reflection of how the initial placement prior to expansion dictates the
    final conformation. This choice was made to provide a standardized target
    geometry that does not require case-by-case manual adjustments --- although
    case-by-case adjustments could be implemented to closely mirror clinical
    planning ---, and because the \gls{web} naturally tends to return to this
    pre-shaped state due to its structural stiffness and the trajectory imposed
    by the delivery catheter. Defining a target geometry that artificially
    encompasses the entire aneurysm sac to force the \gls{web} to completely
    adjust to the sac surface --- especially a lobulated and irregular one ---
    would be physically unrealistic.

    By utilizing a standardized relaxed geometry as the target, our model can
    also capture the potential for incomplete filling, providing
    interventionists with a more realistic prediction of the device's actual
    performance in complex-shaped aneurysms. For example, centering the
    proximal pole at the neck accurately predicted incomplete coverage of the
    lateral blebs in cases C0046 and C0060. Nevertheless, our approach offers
    parameters to adjust the target geometry to better fit the specific anatomy
    of each case if necessary, emulating the manual adjustments an
    interventionist would make during the procedure. This can be done by
    reorienting the target direction and/or, alternatively, to shift the device
    further along the catheter path model, by shifting the \gls{web} south pole
    along the catheter path, thus placing the device more inwards or outwards
    the aneurysm sac. The example shown for case C0052 demostrated how the
    \gls{web} clearly reduced the flow space around the bleb in the dome of
    that case. In this context, the parent-artery catheter model, although
    predicting a similar path within the aneurysm sac, follows the parent
    artery direction, thus being able to capture the approximate path a
    catheter follows when inserted into the aneurysm sac, providing eventually
    a more realistic orientation choice. This distinction is especially
    important in cases with significant parent vessel curvature, and also helps
    identify scenarios where a \gls{web} procedure might not be recommended due
    to severe vascular tortuosity \citep{Pierot2017}.

    Regarding oversized \gls{web} devices, while complete filling is not
    strictly necessary for treatment success \citep{Mouchtouris2022}, our
    results have shown that neck and body coverage did not undergo significant
    changes compared to the recommended device
    (\cref{fig:comparisonOversizedWEB}) while larger configurations promote
    better filling of the dome, as expected. Furthermore, even with oversized
    devices, \gls{avgSignedDistanceDome} was greater than
    \SI{0.45}{\milli\meter} internally, indicating that the \gls{web} did not
    fill the entire sac and, thus, did not occurring protrusion in any of the
    cases. Nonetheless, \citet{dazeo_intra-saccular_2021} noted that decreased
    height still enables a reduction in blood velocity and \gls{wss}.
    Furthermore, a lower height can prevent dome perforation by the
    interventionist during the procedure while still achieving neck occlusion.
    In this sense, devices that occupy the entire aneurysm may not be
    necessary, as sealing the neck remains the primary clinical objective.
    Nonetheless, \gls{web} selection based on aneurysm size remains a difficult
    task. On one hand, undersized devices lead to incomplete neck coverage with
    blood still flowing into the aneurysm sac, thus causing low occlusion rates
    and a higher risk of treatment failure \citep{Cagnazzo2023}. On the
    other hand, excessive oversizing can cause parent artery protrusion.
    Although the algorithm adapted in this work may not be ready to be used in
    clinical practice, its predictions offers a view of the unfilling spaces
    left by the \gls{web}. This could be useful for virtual predictions
    involving novel techniques such as the CUPCAKE technique, proposed by
    \citet{Vainauskaite2025}.

\subsection*{Limitations}

    In terms of modeling, an important limitation was to model the device using
    joined wires, which does not represent the actual braided structure where
    wires are intertwined and subject to relative displacement. Furthermore, as
    the points near the nodes are very close, this is possibly an explanation
    for our observation that the device geometry changes slightly at the poles,
    exhibiting a stiffer physical behavior. Therefore, future versions of this
    code should consider more realistic representation of the device wires,
    such as a true braided structure, to better capture the mechanical behavior
    of the device and its interaction with the aneurysm wall, as well as a more
    realistic deployment path.

    A significant limitation of this study is the lack of validation against
    clinical cases. To validate the employed mechanical model, a similar
    comparison with real imaging is necessary to evaluate the final apposition
    of the device. Although this validation would be mandatory for clinical
    application, the present study focused on evaluating the feasibility of the
    method and its potential especially to generate deployed \gls{web} models
    for \gls{cfd} simulations, with an improved deployed \gls{web} over
    geometry-based methods. In this sense, the present study provides a proof
    of concept for the use of this method to generate \gls{web} geometries for
    \gls{cfd} simulations, which can be used to evaluate the hemodynamic
    effects of the device on the aneurysm and surrounding vasculature.

\section{Conclusion}
\label{sec:conclusion}

    This study demonstrated the feasibility of adapting a physics-based
    stent-deployment algorithm for the virtual implantation of \gls{web}
    devices. Under optimal computational settings established by our parametric
    analysis --- a surface edge length of \SI{0.03}{\milli\meter} and a
    deployment tolerance of \SI{1e-4}{\milli\meter} --- the algorithm
    successfully predicted device apposition without wall penetration in nine of
    ten patient-specific geometries. For the remaining cases, shown to have
    highly irregular surfaces, the algorithm demonstrated that localized
    penetration could be successfully resolved by decreasing the edge length in
    these local regions and adjusting the positioning along the catheter model
    centerline, thereby achieving successful \gls{web} deployment for all cases
    across the three tested target orientations. Furthermore, the simulation of
    oversized devices showed that larger configurations significantly improve
    dome filling and neck-body apposition for three of four cases tested, with
    a reduction in the mean distance to the wall without necessarily causing
    parent artery protrusion, highlighting the algorithm's potential utility in
    surgical planning.

    By simulating the mechanical behavior of device expansion, this approach
    offers the advantage of providing a more realistic representation of the
    deployment compared to traditional geometry-based methods, being inherently
    predictive and, thus, helpful, especially, for hemodynamics investigations
    using numerical methods.

    \subsection*{Acknowledgments}

        This research was supported by grant 2024/19187-8 of
        \gls{fapesp} and grant Universal 406254/2025-7 of Brazil's National
        Council for Scientific and Technological Development (CNPq).

    \subsection*{Funding}

        This research was supported by grant 2024/19187-8 of
        \gls{fapesp} and grant Universal 406254/2025-7 of Brazil's National
        Council for Scientific and Technological Development (CNPq).

    \subsection*{Ethics Statement}

        Ethical approval was not required for this study, as it utilized public dataset repository.

    \subsection*{Conflicts of Interest}

        The authors declare that they have no known competing financial interests or personal relationships that could have appeared to influence the work reported in this paper.

    \subsection*{Data Availability Statement}

    The datasets analyzed during the current study are publicly available in the Aneurisk Repository under the link \url{https://github.com/iagolessa/AneuriskDatabase.git}

	\bibliographystyle{unsrtnat}
    \bibliography{bib/zotero-library,bib/WEB,bib/websites}

\end{document}